\documentclass[12pt]{article}
\usepackage{amsfonts,amssymb,graphicx,fullpage}
\usepackage[tight]{subfigure}
\usepackage{rotating,cancel}
\usepackage{sidecap}
\pdfoutput=1
\usepackage{ifpdf}
\ifpdf
  \usepackage{color}
  \definecolor{darkblue}{rgb}{0.3,0.3,0.6}
    \definecolor{darkgreen}{rgb}{0,0.6,0}
  \usepackage[pdftitle={Model Building on Z2xZ6'},pdfauthor={Gabriele Honecker and Wieland Staessens},pdfsubject={String theory},pdfkeywords={string theory},colorlinks=true,linkcolor=black,urlcolor=darkblue,filecolor=darkblue,citecolor=darkblue,menucolor=darkblue,breaklinks=true,bookmarks=true,anchorcolor=black,unicode=true]{hyperref}
\else
  
\fi
\usepackage[numbers,compress]{natbib}
\usepackage{a4wide}
\usepackage{longtable}
\usepackage{graphicx}
\usepackage{subfigure}
\usepackage[centertags]{amsmath}
\usepackage{multirow}
\usepackage{slashed}

\newcommand{\bCentering}{\centering}
\newcommand{\bCaption}{\caption}

\def\muc{\multicolumn}

\def\Z{\mathbb{Z}}

\def\ov{\overline}
\def\N{\mathbf{N}}
\def\Sym{\mathbf{Sym}}
\def\Anti{\mathbf{Anti}}
\def\Adj{\mathbf{Adj}}
\def\Tr{\text{Tr}}

\def\ov{\overline}
\def\1{{\bf 1}}
\def\2{{\bf 2}}
\def\3{{\bf 3}}
\def\4{{\bf 4}}
\def\6{{\bf 6}}
\def\OR{\Omega\mathcal{R}}

\def\targ#1#2{\genfrac{[}{]}{0pt}{}{#1}{#2}}

\def\targ2#1#2{\genfrac{}{}{0pt}{}{#1}{#2}}

\definecolor{blus}{rgb}{0.1,0.1,0.8}
\definecolor{GreenYellow}{cmyk}{0.15,0,0.69,0}
\definecolor{Yellow}{cmyk}{0,0,1,0}
\definecolor{Goldenrod}{cmyk}{0,0.10,0.84,0}
\definecolor{Dandelion}{cmyk}{0,0.29,0.84,0}
\definecolor{Apricot}{cmyk}{0,0.32,0.52,0}
\definecolor{Peach}{cmyk}{0,0.50,0.70,0}
\definecolor{Melon}{cmyk}{0,0.46,0.50,0}
\definecolor{YellowOrange}{cmyk}{0,0.42,1,0}
\definecolor{Orange}{cmyk}{0,0.61,0.87,0}
\definecolor{BurntOrange}{cmyk}{0,0.51,1,0}
\definecolor{Bittersweet}{cmyk}{0,0.75,1,0.24}
\definecolor{RedOrange}{cmyk}{0,0.77,0.87,0}
\definecolor{Mahogany}{cmyk}{0,0.85,0.87,0.35}
\definecolor{Maroon}{cmyk}{0,0.87,0.68,0.32}
\definecolor{BrickRed}{cmyk}{0,0.89,0.94,0.28}
\definecolor{Red}{cmyk}{0,1,1,0}
\definecolor{OrangeRed}{cmyk}{0,1,0.50,0}
\definecolor{RubineRed}{cmyk}{0,1,0.13,0}
\definecolor{WildStrawberry}{cmyk}{0,0.96,0.39,0}
\definecolor{Salmon}{cmyk}{0,0.53,0.38,0}
\definecolor{CarnationPink}{cmyk}{0,0.63,0,0}
\definecolor{Magenta}{cmyk}{0,1,0,0}
\definecolor{VioletRed}{cmyk}{0,0.81,0,0}
\definecolor{Rhodamine}{cmyk}{0,0.82,0,0}
\definecolor{Mulberry}{cmyk}{0.34,0.90,0,0.02}
\definecolor{RedViolet}{cmyk}{0.07,0.90,0,0.34}
\definecolor{Fuchsia}{cmyk}{0.47,0.91,0,0.08}
\definecolor{Lavender}{cmyk}{0,0.48,0,0}
\definecolor{Thistle}{cmyk}{0.12,0.59,0,0}
\definecolor{Orchid}{cmyk}{0.32,0.64,0,0}
\definecolor{DarkOrchid}{cmyk}{0.40,0.80,0.20,0}
\definecolor{Purple}{cmyk}{0.45,0.86,0,0}
\definecolor{Plum}{cmyk}{0.50,1,0,0}
\definecolor{Violet}{cmyk}{0.79,0.88,0,0}
\definecolor{RoyalPurple}{cmyk}{0.75,0.90,0,0}
\definecolor{BlueViolet}{cmyk}{0.86,0.91,0,0.04}
\definecolor{Periwinkle}{cmyk}{0.57,0.55,0,0}
\definecolor{CadetBlue}{cmyk}{0.62,0.57,0.23,0}
\definecolor{CornflowerBlue}{cmyk}{0.65,0.13,0,0}
\definecolor{MidnightBlue}{cmyk}{0.98,0.13,0,0.43}
\definecolor{NavyBlue}{cmyk}{0.94,0.54,0,0}
\definecolor{RoyalBlue}{cmyk}{1,0.50,0,0}
\definecolor{Blue}{cmyk}{1,1,0,0}
\definecolor{Cerulean}{cmyk}{0.94,0.11,0,0}
\definecolor{Cyan}{cmyk}{1,0,0,0}
\definecolor{ProcessBlue}{cmyk}{0.96,0,0,0}
\definecolor{SkyBlue}{cmyk}{0.62,0,0.12,0}
\definecolor{Turquoise}{cmyk}{0.85,0,0.20,0}
\definecolor{TealBlue}{cmyk}{0.86,0,0.34,0.02}
\definecolor{Aquamarine}{cmyk}{0.82,0,0.30,0}
\definecolor{BlueGreen}{cmyk}{0.85,0,0.33,0}
\definecolor{Emerald}{cmyk}{1,0,0.50,0}
\definecolor{JungleGreen}{cmyk}{0.99,0,0.52,0}
\definecolor{SeaGreen}{cmyk}{0.69,0,0.50,0}
\definecolor{Green}{cmyk}{1,0,1,0}
\definecolor{ForestGreen}{cmyk}{0.91,0,0.88,0.12}
\definecolor{PineGreen}{cmyk}{0.92,0,0.59,0.25}
\definecolor{LimeGreen}{cmyk}{0.50,0,1,0}
\definecolor{YellowGreen}{cmyk}{0.44,0,0.74,0}
\definecolor{SpringGreen}{cmyk}{0.26,0,0.76,0}
\definecolor{OliveGreen}{cmyk}{0.64,0,0.95,0.40}
\definecolor{RawSienna}{cmyk}{0,0.72,1,0.45}
\definecolor{Sepia}{cmyk}{0,0.83,1,0.70}
\definecolor{Brown}{cmyk}{0,0.81,1,0.60}
\definecolor{Tan}{cmyk}{0.14,0.42,0.56,0}
\definecolor{Gray}{cmyk}{0,0,0,0.50}
\definecolor{Black}{cmyk}{0,0,0,1}
\definecolor{White}{cmyk}{0,0,0,0}

\definecolor{mygr}{rgb}{0,0.6,0}
\definecolor{mygrey}{rgb}{0,0.1,0.2}
\definecolor{myblue}{rgb}{0,0.5,0.9}
\definecolor{myblue2}{rgb}{0,0.5,0.5}
\definecolor{myorange}{rgb}{1,0.5,0}
\definecolor{mypurple}{rgb}{0.6,0,1}
\definecolor{mygolden}{rgb}{1,0.8,0.2}

\newcommand{\bCaptionfonts}{\small}
\makeatletter
\long\def\@makecaption#1#2{%
  \vskip\abovecaptionskip
  \sbox\@tempboxa{{\bCaptionfonts #1: #2}}%
  \ifdim \wd\@tempboxa >\hsize
    {\bCaptionfonts #1: #2\par}
  \else
    \hbox to\hsize{\hfil\box\@tempboxa\hfil}%
  \fi
  \vskip\belowcaptionskip}
\makeatother

\makeatletter
\let\ORIGINALlatex@openbib@code=\@openbib@code
\renewcommand{\@openbib@code}{\ORIGINALlatex@openbib@code\setlength{\itemsep}{1ex plus.5ex minus.5ex}\setlength{\parsep}{0pt}}
\makeatother

\def\mathtab#1#2#3{\begin{table}[th]\bCentering$#1$\bCaption{#3}\label{tab:#2}\end{table}}
\def\mathtabfix#1#2#3{\begin{table}[th]\bCentering\resizebox{\linewidth}{!}{$#1$}\bCaption{#3}\label{tab:#2}\end{table}}

\allowdisplaybreaks[4]

\begin{document}
\begin{center}
\begin{flushright}
{\small MITP/13-082\\ 
\today}

\end{flushright}

\vspace{25mm}
{\Large\bf On axionic dark matter in Type IIA string theory}

\vspace{12mm}
{\large Gabriele Honecker${}^{\heartsuit}$ and Wieland Staessens${}^{\spadesuit}$
}

\vspace{8mm}
{
\it PRISMA Cluster of Excellence \&
Institut f\"ur Physik  (WA THEP), Johannes-Gutenberg-Universit\"at, D-55099 Mainz, Germany
\;$^{\heartsuit}${\tt Gabriele.Honecker@uni-mainz.de},~$^{\spadesuit}${\tt wieland.staessens@uni-mainz.de}}

\vspace{24mm}{\bf Abstract}\\[2ex]\parbox{140mm}{
We investigate viable scenarios with various axions in the context of supersymmetric field theory and in globally consistent D-brane models.
The Peccei-Quinn symmetry is associated with an anomalous $U(1)$ symmetry, which acquires mass at the string scale but remains as a perturbative global symmetry at low energies.
The origin of the scalar Higgs-axion potential from F-, D- and soft breaking terms is derived, and two Standard Model examples of global intersecting D6-brane models in Type II orientifolds are presented, which
differ in the realisation of the Higgs sector and in the hidden sector, the latter of which is of particluar importance for the soft supersymmetry breaking terms.    
}
\end{center}

\thispagestyle{empty}
\clearpage 

\tableofcontents

\setlength{\parskip}{1em plus1ex minus.5ex}
\section{Introduction}\label{S:intro}

The nature of dark matter and dark energy remains one of the biggest puzzles of modern physics. Collider searches as well as direct and indirect detection experiments rule out more and more scenarios Beyond the Standard Model. Most of these searches rely on the requirement that new particles have sizable cross-sections involving Standard Model gauge interactions. 
String theory on the other hand contains a variety of particles without such gauge interactions, including in particular the neutral axionic pseudo-scalars in the closed string sector that complexify  the string coupling and compactification moduli. On the other hand, axions carrying some charge under a global $U(1)_{PQ}$ symmetry were proposed in 1977 to solve the strong CP-problem~\cite{Peccei:1977hh,Peccei:1977ur,Weinberg:1977ma,Wilczek:1977pj}, and experimental searches for axions have been intensified recently, see e.g.~\cite{Vogel:2013bta,Bahre:2013ywa,Baker:2013zta,Graham:2013gfa,Irastorza:2013kda}. Over the past decades, a combination of astrophysical observations, laboratory experiments and cosmological considerations,  shrunk the parameter space for the  axion decay constant $f_{\alpha}$ to the region $10^9$ GeV $< f_{\alpha}< $ $10^{12}$ GeV, the so-called `axion window'. The lower bound of the axion window is set by the stellar evolution of red giants, white dwarfs and hot neutron stars. And the higher bound follows from cosmological considerations, when the axion is treated as a dark matter candidate. Current and future experiments are designed to probe (parts of) the axion window, see e.g.~\cite{Ringwald:2013via} for an up-to-date overview.

While global continuous symmetries have been argued to be inconsistent with gravity~\cite{Abbott:1989jw,Coleman:1989zu,Kallosh:1995hi,Banks:2010zn,Banks:1988yz,Hellerman:2010fv}, string theory generically contains Abelian $U(1)_{\text{massive}}$ gauge symmetries whose anomalies are cancelled by the generalised Green-Schwarz mechanism, which also produces string scale mass terms for  the $U(1)_{\text{massive}}$'s. Within the context of Type II superstring theories, such massive gauge symmetries and associated charged states arise in the open string sector, see e.g.~\cite{Blumenhagen:2006ci,Ibanez:2012zz} for reviews. $U(1)_{\text{massive}}$ remains as a perturbative global symmetry at low energies, which is broken by non-perturbative effects such as D-brane instantons~\cite{Blumenhagen:2009qh}. The Green-Schwarz mechanism associates a closed string axion with the longitudinal mode of an open string vector. It is thus natural to combine axions from the closed and open string sector such that one axion solves the strong CP problem, while others account for the dark sector of the universe, see e.g.~\cite{Kim:2009cp,Kim:2013pja}. In contrast to compactifications of the heterotic string,  within Type II string theory different energy scales can easily be decoupled due to the affiliation of gravity to the closed string sector and gauge symmetries to the open string sector.

There exist various supersymmetric or string inspired versions of field theoretic axion models in the literature, see e.g.~\cite{Coriano:2008xa,Higaki:2011bz,Arias:2012az,Chatzistavrakidis:2012bb,Bae:2013hma} and references therein, and axions have been discussed before in the context of the heterotic string, see e.g.~\cite{Choi:2009jt}. Within the Type IIB string theory context, previous work in the LARGE volume scenario concentrated on closed string axions, see e.g.~\cite{Cicoli:2012aq,Higaki:2012ar,Conlon:2013txa,Higaki:2013lra,Angus:2013zfa,Higaki:2013qka,Higaki:2011me,Gao:2013rra}, and the idea of open string axions in intersecting D-brane inspired scenarios has been put foward recently in~\cite{Berenstein:2012eg}, while to our knowledge little is known about the combined effect of closed and open string axions in the context of {\it globally consistent} D-brane configurations with the (supersymmetric) Standard Model or GUT spectrum.  

This article aims at closing this apparent gap by first discussing the appearance of various types of axion within field theory and Type II string theory. We then proceed to investigate two globally consistent D6-brane models with Standard Model spectrum in detail in view of their Higgs-axion potentials and possibilities of supersymmetry breaking. The D-brane set-up and massless spectrum of the first model on $T^6/\Z_6$ has been constructed in~\cite{Honecker:2004kb,Honecker:2004np,Gmeiner:2009fb} and contains a `hidden' $USp(2)$ gauge factor, while the set-up and spectrum of the second model on $T^6/\Z_6'$ with hidden $USp(6)$ gauge group have been presented in~\cite{Gmeiner:2008xq,Gmeiner:2009fb} with a discussion on superpotential couplings in~\cite{Honecker:2012jd,Honecker:2012fn}. 
Throughout the present work, particular attention will be paid to the occurrence of different energy scales such as the Peccei-Quinn, electroweak and supersymmetry breaking scales in dependence of their prevailing origin from the closed or open string sector.

This article is organised as follows: in section~\ref{S:Axions-FT} we briefly review field theoretical axion models and provide a Type II string theory motivated extension to their supersymmetric version. The origin of couplings within global supersymmetry including soft supersymmetry breaking terms is discussed. Section~\ref{S:Axions-STRINGS} contains a discussion of the various types of closed and open string axions in Type IIA string theory and associated massive Abelian gauge symmetries. The concept is illustrated by two examples of Type IIA orientifold compactifications on $T^6/\Z_6$ and $T^6/\Z_6'$ with supersymmetric Standard Model spectrum, and the possibility to break supersymmetry by a hidden sector gaugino condensate is addressed. Section~\ref{S:Conclusions} contains our conclusions. Technical details on the four-dimensional field theory are collected in appendices~\ref{A:ChiralRotation} and~\ref{A:Integration}, and appendix~\ref{A:Spectra} contains the full light matter spectrum of the two examples.

\section{Axions in Field Theory}\label{S:Axions-FT}

In this section, we first briefly review the well-known DFSZ axion model and then proceed to discuss a supersymmetric D-brane inspired extension as well as the origin of soft supersymmetry breaking terms in the scalar Higgs-axion potential.

\subsection{The DFSZ model}\label{Ss:DFSZ}

In 1977, Peccei and Quinn proposed the existence of a spontaneously broken global $U(1)_{PQ}$ symmetry to solve the strong CP-problem~\cite{Peccei:1977hh,Peccei:1977ur}. 
The associated pseudo Nambu-Goldstone boson in the original model consists of an axion that arises from two Higgs doublets~\cite{Peccei:1977hh,Peccei:1977ur,Weinberg:1977ma,Wilczek:1977pj}.
In this original PQWW  axion model, the breaking scales of the electro-weak and  global $U(1)_{PQ}$ symmetry coincide, and consequently non-negligible contributions of axions
 to  hadronic decay products involving heavy quarks in the initial state are expected. Due to non-observation of such effects, the most simple PQWW axion model is ruled out to date, see e.g.~\cite{Zavattini:2005tm,Ehret:2010mh,Barth:2013sma}
 and references therein.
  
 The breaking scale of the global $U(1)_{PQ}$ symmetry can be decoupled from the electroweak scale by introducing an additional scalar field, which is only charged under 
 $U(1)_{PQ}$, but neutral under the Standard Model group $SU(3) \times SU(2)_L \times U(1)_Y$. One distinctive model by Kim, Schifman, Vainshtein and Zakharov (KSVZ)
 uses a vector-like heavy quark pair to which this neutral scalar couples~\cite{Kim:1979if,Shifman:1979if}.   In contrast to the original PQWW model, the KSVZ model contains 
 only one Higgs doublet, and therefore we do not expect any immediately obvious  generalisation to  supersymmetric field theory as low-energy limit of some Type II string theory compactification. 
 
 Another prominent axion model by Dine, Fischler, Sredenicki and Zhitnitsky (DFSZ) contains two Higgs doublets as well as a  Standard Model-singlet complex scalar field~\cite{Dine:1981rt,Zhitnitsky:1980tq}.
Since this model can be generalised to supersymmetric field theory in a natural way, we give here a brief review of the distinctive features such as the scalar potential, axion decay 
constant and axion mass. The two Higgs doublets $(H_u,H_d)$ schematically couple to quarks and leptons via
 \begin{eqnarray}\label{Eq:Yukawas}
{\cal L}_{\text{Yukawa}} = f_u\, Q_L\cdot H_u\, u_R + f_d\,  Q_L \cdot H_d\, d_R + f_e \, L \cdot H_d  \, e_R + f_{\nu} \, L \cdot H_u \, \nu_R
,
\end{eqnarray}
and the coupling to the complex scalar $SU(2)_L \times U(1)_Y$ singlet field $\tilde{\sigma}$ is contained in the scalar Higgs-axion potential,
\begin{eqnarray}
V_{\text{DFSZ}}(H_u, H_d, \tilde{\sigma}) &=&  \lambda_u (H_u^\dagger H_u -  \frac{v_u^2}{2} )^2 + \lambda_d (H_d^\dagger H_d -  \frac{v_d^2}{2} )^2 + \lambda_{\tilde{\sigma}} (\tilde{\sigma}^{\dagger} \tilde{\sigma} -  \frac{v_{\tilde{\sigma}}^2}{2})^2 \nonumber \\
&&+ (a\, H_u^\dagger H_u + b\, H_d^\dagger H_d )\, \tilde{\sigma}^{\dagger} \tilde{\sigma} +  ( c\, H_u \cdot H_d\, \tilde{\sigma}^2 + h.c. ) \label{Eq:DFSZpotential} \\
&& + d\, | H_u \cdot H_d |^2 + e\, |  H_u^\dagger H_d |^2, \nonumber
\end{eqnarray}
where the abbreviation $H_u \cdot H_d = H_u^i \epsilon_{ij} H_d^j = h_u^+ h_d^- - h_u^0 h_d^0$  has been used as well as the standard decomposition of the Higgs doublets into
charged and uncharged components and {\it vev}s,
\begin{equation}\label{Eq:Def-Higgses+vevs}
H_u = \left( \begin{array}{c} h_u^+ \\  h^0_u \end{array} \right), \qquad H_d = \left( \begin{array}{c}  h_d^0 \\ h^-_d \end{array} \right)
,
\qquad \langle h^0_u \rangle = \frac{v_u}{\sqrt{2}}
,
\qquad \langle h^0_d \rangle =  \frac{v_d}{\sqrt{2}}
.
\end{equation}
$\lambda_u, \lambda_d,\lambda_{\tilde{\sigma}}, a \ldots e$ in equation~(\ref{Eq:DFSZpotential}) denote all possible four-point couplings. 
Due to the term $(c\,  H_u \cdot H_d\, \tilde{\sigma}^2 + h.c. )$, also the Higgs doublets have to be charged under the global $U(1)_{PQ}$ symmetry if the scalar $SU(2)_L \times U(1)_Y$ singlet $\tilde{\sigma}$ transforms non-trivially,
\begin{eqnarray}
\tilde{\sigma} \to e^{i\, q_{\tilde{\sigma} \alpha}} \tilde{\sigma},\quad  H_u \to e^{i\, q_u \alpha} H_u,\quad  H_d \to e^{i\, q_d \alpha} H_d
\qquad 
\text{with}
\qquad
q_u+q_d = - 2 q_{\tilde{\sigma}}
.
\end{eqnarray}
From the Yukawa couplings~(\ref{Eq:Yukawas}), one can then also read off relations among the transformations of the quarks and leptons and Higgs bosons under $U(1)_{PQ}$,
\begin{eqnarray}
\left.\begin{array}{r}
Q_L \to e^{i\, q_Q \alpha} Q_L, \quad 
u_R \to e^{i\, \tilde{q}_u \alpha } u_R, \quad 
d_R \to e^{i\, \tilde{q}_d \alpha } d_R
\\
L  \to e^{i\, q_L \alpha} L, \quad 
e_R \to e^{i\, q_e \alpha } e_R, \quad 
\nu_R  \to e^{i\, q_{\nu} \alpha } \nu_R
\end{array}\right\}
\text{ with }
\left\{\begin{array}{c}
q_Q + \tilde{q}_u = L + q_{\nu} = -  q_u
\\
q_Q + \tilde{q}_d = L + q_e = -  q_d
\end{array}\right.
.
\end{eqnarray}
There are two inequivalent consistent choices of charge assignments with either left-handed quarks $Q_L$ uncharged and right-handed quarks $(u_R,d_R)$ charged or vice versa.
The discussion for leptons is completely analogous. In both cases, the (up to overall sign flip) unique choice is $q_u=q_d=1$ for the $U(1)_{PQ}$ charge of the Higgses.
Since in stringy D-brane models such as the examples below in section~\ref{S:Axions-STRINGS}, 
an anomalous and massive $U(1) \subset U(2)_L$  gauge symmetry constitutes a natural candidate for $U(1)_{PQ}$, we summarise 
the associated second choice of charge assignments with neutral right-handed particles in table~\ref{Tab:PQchargesSUSY}.
\begin{table}[h]
\begin{center}
\begin{equation*}
\begin{array}{|c||c|c|c||c|c|c||c|c||c||c|}
\hline
\muc{11}{|c|}{\text{\bf Charge assignments $q_{\text{particle}}$ under } U(1)_{PQ}}
\\\hline\hline
\text{Matter} & Q_L & u_R & d_R & L &  e_R &  \nu_R & H_u & H_d & \tilde{\sigma} & \Sigma \\ 
\hline
U(1)_{PQ} & - 1  & 0  & 0  & - 1 & 0 &  0  &  1 &  1 & - 1 & - 2 \\
\hline
\end{array}
\end{equation*}
\caption{Standard Model particles and axion field  $U(1)_{PQ}$  charge assignment for the choice of uncharged right-handed particles.
$\tilde{\sigma}$ refers to the axion in the original DFSZ model of equation~(\protect\ref{Eq:DFSZpotential}), $\Sigma$ to the axion superfield in the supersymmetric D-brane inspired DFSZ model
of section~\protect\ref{Ss:SUSY-DFSZ}.
\label{Tab:PQchargesSUSY}}
\end{center}
\end{table}

The coupling constants in equation~(\ref{Eq:DFSZpotential}) are constrained by measurements as follows:
\begin{itemize}
\item
The experimentally observed value of the $\rho\equiv  \frac{M_W^2}{M_Z^2 \cos^2 \theta_W}$ parameter is close to one.
In the Standard Model, this is true at tree-level, whereas in D-brane inspired models, at tree-level $\rho$ can deviate from one, see e.g.~\cite{Coriano':2005js},  
 in which case the scale at which new physics appears is firmly constrained, unless higher order or non-perturbative effects become important.
\item
The axion remains invisible at low energies if $U(1)_{PQ}$ is broken at a much higher scale than the electro-weak symmetry, implying a hierarchy of {\it vev}s,
\begin{eqnarray}\label{Eq:DFSZ-sigma_hierarchy}
\langle \tilde{\sigma}  \rangle \equiv v_{\tilde{\sigma}} \gg \sqrt{v_u^2 + v_d^2}
.
\end{eqnarray}
\end{itemize}
The physical axion $\alpha$ is generically a mixture of the argument of the complex scalar $\tilde{\sigma}$ and the neutral CP-odd Higgs bosons. However, imposing the 
hierarchy~(\ref{Eq:DFSZ-sigma_hierarchy}) results in the physical axion stemming primarily from the singlet $\tilde{\sigma}$.
 The axion mass is determined using Bardeen-Tye methods~\cite{Bardeen:1977bd},
\begin{eqnarray}
m^2_{\alpha} = \frac{f_\pi^2}{f_{\alpha}^2} m_\pi^2 N^2 \frac{m_u m_d}{(m_d+m_u)^2} \sim \left(74 \text{ keV } \, \frac{250 \text{ GeV }}{f_{\alpha}}\right)^2,
\end{eqnarray}
with $N$ the number of quark doublets, $m_u,m_d, m_{\pi}$ the masses of up- and down-type quarks and pions and $f_{\pi}$ the pion decay constant. 
The coupling of the axion $\alpha$ to matter is determined by the axion decay constant $f_{\alpha}$, which also sets the strength of 
the axion coupling to gluons,
\begin{eqnarray}\label{Eq:StandardActionAxion}
{\cal L}_{\alpha} \supset \frac{1}{2} \left(\partial_\mu \alpha \right) \left(\partial^\mu \alpha \right) -  \frac{1}{32 \pi^2} \frac{\alpha(x)}{f_{\alpha}}\, \Tr ( G_{ \mu \nu} \tilde G^{\mu \nu}) 
.
\end{eqnarray}
With respect to the PQWW axion model (for which the axion decay constant $f_{\alpha}$ is equal to $\sqrt{v_u^2 + v_d^2}$), the axion-gluon coupling as well as the couplings to ordinary matter are
 suppressed by a factor $r=\sqrt{v_u^2 + v_d^2}/f_{\alpha}$, implying that the production of axions is reduced by a factor $r^2$. For this reason, the DFSZ axion has been dubbed an `invisible axion'.

\subsection{Supersymmetrising the DFSZ model}\label{Ss:SUSY-DFSZ}

Since the DFSZ axion model contains two Higgs doublets in conjugate representations, it can easily be promoted to supersymmetric field theory. The  axion and Higgs scalar potential 
can then be decomposed into three different components,
\begin{eqnarray}\label{Eq:DFSZSusy3parts}
V_{DFSZ} = V_F + V_D +  V_{\text{soft}}.
\end{eqnarray}
One sublety arises concerning the axion-Higgs coupling in the second line of equation~(\ref{Eq:DFSZpotential}). 
On the one hand, any four-point coupling is non-renormalisable and thus suppressed by the cut-off scale, which might be $M_{\text{Planck}}$ in field theoretical models or 
 $M_{\text{string}}$ in string models. Such suppression has been related to the smallness of the $\mu$-term e.g. in~\cite{Kim:1983dt}.
On the other hand, a renormalisable  three-point coupling can be engineered if the axion $\tilde{\sigma}$ is replaced by a chiral superfield $\Sigma$ containing the axionic scalar $\sigma$ with twice the charge,
\begin{eqnarray}\label{Eq:Sigma-charge}
\Sigma \to e^{i\, q_\Sigma \alpha} \, \Sigma
\qquad 
\text{with}
\qquad 
q_u + q_d = - q_{\Sigma}
,
\end{eqnarray}
such that $V_{\text{SUSY-DFSZ}} \supset H_u \cdot H_d \, \sigma$.
This choice is motivated by D-brane models as follows: since in Type II string theory all charged particles arise from open strings, their gauge representations are defined by their two end points
to be of the type $\{(\N_a,\ov{\N}_b), (\Adj_a)\}$, plus in the presence of O-planes $\{(\N_a,\N_b),(\Anti_a), (\Sym_a)\}$, or some hermitian conjugate thereof under $U(N_a) \times U(N_b)$. In particular, the 
Higgs fields have one endpoint on a stack of D-branes supporting either a $U(2)_L$ or $USp(2)_L$ gauge group, and the other endpoint lies on a single D-brane with $U(1)_c$ gauge group\footnote{
In many models,  $U(1)_c$ arises from a spontanous breaking of a right-symmetric group $USp(2)_R$ or $SO(2)_R$.}, which contributes to the hypercharge $U(1)_Y$. Demanding that both the Higgs field 
and some $SU(3) \times SU(2)_L \times U(1)_Y$ singlet carry charge under an anomalous and massive gauge symmetry boils down to two options:
\begin{enumerate}
\item
The Peccei-Quinn symmetry is identified with the anomalous and massive \linebreak \mbox{$U(1)_b \subset U(2)_b \simeq SU(2)_L \times U(1)_{b}$} symmetry, and the Standard Model singlet with $U(1)_b$ charge 
arises as the antisymmetric  representation $(\Anti_b)$ of $U(2)_b$ or its conjugate. This non-Abelian singlet obeys the charge assignment of $\Sigma$ in table~\ref{Tab:PQchargesSUSY} under the 
standard decomposition of non-Abelian/Abelian representations (see e.g.~\cite{Ibanez:2012zz}) 
\begin{equation}
U(N) \simeq SU(N)_{ \times U(1)}
\qquad
\Rightarrow
\qquad
\left\{ \begin{array}{c}
(\N_{U(N)}) \simeq (\N_{SU(N)})_1 \\  (\Anti_{U(N)}) \simeq  (\Anti_{SU(N)})_2
\end{array}\right.
.
\end{equation}
This choice of $U(1)_{PQ}$ thus leads to a three-point coupling among $SU(3) \times SU(2)_L \times U(1)_Y \times U(1)_{PQ}$ charged states.
\item
The $U(1)_{PQ}$ symmetry involves some massive linear combination of Abelian gauge factors including $U(1)_c$. This case can only occur if $U(1)_c$ cannot 
be continuously connected to any right-symmetric group $USp(2)_R$ or $SO(2)_R$ since such a connection enforces $U(1)_c$ to be a  massless and anomaly-free gauge symmetry
by itself. There are again two different options:
\begin{enumerate}
\item
$U(1)_{PQ}$ contains the combination $U(1)_c-U(1)_d$, which is orthogonal to the massless hypercharge defined below in equation~(\ref{Eq:Def-Y+BL}).
In this case, the axion can be identified with the scalar superpartner $\tilde{\nu}_R$ of the neutrino, and it again obeys the charge assignment of $\Sigma$ in equation~(\ref{Eq:Sigma-charge}).
\item
If $U(1)_{PQ}$ contains some combination of $U(1)_c$ with an additional $U(1)_e$ gauge factor  of a D-brane, that is not required to obtain the Standard Model chiral spectrum, yet is a part of the definition for the hypercharge, the axion can arise from strings stretched between D-branes $c$ and $e$. In this case, the axion carries charge $\pm 1$ in accordance with  $\tilde{\sigma}$ in table~\ref{Tab:PQchargesSUSY}.
 \end{enumerate}
\end{enumerate}

In section~\ref{Ss:Z2N},  two supersymmetric sample  DFSZ models are presented, which display typical features in the context of intersecting D6-brane model building. In both cases, the models are intrinsically left-right symmetric with a spontaneously broken right-symmetric gauge group $USp(2)_c \to U(1)_c$. Hence,  in the remainder of the article we will focus on case 1  with $U(1)_b \simeq U(1)_{PQ}$ and the r\^ole of the singlet $\Sigma$ fulfilled by a chiral supermultiplet in the antisymmetric representation of $U(2)_b$. 

After choosing this configuration, the next step consists in mapping the various terms in the DFSZ potential from equation~(\ref{Eq:DFSZpotential}) to a specific  origin in supersymmetric field theory based on the decomposition into F-terms, D-terms and soft supersymmetry breaking terms as in equation~(\ref{Eq:DFSZSusy3parts}):
\begin{enumerate}
\item \underline{F-term potential:}
The terms of the form $ H_u^\dagger H_u |\sigma|^2 $, $H_d^\dagger H_d |\sigma|^2$ and $ | H_u \cdot H_d |^2$ all arise from a single superpotential term of the form:
\begin{equation}\label{Eq:SusyCouplingHiggsSinglet}
{\cal W}_{\text{DFSZ}} = \mu \; H_u \cdot H_d \; \Sigma,
\end{equation}
where the various coupling constants are now unified, i.e.~$a=b=d=|\mu|^2$, due to supersymmetry. Note that this cubic coupling already requires the Higgses to be charged under the Peccei-Quinn symmetry
if the Standard Model singlet superfield $\Sigma$ with scalar component $\sigma$ transforms non-trivially under  $U(1)_{PQ}$.
\item \underline{D-term potential:}
The quartic terms $ (H_u^\dagger H_u)^2$, $(H_d^\dagger H_d)^2$ and $|  H_u^\dagger H_d |^2$ arise from the K\"ahler potential $K^{\text{SUSY}}$, and more explicitly from the gauge-invariant coupling of the chiral Higgs  superfields to the vector superfields associated with the $SU(2)_L\times U(1)_Y$ gauge symmetry, with takes the following, generic form in global supersymmetry: 
\begin{equation}
K^{\text{SUSY}} (\Phi^\dagger e^{2g V} \Phi) = \Phi^\dagger e^{2g V} \Phi,
\end{equation}
where $\Phi$ represents a generic chiral superfield and $V$ a vector superfield.
The $U(1)_b \subset U(2)_b$ Peccei-Quinn symmetry originates from a massive gauge boson in string theory, as explained below in detail in section~\ref{Sss:GSMechanismOpenStringAxions}. This implies that the K\"ahler potential  $K^{\text{SUSY}}$ will also contain gauge-invariant couplings between the vector multiplet associated to $U(1)_b$ and the matter multiplets charged under the symmetry. In this way, also the quartic term $|\sigma|^4$ will be generated in the D-term potential. 
\item \underline{Soft supersymmetry breaking terms:}
The remaining terms have (in field theory) to be added by hand as soft supersymmetry breaking terms that are invariant under the Standard Model gauge group and the Peccei-Quinn symmetry, such as the mass terms $v_u^2 \, H_u^\dagger H_u$, $v_d^2 \, H_d^\dagger H_d$ and $v_{\sigma}^2 \, |\sigma|^2$. The cubic coupling $H_u\cdot H_d\; \sigma$ can be realised through trilinear ${\cal A}$-terms. The soft supersymmetry breaking terms can be written in a manifestly supersymmetric way through the introduction of a spurion superfield $\eta$. The trilinear ${\cal A}$-term can then be captured by the superpotential: 
\begin{equation}
{\cal W}_{\text{soft}} = \eta\, c H_u\cdot H_d\, \Sigma .
\end{equation}
The soft supersymmetry breaking mass terms on the other hand can be generated through a K\"ahler potential of the following form:
\begin{equation}
K_{\text{soft}} = \eta \ov \eta\; m^2_{\Phi} \, \Phi^\dagger e^{2 g  V} \Phi.
\end{equation}
In section~\ref{Ss:Soft-SUSY} we briefly discuss how these soft supersymmetry breaking terms can arise through gravity mediation of spontaneous supersymmetry breaking in a hidden sector. 
\end{enumerate}
%

\subsection{Gravity mediation and gaugino condensation}\label{Ss:Soft-SUSY}

For realistic globally supersymmetric models, soft supersymmetry breaking terms form a hands-on way to break supersymmetry without spoiling its abilities to solve the hierarchy problem and the related naturalness problem concerning the Higgs mass in the Standard Model. One way to generate these soft supersymmetry breaking terms is by coupling the supersymmetric field theory to gravity and allowing gravity to mediate spontaneous supersymmetry breaking in a hidden sector to the visible sector. For a Type II orientifold compactification, the low energy effective field theory of the massless string modes reduces to an ${\cal N}=1$ supergravity theory, so that gravity mediated supersymmetry breaking appears as a natural way to generate the soft supersymmetry breaking terms for the supersymmetrized DFSZ model.   

Coupling the supersymmetrised DFSZ model to gravity leads to an ${\cal N} = 1$ supergravity theory with chiral multiplets and vector multiplets, whose bosonic sector is characterised by the following generic action, cf. e.g.~\cite{Blumenhagen:2006ci,Cerdeno:1998hs}: 
\begin{eqnarray}
{\cal S} &=& - \frac{1}{2 } \int  \left[ M_{\text{Planck}}^2 R^{(4)} \ast {\bf 1} + 2 \, {\cal K}_{M \ov N} \; dX^M \wedge \ast d X^{\ov N} + \frac{1}{2} ({\rm Re}f)_{ab}\,  {\cal F}^a\wedge \ast {\cal F}^{b}\right. \nonumber \\
&& \hspace{20mm} \left.  -\frac{i}{2} ({\rm Im}f)_{ab} \, {\cal F}^a\wedge {\cal F}^{b}   + 2 \, (V_F + V_D) \ast {\bf 1} \vphantom{R^{(4)} \ast {\bf 1}}\right],
\end{eqnarray}
where $X^M$ denote the bosonic components of the chiral multiplets, ${\cal F}^a$ the field strength associated to a gauge group $G_a$ and $f_{ab}$ the holomorphic gauge kinetic function. The F-term scalar potential $V_F$ can be expressed in terms of the supergravity K\"ahler potential ${\cal K}^\text{SUGRA} \equiv {\cal K}$ and its derivatives, the K\"ahler metrics ${\cal K}_{ M \ov N},$  and the superpotential ${\cal W}$:
\begin{equation}\label{Eq:FtermSugra}
V_F = e^{{\cal K}/M_{\text{Planck}}^2 }\left( {\cal K}^{ M \ov N} D_M {\cal W}\, D_{\ov N}  \ov{\cal W} - 3 M_{\text{Planck}}^{-2} |{\cal W}|^2 \right) , \hspace{0.4in} D_M {\cal W} \equiv \partial_M {\cal W} + M_{\text{Planck}}^{-2} (\partial_M {\cal K})\; {\cal W},
\end{equation} 
while the D-term scalar potential $V_D$ is given by:
\begin{equation}
V_D = \frac{1}{2} ({\rm Re}f^{-1})^{ab} \; {\cal D}_a {\cal D}_b,
\end{equation}
where ${\cal D}_a$ represent the auxiliary fields in the vector multiplets. An explicit expression for ${\cal D}_a$ is not given here, as they will play no r\^ole in the following. 

From a low energy perspective, the massless chiral multiplets can be split up into a set of observable chiral matter ($C^\alpha$) charged under the Standard Model group and a set of hidden matter (${\cal H}^m$). 
This decomposition allows for an expansion of the K\"ahler potential and superpotential about the stabilised vacua for the hidden matter fields:
\begin{equation}
\begin{array}{ccl}
{\cal K}(C^\alpha, C^{\dagger \ov \alpha}, {\cal H}^m, {\cal H}^{\dagger \ov m} ) &= & {\cal K}^{(0)} ({\cal H}^m, {\cal H}^{\dagger \ov m} ) + \tilde {\cal K}_{\alpha \ov\beta}  ({\cal H}^m, {\cal H}^{\dagger \ov m} ) \; C^\alpha C^{\dagger \ov \beta} \\
&& 
+ \left[  \frac{1}{2}  Z_{\alpha \beta} ({\cal H}^m, {\cal H}^{\dagger \ov m} )  \; C^\alpha C^\beta + h.c. \right] + \ldots, \\
{\cal W} (C^\alpha, {\cal H}^m) &= & {\cal W}_0 (h^m) + \frac{1}{2}  \mu_{\alpha \beta}({\cal H}^m) C^\alpha C^\beta + \frac{1}{6} Y_{\alpha \beta \gamma} C^\alpha C^\beta C^\gamma + \ldots.
\end{array} \label{Eq:SUGRAEffPot}
\end{equation}
By inserting these expansions into the F-term scalar potential of equation (\ref{Eq:FtermSugra}) and replacing the hidden scalars by their vacuum expectation values, one obtains an effective field theory for the visible sector, which in the flat limit (i.e.~$M_{\rm Planck} \rightarrow \infty$ while keeping the gravitino mass $m_{3/2}$ fixed) reduces to a globally supersymmetric theory with soft supersymmetry-breaking terms~\cite{Kaplunovsky:1993rd,Brignole:1997dp} of the following form for the scalar fields:
 \begin{equation}
V_{\rm soft} = m_{\alpha \ov \beta}^2 C^\alpha C^{\dagger \ov \beta} + \left[ \frac{1}{6} A_{\alpha \beta \gamma} C^\alpha C^\beta C^\gamma + \frac{1}{2} B_{\alpha \beta} C^\alpha C^\beta + h.c \right], 
\end{equation} 
with soft supersymmetry-breaking parameters given by:
\begin{equation}
\begin{array}{ccl}
 m_{\alpha \ov \beta}^2 & = & (m_{3/2}^2 + V_0) \tilde{\cal K}_{\alpha \ov \beta}  - \ov F^{\ov m} \left( \partial_{\ov m} \partial_n \tilde{\cal K}_{\ov \alpha \beta} - \partial_{\ov m} \tilde{\cal K}_{\ov \alpha \gamma} \tilde{\cal K}^{\gamma \ov \delta} \partial_n \tilde{\cal K}_{\ov \delta \beta}  \right) F^n  \\
 A_{\alpha \beta \gamma}  & = & \frac{\ov{\cal W}_0}{|{\cal W}_0|} e^{{\cal K}^{(0)}/2M_{\text{Planck}}^2} F^m \left[ \partial_m  {\cal K}^{(0)}\,  Y_{\alpha \beta \gamma} + \partial_m Y_{\alpha \beta \gamma} - \left( \tilde{\cal K}^{\delta \ov \rho} \partial_m \tilde{\cal K}_{\ov \rho \alpha} Y _{\delta \beta \gamma} + (\alpha \leftrightarrow \beta) + (\alpha \leftrightarrow \gamma)   \right) \right] \\
 B_{\alpha \beta} & = & \frac{\ov{\cal W}_0}{|{\cal W}_0|} e^{{\cal K}^{(0)}/2M_{\text{Planck}}^2} \left\{  F^m \left[ \partial_m  {\cal K}^{(0)}\, \mu_{\alpha \beta} + \partial_m \mu_{\alpha \beta}  - \left( \tilde{\cal K}^{\delta \ov \rho} \partial_m \tilde{\cal K}_{\ov \rho \alpha} \mu_{\delta \beta} + (\alpha \leftrightarrow \beta) \right) \right] - m_{3/2}  \mu_{\alpha \beta}  \right\} \\
 &&  + (2m^2_{3/2} +V_0) Z_{\alpha \beta}  - m_{3/2} \ov F^{\ov m} \partial_{\ov m} Z_{\alpha \beta} \\
 && + m_{3/2} F^m \left[ \partial_m Z_{\alpha \beta} - \left( \tilde{\cal K}^{\delta \ov \rho} \partial_m \tilde{\cal K}_{\ov \rho \alpha} Z_{\delta \beta} + (\alpha \leftrightarrow \beta)  \right) \right] \\
&& - \ov F^{\ov m} F^n \left[ \partial_{\ov m} \partial_n Z_{\alpha \beta} - \left( \tilde{\cal K}^{\delta \ov \rho} \partial_n \tilde{\cal K}_{\ov \rho \alpha} \partial_{\ov m} Z_{\delta \beta} + (\alpha \leftrightarrow \beta)  \right)  \right]
\end{array},
\end{equation}
with the tree-level cosmological constant  $V_0$ and the gravitino mass $m_{3/2}$:
\begin{equation}
V_0 = M_{\text{Planck}}^{-2} \ov F^{\ov m} {\cal K}^{(0)}_{\ov m n} F^n - 3 m_{3/2}^2 , \qquad m_{3/2}^2 = e^{{\cal K}^{(0)}/M_{\text{Planck}}^2} |{\cal W}_0|^2  M_{\text{Planck}}^{-4}, 
\end{equation} 
and the auxiliary field $F^m$ associated to the hidden field ${\cal H}^m$ given by:
\begin{equation}\label{Eq:FAuxiliaryHidden}
F^m = e^{{\cal K}^{(0)}/ 2M_{\text{Planck}}^2 }  {\cal K}^{(0) m \ov n} \left( \partial_{\ov n} {\cal W}_0 + M_{\text{Planck}}^{-2}  \partial_{\ov n} {\cal K}^{(0)}   {\cal W}_0  \right).
\end{equation}
When one of the hidden chiral matter fields ${\cal H}^m$ acquires a non-vanishing $F$-term \mbox{($\langle F^m\rangle \neq 0$)}, not only supersymmetry will be spontaneously broken, but also the soft supersymmetry breaking terms for the visible sector are expected to emerge by virtue of the gravitational coupling between the hidden and visible sector. 

These considerations shift the spotlight to the mechanism by which the hidden chiral matter is stabilised and its auxiliary field simultaneously obtains a non-zero {\it vev}. If the hidden sector contains a strongly coupled gauge sector, condensing gaugini~\cite{Ferrara:1982qs,Nilles:1983ge} are expected to induce a non-perturbative superpotential for (some of) the hidden chiral matter, such that the latter are stabilised and no longer correspond to flat directions in the scalar potential. More explicity, assume that the model contains a hidden chiral superfield ${\cal H}$ that couples to the supergauge-invariant field strength $W_\alpha$ of the strongly coupled gauge symmetry ${G}_{\text{hidden}}$ through the superpotential term:
\begin{equation} 
{\cal W}_0 (h) \supset f(h) W^\alpha W_\alpha .
\end{equation}
The auxiliary field (\ref{Eq:FAuxiliaryHidden}) of ${\cal H}$ should then be generalised to:
\begin{equation}
F^{\cal H} = e^{{\cal K}^{(0)}/2M_{\text{Planck}}^{2} }  {\cal K}^{(0) {\cal H} \ov m} \left( \partial_{\ov m} {\cal W}_0 + M_{\text{Planck}}^{-2}  \partial_{\ov m} {\cal K}^{(0)}   {\cal W}_0  \right) + \frac{1}{4}   {\cal K}^{(0) {\cal H} {\cal H}}
 \partial_{\cal H} f  \, \lambda \lambda,
 \end{equation}
 where $f$ represents the gauge kinetic function and $\lambda$ the gaugino of the strongly coupled gauge theory. At a mass scale $\Lambda_c $, the gauge symmetry becomes strongly coupled, and the gaugini condense with the characteristic scale $\langle \lambda \lambda \rangle = \Lambda_c^3$. The auxiliary field $F^{\cal H}$ then acquires a non-zero {\it vev} that sets the supersymmetry breaking scale $M_{\cancel{SUSY}}^2 = \langle F^{\cal H} \rangle \sim \Lambda_c^3/ M_{\text{Planck}}$. The gravitino will absorb the Weyl-fermion of the superfield $h$ through the super-Higgs effect and acquire a mass of the order $m_{3/2} \sim \Lambda_c^3/ M_{\text{Planck}}^2$.

In type II superstring compactifications, the r\^ole of the hidden matter is played both by truly hidden gauge sectors as well as by the complex structure and K\"ahler moduli of the internal Calabi-Yau orientifold. In type IIA compactifications gaugino condensation will generate a non-perturbative superpotential for the complex structure moduli, provided that the hidden sector comes with a strongly coupled gauge symmetry
as in the examples below in section~\ref{Sss:SoftSUSYbreakingModels}.

\section{Axions in Type IIA String Theory}\label{S:Axions-STRINGS}

\subsection{Massive gauge symmetries and open and closed string axions}\label{Ss:Open+closed}

In Type II superstring theory, axions and axion-like particles as well as Abelian gauge bosons can arise from various sectors. We first briefly discuss axions arising in the closed string sector in section~\ref{Sss:Closed-axions} and then focus on axions from the open string sector and massive Abelian gauge symmetries in section~\ref{Sss:GSMechanismOpenStringAxions}.

\subsubsection{Closed string axions}\label{Sss:Closed-axions}

The most explored scenarios focus on closed string axions emerging from the Kaluza-Klein reduction of massless $p$-forms appearing in the closed string spectrum such as the NS-NS 2-form and the RR-forms, see e.g.~\cite{Svrcek:2006yi} for an overview. The shift symmetry of these CP-odd real scalars is a remnant of the gauge invariance of the $p$-forms, while the axion decay constant is set by the non-canonical prefactor in the kinetic term of the axion appearing in the low energy effective action upon dimensional reduction. For D-brane models in Type II superstring theory, the linear coupling of the axion to the topological QCD charge density $\Tr (G_{\mu \nu} \tilde G^{\mu \nu})$ follows from the dimensional reduction of the D-brane Chern-Simons action, and more explicitly from the term (see e.g.~\cite{Ibanez:2012zz,Blumenhagen:2006ci}):
\begin{equation}\label{Eq:CS-coupling}
{\cal S}_{D3+q}^{\text{CS}} \supset \frac{M_{\text{string}}^{q}}{4 \pi }\int_{\mathbb{R}^{1,3} \times  \Pi^q} C_q \wedge \Tr( F\wedge F),
\end{equation}
with  $\Pi^q$ the $q$-cycle wrapped by the D$(q+3)$-brane along the internal directions and $C_q$ the $q$-form from the RR-sector.

Let us briefly review some of these aspects for intersecting D6-brane scenarios in Type IIA string theory, where the closed string axions arise from the dimensional reduction of the RR 3-form $C_3$.\footnote{The external and internal degrees of freedom of the RR 1-form $C_1$ are projected out on a Calabi-Yau orientifold compactification (see e.g.~\cite{Grimm:2004ua}), and $C_3$ expanded along orientifold invariant $(1,1)$-forms provides four-dimensional vectors related to isometries of the compact space.} The first step to obtain the low energy effective action for the closed string axions consists in expanding this 3-form with respect to a basis of real $\OR$-even 3-forms $\Lambda_i$,
\begin{equation}
C_3  \supset \sum_{i=0}^{h_{21}} \xi^i (x) \Lambda_i (y),
\end{equation}
where $x$ denote the coordinates along $\mathbb{R}^{1,3}$ and $y$ the coordinates on the internal Calabi-Yau orientifold. The dimensional reduction of the kinetic term for the four-form flux will then reduce to the kinetic terms for the closed string axions $\xi^i$, while the coupling $C_{3}\wedge \Tr (F \wedge F)$  in the Chern-Simons action~(\ref{Eq:CS-coupling}) for the D6-brane reduces to the instanton term:
\begin{equation}
{\cal L}_\xi \supset \frac{\pi M_{\text{string}}^8}{4} \; \text{Vol}_6 \; (\partial_\mu \xi^i)( \partial^\mu \xi^i)  - \frac{M_{\text{string}}^3}{4 \pi} \; \text{Vol}_3 \; \xi^i\, \Tr(F\wedge F). 
\end{equation}
In this expression,  $\text{Vol}_6$ is the volume of the six-dimensional internal space and $\text{Vol}_3$ the volume of the compact three-cycle $\Pi$ wrapped by the D6-brane. A simple rescaling of the closed string axion $\xi^i$ brings the low energy effective action to the conventional form of equation (\ref{Eq:StandardActionAxion}), from which we can read off the axion decay constant\footnote{Similar relations exist for other D$(3+q)$-brane model building scenarios involving $\text{Vol}_{q}$.}:
\begin{equation}
f_\xi = \frac{1}{8 \sqrt{2 \pi} }   \frac{\sqrt{ \text{Vol}_6}}{\text{Vol}_3} \; M_{\text{string}}.
\end{equation}
One immediately notices that the axion decay constant for closed string axions is proportional to $M_{\text{string}}$, which makes it challenging to identify the QCD axion as a closed string axion if the string scale is too high or too low. This consideration extends to other model building scenarios in string theory.

Another obstacle~\cite{Conlon:2006tq} for closed string axions to solve the strong CP-problem appears when involving moduli stabilisation of their CP-even scalar partners within ${\cal N}=1$ multiplets, i.e. the dilaton and complex structure and K\"ahler moduli (the `saxions'). For a saxion stabilised supersymmetrically by non-perturbative corrections,  its associated axion is also stabilised with the same mass. The no-go theorem in~\cite{Conlon:2006tq} further indicates that the presence of massless axions implies tachyonic directions in the scalar potential. However, if some of the saxions are stabilised by perturbative effects in $\alpha'$ or $g_{\text{string}}$, the no-go theorem can be circumvented, as realised explicitly in the context of the  LARGE Volume Scenario
in~\cite{Cicoli:2012sz}. Yet for unfixed closed string axions, their axion decay constant is still proportional to $M_{\text{string}}$, such that their appropriateness to serve as the QCD axion is strongly correlated with an intermediate string scale ($M_{\text{string}}\sim 10^{12}$ GeV).

In order to disentangle the axion decay constant from $M_{\text{string}}$, one has to turn to other axion sources than the closed string sector. In section~\ref{Ss:SUSY-DFSZ} we argued that there exists a natural way to embed a supersymmetric DFSZ-type axion model into D-brane model building scenarios, in which case the axion then resides in a chiral matter multiplet from the open string sector. Moreover, as the open string saxion is assumed be stabilisable through standard field theory mechanisms (i.e.~spontaneous symmetry breaking), open string axions represent an alternative loophole to the no-go theorem of~\cite{Conlon:2006tq}.

\subsubsection{The Green-Schwarz mechanism  and open string axions}\label{Sss:GSMechanismOpenStringAxions}

In D-brane model building scenarios,  $U(1)$ symmetries appear as the centers of unitary gauge groups supported on the respective D-brane worldvolumes, and  gauge anomalies are canceled by the generalized Green-Schwarz mechanism, in which the local $U(1)$ symmetry acquires a St\"uckelberg mass proportional to $M_{\text{string}}$ by eating a closed string axion. In that case, the $U(1)$ survives perturbatively as a global anomalous $U(1)$ symmetry that is only broken by non-perturbative effects~\cite{Blumenhagen:2009qh} with at most a discrete Abelian symmetry as remnant~\cite{BerasaluceGonzalez:2011wy,Anastasopoulos:2012zu,Ibanez:2012wg,Honecker:2013hda,Honecker:2013kda}. Such perturbative global symmetries are very suitable to serve as Peccei-Quinn symmetries, as the discussion in the preceding section~\ref{Ss:SUSY-DFSZ} and the two explicit examples in section~\ref{Ss:Z2N} below clearly show. 

In section~\ref{Ss:SUSY-DFSZ}, it was already pointed out that the open string axion resides in an ${\cal N}$=1 matter multiplet arising at some D-brane intersection involving the D-brane supporting the anomalous $U(1)$ symmetry. More explicitly, an open string axion corresponds to the phase of a complex scalar field charged under the anomalous $U(1)$ symmetry, similarly to the field theory set-up discussed in section \ref{S:Axions-FT}. 
At the string scale, the bosonic part is described by the following Lagrangian:
\begin{eqnarray}\label{Eq:GS-coupling}
{\cal L}_{U(1)+\text{axion}} = \left|(\partial_\mu + i\,  q B_\mu) \sigma \right|^2 + \frac{1}{2} \left( \partial_\mu \xi + M_{\text{string}} B_\mu \right)^2,
\end{eqnarray}
where $\sigma$ denotes the complex scalar field with charge $q$ under the anomalous $U(1)$ symmetry with gauge potential $B_\mu$, while $\xi$ represents the closed string axion eaten by $B_\mu$ in the St\"uckelberg mechanism. The open string axion $\rho$ arises as the phase of the complex field $\sigma$:
\begin{eqnarray}
\sigma = \frac{v+s(x)}{\sqrt{2}} e^{i\, \rho(x)/v},
\end{eqnarray}
where $s(x)$ describes the fluctuations of the open string saxion about its vacuum expectation value $v$. After inserting the expression for $\sigma$ back into the Lagrangian, one obtains the following action by keeping only track of the CP-odd scalars:
\begin{eqnarray}
{\cal L}_{\text{CP-odd}} = \frac{1}{2} (\partial_\mu \rho)^2 + \frac{1}{2} (\partial_\mu \xi)^2 + ( q \, v \,  \partial^\mu \rho + M_{\text{string}} \, \partial^\mu \xi)  \, B_\mu   + \left( \frac{ q^2 v^2}{2} + \frac{M_{\text{string}}^2}{2} \right) B_\mu B^\mu.
\end{eqnarray}
This action can be brought back to the standard form: 
\begin{eqnarray}\label{Eq:StuckClosedOpenAxion}
{\cal L}_{\text{CP-odd}} = \frac{1}{2} \left( \partial_\mu \zeta + m_B B_\mu   \right)^2 + \frac{1}{2} (\partial_\mu \alpha)^2,
\end{eqnarray}
by an $SO(2)$ transformation on the two CP-odd scalars $(\xi, \rho)$:
\begin{equation}
\zeta = \frac{M_{\text{string}}\, {\xi} +  q v\, \rho}{\sqrt{M_{\text{string}}^2 + q^2 v^2}}, \qquad  \alpha =  \frac{M_{\text{string}}\, \rho  -   q v\, {\xi}}{\sqrt{M_{\text{string}}^2 +  q^2 v^2}}.
\end{equation}
Hence, it is in fact the linear combination $\zeta$ that turns into the longitudinal component of the $U(1)$ gauge boson with mass 
\begin{equation}
m_B^2 =  M_{\text{string}}^2 + q^2 v^2, 
\end{equation}
while the orthogonal linear combination $\alpha$ remains as massless axion.

Analogously to field theory axions, the open string axion couplings to Higgses and matter can be chirally rotated away in favour of the linear coupling to $\Tr(G\wedge G)$, see
 appendix~\ref{A:ChiralRotation} for technical details. Thus, both open and closed string axions will provide for an axion coupling to the topological QCD charge density:
\begin{eqnarray}
{\cal L}_{\rm anom} = \frac{1}{32 \pi^2} \frac{\xi(x)}{f_\xi} \Tr (G_{\mu \nu} \tilde G^{\mu \nu})  + \frac{1}{32 \pi^2} \frac{\rho(x)}{f_\rho} \Tr (G_{\mu \nu} \tilde G^{\mu \nu}),  
\end{eqnarray}
with the closed string axion decay constant $f_\xi \sim M_{\text{string}}$ and the open string axion decay constant $f_\rho =  q v$.\footnote{The axion-gluon couplings are proportional to the $U(1)_{PQ}-G^2$ anomaly coefficient due to the standard string theoretic anomaly cancellation via the generalised Green-Schwarz mechanism.}
Performing the $SO(2)$ rotation also in this part of the Lagrangian yields the axion-gluon couplings in the $(\zeta, \alpha)$-basis: 
\begin{eqnarray}
{\cal L}_{\rm anom} =\frac{1}{16 \pi^2} \frac{\zeta (x)}{ f_\zeta}\Tr (G_{\mu \nu} \tilde G^{\mu \nu})  + \frac{1}{32 \pi^2} \frac{\alpha (x)}{f_{\alpha}}  \Tr (G_{\mu \nu} \tilde G^{\mu \nu}),
\end{eqnarray}
for which the decay constants are now given by
\begin{equation}
f_{\zeta} = \frac{\sqrt{M_{\text{string}}^2 + (qv)^2}}{2}, \qquad f_{\alpha} = \frac{M_{\text{string}}\, q v \sqrt{M_{\text{string}}^2 + (qv)^2}}{ (M_{\text{string}}^2 - (qv)^2)}.
\end{equation}
For models where the St\"uckelberg mass is much heavier than the scale at which $\sigma$ acquires a {\it vev}, i.e.~$M_{\text{string}}\gg v$, the axion $\zeta$ eaten by the gauge boson consists primarily of the closed string axion $\xi$. The orthogonal massless state $\alpha$ on the other hand will then be mostly composed of the open string axion $\rho$. Notice that this will also be reflected in the decay constants of the respective axions: $f_\zeta \sim M_{\text{string}} = f_{\xi}$ and $f_\alpha \sim q v = f_\rho$. Hence, in this configuration the string scale $M_{\text{string}}$ can be much higher than $10^{12}$ GeV, as the presence of an open string axion provides another candidate for the QCD axion.

Up to this point, both closed and open string axions have been considered, but only $U(1)$ symmetries from the open string sector have been taken into account. Generically, orientifold compactifications of Type II string theory contain some number of closed string vectors, also called `RR photons'.  In Type IIA string theory, these arise from the dimensional reduction $C_3 \supset \sum_{i=1}^{h_{11}^+}  A^{\mu}_i(x) \, \omega^i (y)$ with  orientifold-even (1,1)-forms $\omega^i$. However, as shown in~\cite{Camara:2011jg}, these closed string $U(1)$'s are completely decoupled from the open string sector
unless closed string fluxes significantly distort the Calabi-Yau geometry.

\subsection{Standard Models on $T^6/\Z_{6}$ and $T^6/\Z_{6}^{\prime}$  orbifolds}\label{Ss:Z2N}

Two supersymmetric  intersecting D6-brane models, in which the distinguished features presented in the previous sections are explicitly realised, have been constructed on $T^6/\Z_6$~\cite{Honecker:2004kb} and $T^6/\Z_6'$~\cite{Gmeiner:2008xq}. 
Both models are globally defined (i.e. they satisfy RR tadpole cancellation and K-theory constraints) and contain a spontaneously broken right-symmetric group,
 \begin{equation}
 \begin{aligned}
&  U(3)_a \times U(2)_b \times USp(2)_c \times U(1)_d \times G_{\text{hidden}} 
\\
 \stackrel{\text{Green Schwarz mechanism}}{\longrightarrow} \quad & 
 SU(3)_a \times SU(2)_b \times USp(2)_c \times U(1)_{B-L} \times U(1)_{\text{massive}}^2  \times G_{\text{hidden}}  \\
 \stackrel{\text{continuous displacement/Wilson line}}{\longrightarrow} \quad & 
 SU(3)_a \times SU(2)_b \times U(1)_Y \times U(1)_{B-L} \times U(1)_{\text{massive}}^2 \times G_{\text{hidden}} 
 ,
\end{aligned}
 \end{equation}
with  the following $B-L$ and hypercharge assignments in terms of original $U(1)_{a,b,c,d}$ charges,
\begin{equation}\label{Eq:Def-Y+BL}
 Q_{B-L}=\frac{Q_a}{3} + Q_d
 ,
\qquad 
Q_Y=\frac{Q_{B-L}+Q_c}{2} =\frac{Q_a}{6} + \frac{Q_c+Q_d}{2}
.
\end{equation}
On $T^6/\Z_6$ the `hidden' gauge group is $G_{\text{hidden}} =USp(2)_e$, and it can be broken to its center $U(1)_e$ in the same way as the right-symmetric group $USp(2)_c$ is broken by switching on a ${\it vev}$ of the complex scalar encoding the displacement and Wilson line of the respective D6-brane. The hidden gauge group $G_{\text{hidden}} = USp(6)_h$ on $T^6/\Z_6'$ on the other hand cannot be broken by such a {\it vev} since the corresponding stack of D6-branes is perpendicular to the O6-plane orbits while the other special D6-branes with $USp(2)$ gauge groups in both models are parallel to some O6-plane orbit.

The chiral spectra of both models are listed in tables~\ref{tab:5stackLRSChiral} and~\ref{tab:5stack-Z6p-Chiral} in appendix~\ref{A:Spectra}, and the corresponding non-chiral spectra are given in tables~\ref{tab:5stackLRSNonChiral} and~\ref{tab:5stack-Z6p-NonChiral} for $T^6/\Z_6$ and $T^6/\Z_6'$, respectively. In both cases, the three left-handed quarks $Q_L$ stem from  $a(\theta^k b')_{k\in \{0 \ldots 2\}}$ sectors, where $(\theta^k b')$ denotes the $k^{\rm th}$ orbifold image of $b'$, which in turn denotes the orientifold image of D6-brane $b$. 
The right-handed down-type quarks $d_R$ are localised in the $a(\theta^k c)$ sectors, and the right-handed up-type quarks $u_R$ in the $a(\theta^k c')$ sectors.
The charges under the anomalous and massive $U(1)_b$ in the quark sector thus coincide with those given for the global $U(1)_{PQ}$ in supersymmetric field theory in table~\ref{Tab:PQchargesSUSY}. The $c (\theta^k d)'$ sectors produce the right-handed electrons $e_R$, while the right-handed neutrinos $\nu_R$ emerge from intersections in the $c (\theta^k d)$ sectors.
In both models, four vector-like pairs of states in the antisymmetric representation of $U(2)_b$ can be found in the $b (\theta^k b)'$ sectors. These states provide exactly the degrees of freedom for the singlet superfield $\Sigma$ required to establish the supersymmetric version of the DFSZ axion-model as discussed in section~\ref{Ss:SUSY-DFSZ}. Moreover, due to the spontaneously broken underlying left-right symmetric gauge structure in both models, the $U(1)_b$ symmetry provides the only viable realisation of a $U(1)_{PQ}$ symmetry.

The first noteworthy difference between the massless spectra of the two models occurs in the left-handed leptonic sector: the model on $T^6/\Z_6$ has exactly three left-handed leptons in the $b (\theta^k d)'$ sectors, while the chiral part of the spectrum on $T^6/\Z_6'$ contains six left-handed leptons in the $b (\theta^k d)$ sectors and three anti-leptons in the $b (\theta^k d)'$ sectors. A second difference concerns the Higgs-sector for the models: in the $T^6/\Z_6$ case, the Higgses $(H_u,H_d)$ arise as vector-like pairs with opposite $U(1)_b$ charge, while on $T^6/\Z_6'$ the Higgses have identical $U(1)_b$ charge and belong to the chiral sector.   As a consequence, the way in which the supersymmetric DFSZ model of section \ref{Ss:SUSY-DFSZ} is realised for both D-brane models will differ. Further details are given in sections~\ref{Ss:Z6} and~\ref{Ss:Z6p} for the $T^6/\Z_6$ and $T^6/\Z_6'$ model, respectively. 
Last but not least, the $T^6/\Z_6$ model contains the weakly coupled `hidden' gauge group $USp(2)_e$, whereas for the model on $T^6/\Z_6'$ the hidden gauge group $USp(6)_h$ runs to strong coupling below the string scale, and therefore a hidden sector gaugino condensate can form and break supersymmetry spontaneously. As anticipated in section~\ref{Ss:Soft-SUSY},
 this will by gravity mediation result in soft breaking terms in the Standard Model sector as further discussed in section~\ref{Sss:SoftSUSYbreakingModels}.

\subsubsection{A supersymmetric DFSZ model on $T^6/\Z_6$}\label{Ss:Z6}

We briefly review here the Higgs sector of the global 5-stack D6-brane model on $T^6/\Z_6$, that was first constructed in~\cite{Honecker:2004kb,Honecker:2004np} with the non-chiral spectrum, 
beta function coefficients and 1-loop gauge threshold corrections explicitly given in~\cite{Gmeiner:2009fb}, before proceeding to the new discussion of the Higgs-axion potential. The complete massless  chiral and vector-like matter states are for convenience listed in appendix~\ref{A:Spectra} in tables~\ref{tab:5stackLRSChiral} and~\ref{tab:5stackLRSNonChiral} after spontaneous breaking of the right-symmetric group $USp(2)_c \to U(1)_c$.

In the left-right symmetric phase, the Higgs-sector of this model consists of a non-chiral pair of bifundamental states located at the intersection points of the $bc=bc'$-sector. Under a continuous displacement $\sigma^3_c$ of (or a Wilson line $\tau^3_c$ along) the $c$-brane on the third two-torus
without $\Z_2$ action, the $USp(2)_c$ gauge group is broken to an Abelian gauge group $U(1)_c$, and the non-chiral pair $[(\2,\2) +h.c.]$ of $U(2)_b \times USp(2)_c$ splits up into a non-chiral pair $[(\2)^{(-1)} + h.c.]$ of $U(2)_b^{(\times U(1)_c)}$ in the $bc$-sector and another non-chiral pair $[(\2)^{(1)} + h.c.]$  in the $bc'$-sector as listed in table~\ref{tab:5stackLRSNonChiral}. This implies that the model possesses two up-type Higgs-doublets $H_u^{(i)}$ and two down-type Higgs-doublets $H_d^{(i)}$ with $i=1$ for the $bc$-sector and $i=2$ for the $bc'$-sector.

Upon a spontaneous breaking $USp(2)_c \to U(1)_c$ of the $c$-brane gauge group, the Higgs multiplets acquire a mass 
$\sqrt{\alpha' } \; m_{\text{Higgs}} \propto \sqrt{ v_3(\sigma^3_c)^2 + v_3^{-1}(\tau^3_c)^2}$ 
(with the two-torus volume $v_3$ in units of $\alpha'$) as denoted by the lower index $m$ of the multiplicities in table~\ref{tab:5stackLRSNonChiral}. If also the D-branes $a$ and $b$ are displaced from the origin and relatively to each other, i.e. $ \sigma_a \neq \sigma_b \neq \sigma_c \neq \sigma_a$ and $\sigma_{x, x \in \{a,b,c\}} \neq 0$, a large part of the vector-like matter states is rendered massive as displayed in table~\ref{tab:5stackLRSNonChiral}. 
The open string axion candidates $\Sigma_{0 \ldots 3}$, $\tilde{\Sigma}_{0 \ldots 3}$ with $U(1)_b \simeq U(1)_{PQ}$ charge $+2$ and $-2$, respectively, arise as antisymmetric representations of $U(2)_b$ as anticipated.
For $\sigma_b \neq 0$, one vector-like pair $\Sigma_0+\tilde{\Sigma}_0$ residing in the $bb'$ sector becomes massive as denoted by $\1_m$, while the other three vector-like pairs $(\Sigma_i + \tilde{\Sigma}_i)_{i=1,2,3}$ arise in the  $b (\theta^k b)'_{k=1,2}$ sectors at non-vanishing angles and thus stay massless. For the sake of simplicity in the discussion below, $(\sigma^3_c,\tau^3_c) \neq (0,0)$ is required to break the right-symmetric group, but we assume either $(\sigma^3_b, \tau^3_b) \approx (0,0)$ and therefore $m_{H_d^{(1)}} = m_{H_u^{(1)}} \approx m_{H_u^{(2)}} = m_{H_d^{(2)}}$
or $(\sigma^3_b, \tau^3_b) \approx \pm (\sigma^3_c,\tau^3_c)$ and thus one massless and one heavy Higgs pair $(H_d^{(i)},H_u^{(i)})$.

 The {\it a priori} massless states $(\Sigma_i + \tilde{\Sigma}_i)_{i=1,2,3}$ 
  play  the r\^ole of the axion superfields in the supersymmetric DFSZ model. The mass for the singlet superfield is thus expected to be independent of the scale at which the right-symmetric gauge group $USp(2)_c$ is broken. For the model at hand on $T^6/\Z_6$, the
  stringy coupling selection rule of closed polygons (with D-branes along the edges and charged matter at the apexes) implies that the superpotential coupling in equation~(\ref{Eq:SusyCouplingHiggsSinglet}) between the Higgses and the singlet superfield is only perturbatively present in the form of a quartic coupling with a multiplet $B_1$ in the `adjoint' representation of $U(1)_b$:\footnote{A state in the adjoint representation of a $U(2)$ gauge group decomposes into one adjoint state under $SU(2)$ and one `adjoint' state under the centre $U(1)$:
\begin{equation}
\begin{array}{ccc}
U(2) \simeq SU(2) \times U(1),
\qquad
 \4_\Adj  = (\3_\Adj)_0 \oplus (\1_\Adj)_0 
\end{array}.
\end{equation}
For our purposes, the singlet `adjoint' state is considered since by giving a {\it vev} to it, at low energies effectively the relevant three-point Yukawa couplings are obtained.}    
\begin{equation} \label{Eq:4ptCouplingDFSZ}
   {\cal W}_{\text{4-point}} = \frac{\mu_1}{M_{\text{string}}}\, H_d^{(1)} \cdot H_u^{(2)}  \, \Sigma_1 \, B_1+  \frac{\mu_2}{M_{\text{string}}}\, H_d^{(2)} \cdot H_u^{(1)}  \, \tilde \Sigma_{1} \, B_1 , 
\end{equation}
for which the corresponding worldsheet instantons take the form of closed quadrilaterals, whose volume depends on the relative displacements of the D-brane stacks.
The geometric configuration of the D-branes also allows for a cubic coupling between the two singlet superfields and the `adjoint' singlet state:
\begin{equation}\label{Eq:3ptCouplingDFSZ}
 {\cal W}_{\text{3-point}} = \kappa \, \Sigma_1 \, \tilde \Sigma_1 \, B_1.
\end{equation}
For simplicity, the two types of superpotential contributions above have been restricted to the states located at the origin on the third two-torus labeled by the index $1$. Similar expressions can be written down for the states situated at the two other $\Z_3$ fixed points on the third two-torus with indices $2$ and $3$.

\begin{figure}[h]
\begin{center}
\vspace{0.2in}
\begin{tabular}{ccc}
\includegraphics[width=5cm]{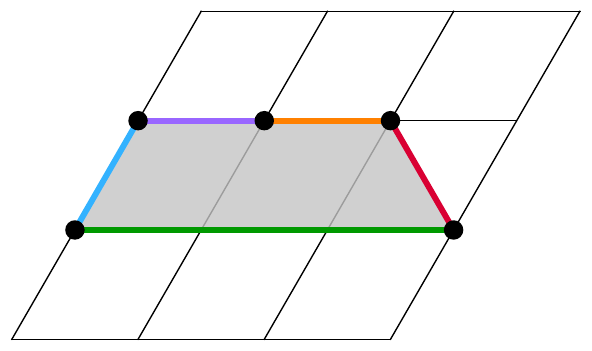} \begin{picture}(0,0) \put(-60,90){$T_{(1)}^2$} \put(-150,28){$Q_L$} \put(-30,28){$B_j$} \put(-130,54){$A_k$} \put(-86,64){$q_R$} \put(-46,60){$H^{(2|1)}_{u|d}$} \end{picture}
& \includegraphics[width=5cm]{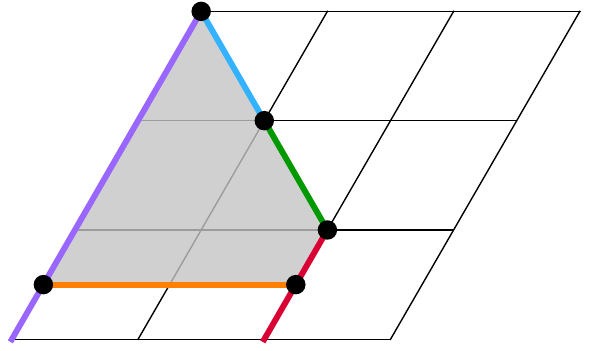} \begin{picture}(0,0) \put(-60,90){$T_{(2)}^2$} \end{picture}
&\\
(a) & (a) &\\
\vspace{-1.2in}&&\\
&&  \includegraphics[width=5.5cm]{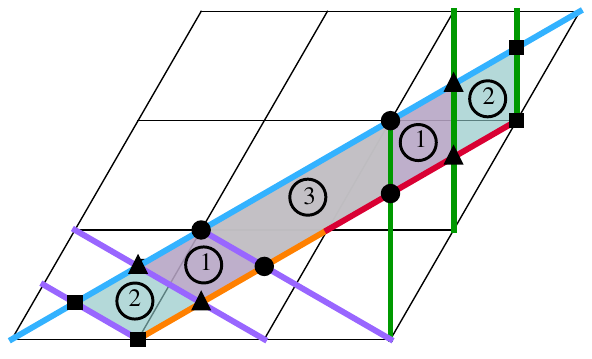} \begin{picture}(0,0) \put(-60,105){$T_{(3)}^2$} \end{picture}\\
&&\\
\vspace{-0.8in}
&&\\
\includegraphics[width=5cm]{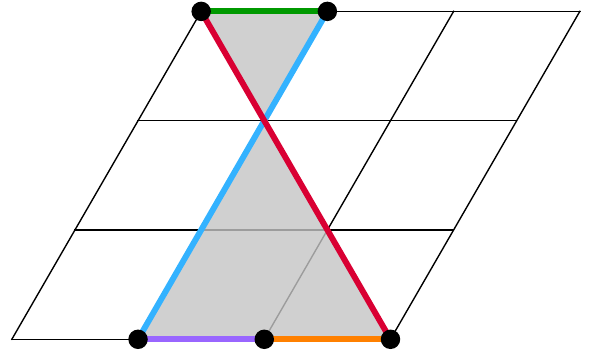} \begin{picture}(0,0) \put(-75,90){$Q_L$} \put(-115,88){$B_j$} \put(-120,-10){$A_k$} \put(-90,-7){$q_R$} \put(-50,-7){$H^{(2|1)}_{u|d}$} \end{picture}
&\includegraphics[width=5cm]{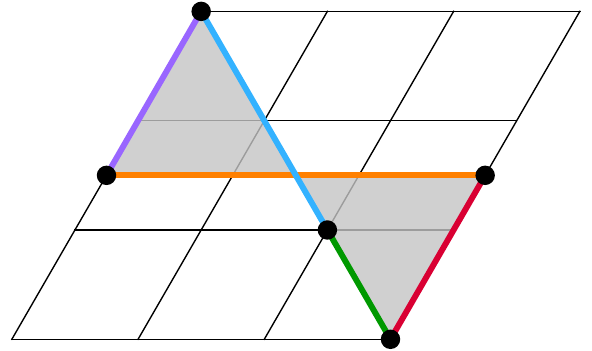}
&\\
(b) & (b) & \\
\end{tabular}
\caption{Pictorial overview of the five-point couplings $B_j A_k \,  Q_L^{(i)}\cdot H_u^{(2)} u_R^{(i)}$ $ [{\color{myblue}a}, {\color{mygr}(\omega^2 b)'}, {\color{red}b'} , {\color{myorange}c}, {\color{mypurple}(\omega^2 a)}] $ and $ B_j A_k \, Q_L^{(i)}\cdot H_d^{(1)} d_R^{(i)}$ $ [{\color{myblue}a}, {\color{mygr}(\omega^2 b)'}, {\color{red}b'} , {\color{myorange}c'}, {\color{mypurple}(\omega^2 a)}] $ from table~\ref{tab:Yuk3pt4pt5ptCoup} on the covering space of the two-tori. One parallelogram represents a cell in the complex torus-lattice,
and for illustrational purposes non-minimal polyangular worldsheets are depicted. $q_R$  denotes either $u_R$ or $d_R$. On $T_{(1)}^2$ and $T_{(2)}^2$, the worldsheet instanton can correspond to a simple (a) or self-intersecting (b) holomorphic curve. On the third two-torus, the shape of the worldsheet instanton is a trapezoid whose area is generation-dependent: the 
coupling of the heaviest generation $(i=3)$ is characterized by the trapezoid 3 with $\bullet$ as vertices, the coupling of the second generation $(i=2)$ by trapezoid 2 with $\blacksquare$ as vertices and the coupling of the first generation $(i=1)$ by trapezoid 1 with $\blacktriangle$ as vertices. The intersection points are chosen such that the first generation has the largest minimal area on $T_{(3)}^2$, as indicated in table \ref{tab:Yuk3pt4pt5ptCoup}. The five-point couplings are given for the left-right symmetric phase of the model. If the right-symmetric group $USp(2)_c$ is broken by a continuous displacement of the $c$-brane along $T^3_{(2)}$, the corresponding trapezoids are deformed. The exact same drawings can be made for the lepton-Yukawa couplings, by substituting brane $a$ for brane $d$.  \label{fig:5ptCouplingsYukQuarks}}
\end{center}
\end{figure}

As detailed in section~\ref{Ss:SUSY-DFSZ}, the Higgses couple to the quarks and leptons through Yukawa interactions in the superpotential, with the left-handed quarks and leptons charged under the $U(1)_b$ Peccei-Quinn symmetry as well. In the $T^6/\Z_6$ model at hand, the Yukawa couplings can only be  realised through five-point couplings involving two additional `adjoint' multiplets of the $U(1)$ gauge factors to complete the apexes of the pentagon-like worldsheet instantons enclosed by the D-brane stacks.\footnote{In the left-right symmetric phase, one could naively expect cubic Yukawa couplings to exist, in which case they would correspond to pointlike worldsheet instantons as the D-brane stacks intersect at a single point. However, upon displacing the D-brane stacks along the third two-torus, the triple intersections points are no longer present. A more profound investigation of the existence of this type of Yukawa coupling is required, but goes well beyond the scope of this article.
\label{Fn:Existence-Yukawas}} 
For the Yukawa couplings with the quarks, an `adjoint' multiplet $A_j$ of $U(1)_a \subset U(3)_a$ and an `adjoint' multiplet multiplet $B_k$ of $U(1)_b \subset U(2)_b$ complete the D-brane sequence:
\begin{equation}
{\cal W}_{\text{quarks}} = \frac{f^{ijk}_u}{M_{\text{string}}^2} A_j B_k\,  Q_L^{(i)} \cdot H_u^{(2)} u_R^{(i)} +  \frac{f^{ijk}_d}{M_{\text{string}}^2} A_j B_k \,  Q_L^{(i)} \cdot H_d^{(1)} d_R^{(i)},
\end{equation}
while the Yukawa couplings to the leptons require an `adjoint' multiplet $B_j$ of $U(1)_b$ and an `adjoint' multiplet $D_k$ of $U(1)_d$,
\begin{equation}
{\cal W}_{\text{leptons}} =   \frac{f^{ijk}_e}{M_{\text{string}}^2} B_j D_k\,  L^{(i)}\cdot H_d^{(2)} e_R^{(i)} +   \frac{f^{ijk}_{\nu}}{M_{\text{string}}^2} B_j D_k\,  L^{(i)}\cdot H_u^{(1)} \nu_R^{(i)}.
\end{equation}
In both superpotential contributions, the indices $i,j,k \in \{1, 2, 3 \}$ denote the fixed points on $T^2_{(3)}$ at which the various states are localised.\footnote{There also exist five-point couplings with mixings among two particle generations as can e.g. be seen on $T^2_{(3)}$ in  figure~\protect\ref{fig:5ptCouplingsYukQuarks} with different choices of trapezoid edges on the left and right. However, a detailed study of the Yukawa sector goes well beyond the scope of the present article and will be addressed in separate work \cite{HoneckerStaessens:2014}. The axion-Higgs potential considered in this article is independent of possible quark or lepton generation mixing terms.} The supersymmetric DFSZ model can thus be fully realised on the orbifold $T^6/\Z_6$, with the superpotential given by:
\begin{equation}\label{Eq:FullSuperpotentialDFSZ}
{\cal W}_{\text{DFSZ}} = {\cal W}_{\text{3-point}} + {\cal W}_{\text{4-point}} + {\cal W}_{\text{quarks}} + {\cal W}_{\text{leptons}}.
\end{equation} 
Table~\ref{tab:Yuk3pt4pt5ptCoup} provides a summary of the various family-diagonal couplings in the DFSZ superpotential, and figure~\ref{fig:5ptCouplingsYukQuarks} gives a pictorial representation of the five-point Yukawa couplings.
\mathtab{
\begin{array}{|c|c|c|c|}
\hline \multicolumn{4}{|c|}{\text{\bf Couplings for the SUSY DFSZ model on $T^6/\Z_6$ in L-R symmetric phase}} \\
\hline \hline
 \text{ \bf Coupling } &  \text{ \bf Sequence } & \text{ \bf Enclosed Area } & \text{ \bf Parameters}\\
\hline \hline 
 M_{\text{string}}^{-2} B_1 A_2 \,  Q_L^{(1)} H Q_R^{(1)} &  & \frac{1}{8}  v_2 + \frac{1}{2} v_3   & f_{u}^1 = f_{d}^1 \sim {\cal O} \left( e^{-  \frac{v_2}{8} - \frac{ v_3}{2} } \right) \\
  M_{\text{string}}^{-2} B_3 A_3 \,  Q_L^{(2)} H Q_R^{(2)} & [a,(\omega^2 b)', b' , c, (\omega^2 a)] &\frac{1}{8} v_2 + \frac{1}{6} v_3   &  f_{u}^2 = f_{d}^2 \sim {\cal O} \left( e^{ - \frac{v_2}{8} - \frac{ v_3}{6} } \right) \\
    M_{\text{string}}^{-2} B_2 A_1 \,  Q_L^{(3)} H Q_R^{(3)} & & \frac{1}{8}  v_2 + \frac{1}{6} v_3  &   f_{u}^3 = f_{d}^3 \sim {\cal O} \left( e^{ -\frac{v_2}{8} - \frac{ v_3}{6} } \right) \\
\hline
 M_{\text{string}}^{-2} B_1 D_2 \,  L^{(1)} H R^{(1)} & & \frac{1}{8}  v_2 + \frac{1}{2} v_3   &f_{e}^1 = f_{\nu}^1 \sim {\cal O} \left( e^{ - \frac{v_2}{8} - \frac{ v_3}{2} } \right)  \\
  M_{\text{string}}^{-2} B_3 D_3 \,  L^{(2)} H R^{(2)} & [d,(\omega^2 b)', b' , c, (\omega^2 d)] & \frac{1}{8} v_2 + \frac{1}{6} v_3   &  f_{e}^2 = f_{\nu}^2 \sim {\cal O} \left( e^{ - \frac{v_2}{8} - \frac{ v_3}{6} } \right) \\
    M_{\text{string}}^{-2} B_2 D_1 \,  L^{(3)} H R^{(3)} & & \frac{1}{8}  v_2 + \frac{1}{6} v_3  &f_{e}^3 = f_{\nu}^3 \sim {\cal O} \left( e^{- \frac{v_2}{8} - \frac{ v_3}{6} } \right) \\
\hline
M_{\text{string}}^{-1} H_d^{(1)}\cdot  H_u^{(2)} \,  \Sigma_1 \, B_1&[b,c,b', (\omega^2 b)]& \frac{1}{2} v_1 + \frac{1}{8} v_2 & \mu_1 \sim {\cal O} \left(e^{-\frac{v_1}{2} - \frac{  v_2}{8} }\right) \\
M_{\text{string}}^{-1}  H_d^{(2)}\cdot  H_u^{(1)} \, \tilde\Sigma_1 \, B_1&[b,c',b', (\omega b)] & \frac{1}{2}  v_1 + \frac{1}{8}  v_2  & \mu_2 \sim {\cal O}\left(e^{-\frac{v_1}{2}-\frac{v_2}{8}}\right)\\
 \Sigma_1 \tilde \Sigma_{1} B_1 &[b,(\omega b)',(\omega b)] & 0 & \kappa \sim {\cal O} (1) \\
 \hline
\end{array}
}{Yuk3pt4pt5ptCoup}{Overview (left column) of the relevant perturbatively allowed three-, four- and five-point couplings appearing in the superpotential (\ref{Eq:FullSuperpotentialDFSZ}) of the supersymmetric DFSZ model on $T^6/\Z_6$. The couplings are presented in the left-right symmetric phase with the Higgs-doublet $(H_u, H_d)$ denoted by $H$, the right-handed quarks by $Q^{(i)}_R$ and the right-handed leptons by $R^{(i)}$. The corresponding D-brane sequence of the coupling is given in the second column from the left, and the area of the minimal holomorphic curve enclosed by the worldsheet instanton is given in the third column from left, where $v_i$ denotes the volume of $T^2_{(i)}$ in units of $\alpha'$. 
The fourth column indicates the order of magnitude for the coupling assigned to the respective $n$-point coupling.}

The `adjoint' multiplets $A_i, B_j$ and $D_k$ encode the geometric deformation moduli of the intersection points of $x(\theta^k x)_{k=1,2}$ for $x\in \{a,b,d\}$, whose {\it vev}s describe brane recombinations of orbifold images for the corresponding stack $x$.
Since only {\it vev}s in the singlet fields are considered, the gauge representations are not affected. Following the discussion in section~\ref{Ss:Soft-SUSY}, the singlet fields fall under the category of `hidden' matter. As such, they are assumed to be stabilised independently from the vacuum configuration of the charged matter, and the {\it vev}s of the scalar fields inside the multiplets determine the scale of the couplings between the charged matter. Integrating out these `adjoint' multiplets $A_i, B_j$ and $D_k$ molds the cubic superpotential ${\cal W}_{\text{3-point}}$ into a supersymmetric mass term for $\Sigma_1$ and $\tilde \Sigma_1$ with mass parameter $\tilde \kappa \equiv \kappa \langle  B_1  \rangle$, while all other couplings in the superpotenial ${\cal W}_{\text{DFSZ}}$ of equation (\ref{Eq:FullSuperpotentialDFSZ}) reduce to the desired cubic couplings with respective coupling parameters:  
\begin{equation}
\tilde \mu_i \equiv \frac{\mu_i}{M_{\text{string}}} \langle B_1  \rangle, \qquad \tilde f_{u|d}^{ijk} \equiv \frac{f^{ijk}_{u|d}}{M^2_{\text{string}}} \langle A_j  B_k \rangle, \qquad \tilde f_{e|\nu}^{ijk} \equiv \frac{f^{ijk}_{e|\nu}}{M^2_{\text{string}}} \langle  B_j D_k \rangle. 
\end{equation}
Due to the presence of string scale suppressed couplings, a distinct hierarchy between the couplings can emerge depending on the value of $M_{\text{string}}$ and {\it vev}s of the `adjoint' multiplets, such as e.g.~$\tilde \kappa \gg \tilde \mu_i,  \tilde f_{u|d}^{ijk}, \tilde f_{e|\nu}^{ijk}$. This observation might be of particular interest to address the $\mu$-problem in a similar manner to the Kim-Nilles proposal of non-renormalisable Planck-suppressed couplings~\cite{Kim:1983dt}. For a string scale of the order $10^{16}$ GeV, the supersymmetric mass term for the Higgses is of the order of the TeV scale, provided that the $\tilde \kappa$-parameter and the {\it vev}s of the saxions are of the order \mbox{$10^{10}$ GeV} (which lies within the axion window). Higher values of the latter two parameters are allowed in situations where the $\mu_i$ parameters are exponentially suppressed e.g. by extended instantonic worldsheets.

When integrating out the `adjoint' multiplet $B_1$ from the cubic superpotential~(\ref{Eq:3ptCouplingDFSZ}) and the quartic superpotential~(\ref{Eq:4ptCouplingDFSZ}), the F-term contributions to the axion-Higgs potential simplify to the following expression:
\begin{equation}
\begin{array}{ccl}
V_F &=& \left| \tilde \kappa  \tilde \sigma_{1} + \tilde \mu_1 H_d^{(1)} \cdot H_u^{(2)}  \right|^2 +  \left| \tilde \kappa \sigma_{1} + \tilde \mu_2 H_d^{(2)} \cdot H_u^{(1)}  \right|^2\\
&& +  |\tilde \mu_1|^2 \left( \left|H_u^{(2)}\right|^2 + \left|H_d^{(1)}\right|^2   \right) | \sigma_1|^2 + | \tilde \mu_2|^2 \left( \left|H_u^{(1)}\right|^2 + \left|H_d^{(2)}\right|^2   \right) |\tilde \sigma_{1}|^2,
\end{array}
\end{equation}
where the symbols $\sigma_1 (\tilde \sigma_1)$ denote the complex scalar field component of the superfields $\Sigma_1 (\tilde \Sigma_1)$, and should not be confused with the displacement parameter $\sigma^3_{x, x\in \{a,b,c\}}$, nor with the singlet field $\tilde \sigma$ in the original, non-supersymmetric  DFSZ-model.\footnote{The scalar components of the Higgs superfields are denoted with capital letters, in the hope that it will not confuse the reader. Small letters for the Higgses are used later on to denote the components of the doublet structure.} 

As anticipated in section~\ref{Ss:SUSY-DFSZ}, the coupling of the Higgses to the Standard Model gauge group leads to a D-term potential consisting of quartic terms for the Higgses:
\begin{eqnarray}
V_D &=& \frac{(g_Y^2 + g_2^2)}{8} \left(  \left|H_u^{(1)}\right|^2 - \left|H_d^{(2)}\right|^2  +  \left|H_u^{(2)}\right|^2 -  \left|H_d^{(1)}\right|^2 \right)^2 \nonumber\\
&& + \frac{g_2^2}{2} \left(  \left|{H_u^{(1)}}^\dagger H_u^{(2)} \right|^2 +  \left|{H_d^{(1)}}^\dagger H_d^{(2)} \right|^2 +  \left|{H_u^{(1)}}^\dagger H_d^{(1)} \right|^2 +  \left|{H_u^{(1)}}^\dagger H_d^{(2)} \right|^2  \right. \nonumber \\
&& \hspace{0.5in}  \left. +  \left|{H_u^{(2)}}^\dagger H_d^{(1)} \right|^2  +  \left|{H_u^{(2)}}^\dagger H_d^{(2)} \right|^2 
-  \left|H_u^{(1)}\right|^2  \left|H_u^{(2)}\right|^2 - \left|H_d^{(1)}\right|^2  \left|H_d^{(2)}\right|^2   \right),  
\label{Eq:DFSZDTerm1}
\end{eqnarray}
where $g_Y$ and $g_2$ denote the $U(1)_Y$ and $SU(2)_b$ gauge couplings, respectively.

Furthermore, as the Higgses and the Standard Model singlets $\Sigma_1$ and $\tilde \Sigma_1$ couple locally to the anomalous $U(1)_b$ gauge symmetry, an additional D-term contribution to the scalar potential has to be taken into account, generating a quartic couplings for $\sigma_1$ and $\tilde \sigma_1$:
\begin{eqnarray}  \label{Eq:DFSZDTerm2}
V^{U(1)_b}_D= \frac{g_2^2}{8} \left( \left|  H_d^{(2)} \right|^2 -\left|  H_d^{(1)} \right|^2 + \left|  H_u^{(1)} \right|^2 -\left|  H_u^{(2)} \right|^2 +  2 | \sigma_1|^2 - 2 |\tilde \sigma_{1}|^2   \right)^2.
\end{eqnarray}

The scalar potential is completed by adding soft supersymmetry breaking terms consistent with the gauge structure of the model:
\begin{equation}
\begin{array}{rcl}
V_{\text{soft}} &=& m^2_{H_u^{(1)}}  \left|H_u^{(1)}\right|^2 + m^2_{H_u^{(2)}}  \left|H_u^{(2)}\right|^2 + m^2_{H_d^{(1)}}  \left|H_d^{(1)}\right|^2 + m^2_{H_d^{(2)}}  \left|H_d^{(2)}\right|^2 \\
&& + m^2_{\sigma_1}  |\sigma_1 |^2  +   m^2_{\tilde\sigma_1}  |\tilde\sigma_1 |^2 - \left( m_{12}^2\, \sigma_1 \tilde\sigma_{1} + h.c. \right)   \\
&& + \left(c_1 H_d^{(1)} \cdot H_u^{(2)}  \sigma_1 + h.c. \right) + \left(c_2 H_d^{(2)} \cdot H_u^{(1)} \tilde\sigma_1 + h.c. \right)\\
&&  + ( m_{11}^2 H_d^{(1)} \cdot  H_u^{(1)} + h.c. ) + ( m_{22}^2 H_d^{(2)} \cdot  H_u^{(2)} + h.c. ). 
\end{array}
\end{equation}
These soft superysmmetry breaking terms include mass terms for the various fields, as well as ${\cal A}$-terms (third line) and ${\cal B}$-terms (fourth line) whose shapes are constrained by gauge invariance. Note that the only interactions between the Higgses and the axions in the scalar potential arise from the ${\cal A}$-terms. As already suggested in section~\ref{Ss:Soft-SUSY}, these soft supersymmetry breaking terms can originate from gravity mediated supersymmetry breaking. In section~\ref{Sss:SoftSUSYbreakingModels} we will come back to this point and discuss how the
gravity mediated soft supersymmetry breaking scenario might work in practice for the two explicit examples. If the antisymmetric states $\Sigma_{2,3}$ and $\tilde \Sigma_{2,3}$ located at the other two $\Z_3$ fixed points on $T_{(3)}^2$ are also included, the F-terms potential, the D-term potential associated to $U(1)_b$ and the soft supersymmery breaking potential have to be generalised, leading to an even more complicated scalar potential.   
 
The full scalar potential for the axion-Higgs sector is given by:
\begin{eqnarray} \label{Eq:FullScalarPotential}
V = V_F + V_D + V^{U(1)_b}_D + V_{\text{soft}},
\end{eqnarray}
whose vacuum structure determines both the spontaneous breaking of the anomalous $U(1)_{PQ}$ symmetry and the electroweak symmetry breaking. In this respect, it is crucial to determine the vacuum configuration for the Higgses and the saxions.
As anticipated in equation~(\ref{Eq:Def-Higgses+vevs}), invariance under $U(1)_{\text{el-mag}}$ requires zero {\it vev}s for the charged Higgs components, and therefore the $H_u^{(i)}$ and $H_d^{(j)}$ vacua are orthogonal, i.e. the terms $\left|{H_u^{(1)}}^\dagger H_d^{(1)} \right|^2$, $  \left|{H_u^{(1)}}^\dagger H_d^{(2)} \right|^2$, $\left|{H_u^{(2)}}^\dagger H_d^{(1)} \right|^2$, and $ \left|{H_u^{(2)}}^\dagger H_d^{(2)} \right|^2 $ vanish in the vacuum. A reasonable guess for the minimum of the D-term potential preserving $U(1)_{\text{el-mag}}$ invariance is thus:
\begin{eqnarray}
\langle h_u^{0,(1)} \rangle = \frac{v_u}{\sqrt{2}}  = \langle h_u^{0,(2)} \rangle, \quad 
\langle h_d^{0,(1)} \rangle = \frac{v_d}{\sqrt{2}}  =  \langle h_d^{0,(2)} \rangle, \quad
\langle \sigma_1\rangle =  \frac{v_{\sigma_1}}{\sqrt{2}} \, e^{i\, \phi_1}, \quad  
\langle \tilde \sigma_1\rangle = \frac{v_{\tilde \sigma_1}}{\sqrt{2}}\, e^{i\, \tilde \phi_1}. 
\label{Eq:MinimumPotential}
\end{eqnarray}
The {\it vev}s have to satisfy additional constraints in order for them to minimise the scalar potential. The discussion for the Higgs-sector will be postponed to section~\ref{Sss:HiggsSector}, as we will first investigate the spontaneous gauge symmetry breaking for this vacuum configuration.

\subsubsection{Symmetry breaking and the QCD axion}\label{Sss:SymmetryBreaking}

The gauge bosons of the $SU(2)_b\times U(1)_Y$ symmetry acquire their mass through the Brout-Englert-Higgs mechanism, while 
the $U(1)_{B-L}$ has to be broken by the {\it vev} of some right-handed sneutrino $\tilde{\nu}_R$.
The St\"uckelberg mechanism supplies a mass for the gauge boson of the $U(1)_b$ gauge symmetry as reviewed in section~\ref{Sss:GSMechanismOpenStringAxions}, leaving only the $SU(3)_{a}\times U(1)_{\text{el-mag}}$ gauge  group as unbroken gauge symmetry for the observable matter sector. 

Besides identifying the symmetry breaking mechanisms for the electroweak and \mbox{$U(1)_{PQ} \simeq U(1)_b$} symmetry, the phenomenological viability of the model also requires to indicate at which scales the symmetry breaking occurs. To this end, we take a closer look at the fields involved in the symmetry breaking mechanisms and write down the gauge invariant kinetic terms for the Higgses, the antisymmetric states $\sigma_1$  and $\tilde \sigma_1$ of $U(2)_b$ and the closed string axion $\xi$, which in the present example stems from the $\Z_2$ twisted sector:
\begin{equation}\label{Eq:KineticHiggs-axion}
{\cal L}_{\text{kin}}^{\text{Higgs-axion}} = \sum_{i=1,2}\left| D_\mu H_u^{(i)} \right|^2 + \sum_{i=1,2}\left| D_\mu H_d^{(i)} \right|^2 +  \left| D_\mu \sigma_{1} \right|^2 + \left| D_\mu \tilde \sigma_{1} \right|^2  + \frac{1}{2} (\partial_\mu \xi + M_{\text{string}} B_\mu)^2,
\end{equation}
where the covariant derivatives acting on the various fields are given explicitly by: 
\begin{equation}
\begin{array}{lcl}
D_\mu H_u^{(i)} &=& \partial_\mu H_u^{(i)} + i \frac{g_2}{2} \vec{\tau} \cdot \vec{W}_\mu H_u^{(i)}  + i \frac{g_Y}{2} Y_\mu H_u^{(i)}  \mp  i  B_\mu H_u^{(i)}, \\
D_\mu H_d^{(i)} &=& \partial_\mu H_d^{(i)} + i \frac{g_2}{2} \vec{\tau} \cdot \vec{W}_\mu H_d^{(i)}  - i \frac{g_Y}{2} Y_\mu H_d^{(i)}  \pm  i  B_\mu H_d^{(i)},\\
D_\mu \sigma_1 &=& \partial_\mu \sigma_1 +  i\, q_{\sigma_1}  B_\mu \sigma_1, \qquad D_\mu \tilde \sigma_1 = \partial_\mu \tilde \sigma_1 + i\, q_{\tilde \sigma_1}  B_\mu \tilde \sigma_1, 
\end{array}
\end{equation}
with the Pauli three matrices $\vec{\tau}$, the upper signs for $i=1$ and the lower signs for $i=2$, and $q_{\sigma_1} = - q_{\tilde \sigma_1} =2$. We also opted to rescale the $U(1)_b$ gauge field $B_\mu$ such that the gauge coupling constant does not appear in these relations, which will simplify the expressions later on in this subsection.

For the vacuum configuration of equation (\ref{Eq:MinimumPotential}), the mass for the charged $W$-bosons is given by the expression:
\begin{equation}
m_W^2 = \frac{g_2^2}{2} (v_u^2 + v_d^2),
\end{equation}
while the masses for $Z^0$ and $B_{\mu}$ and the massless $A^\gamma$ follow from the eigenvalues of the following mass matrix:
\begin{equation}
M^2_{\text{gauge}} = \left( \begin{array}{ccc}  \frac{g_Y^2}{2} (v_d^2 + v_u^2) & -\frac{g_2 g_Y}{2} (v_d^2 + v_u^2) & 0\\
-\frac{g_2 g_Y}{2} (v_d^2 + v_u^2) &  \frac{g_2^2}{2} (v_d^2 + v_u^2) & 0 \\
 0 & 0 & M^2_{\text{string}} + q_{\sigma_1}^2 v_{\sigma_1}^2 +q_{\tilde \sigma_1}^2 v_{\tilde \sigma_1}^2  + 2 (v_d^2 + v_u^2)
    \end{array} \right).
\end{equation}
Due to the presence of two $H_u^{(i)}$ doublets with opposite charges under $U(1)_b$ (and similarly for the two $H_d^{(i)}$ doublets), there are no mixing terms between the $B_\mu$ gauge boson and the neutral Standard Model bosons $Z^0$ and $A^{\gamma}$. The eigenvalues of the mass matrix reduce to the usual expressions for the masses of $Z^0$ and $\gamma$: 
\begin{equation}
\begin{array}{l}
m_\gamma^2 = 0, \hspace{0.4in} m_{Z^0}^2 = \frac{g_Y^2 + g_2^2}{2} \left( v_u^2 + v_d^2 \right), \\ 
m_B^2 = M^2_{\text{string}} + q_{\sigma_1}^2 v_{\sigma_1}^2 +q_{\tilde \sigma_1}^2 v_{\tilde \sigma_1}^2  + 2 (v_d^2 + v_u^2).
\end{array}
\end{equation}
The eigenvectors corresponding to the massless $\gamma$ and massive $Z^0$ gauge bosons are related to $W^3_\mu$ and $Y_\mu$ through an $O(2)$ rotation over the Weinberg angle $\theta_W$:\footnote{For completeness, we also give the relations between the Weinberg angle and the various gauge coupling constants:
\begin{equation}
g_2 \sin \theta_W = g_Y \cos \theta_W = e, \hspace{0.2in} \cos \theta_W = \frac{g_2}{\sqrt{g^2_Y + g_2^2}}, \hspace{0.1in}  \sin \theta_W = \frac{g_Y}{\sqrt{g^2_Y + g_2^2}}.
\end{equation}
}
\begin{equation}
\begin{array}{l}
Z_\mu = \cos \theta_W W^3_\mu - \sin \theta_W Y_\mu,\\
A_\mu =  \sin \theta_W W^3_\mu + \cos \theta_W Y_\mu.
\end{array}
\end{equation}
From the mass relations for the $W$-bosons and the $Z^0$-boson, one can immediately deduce  that the $\rho$-parameter remains equal to one at tree-level (similar to the Standard Model and the MSSM):
\begin{equation}
\rho \equiv \frac{m^2_W}{m^2_Z \cos^2 \theta_W} = 1.
\end{equation}
The non-mixing of the Standard Model gauge bosons and the $U(1)_b$ gauge boson in this vacuum configuration also imply that the tree-level analysis does not provide for lower bounds on the string mass scale or the {\it vev}s $v_{\sigma_1}$ and $v_{\tilde\sigma_1}$, in contrast to scenarios with a single closed string axion~\cite{Coriano':2005js} or with a combination of closed string axions and a single open string axion~\cite{Coriano:2008xa,Berenstein:2010ta}. It remains to be checked how quantum corrections affect the $\rho$-parameter and how the experimental value constrains new physics in this model.  

In the mass acquisition process for the gauge boson $B_\mu$, the closed string axion as well as both open string axions and the CP-odd neutral Higgses all take part.  The next step of the analysis then consists of figuring out which CP-odd scalar is eaten by the $U(1)_b$ gauge boson and which CP-odd scalar can serve as the QCD axion. The QCD axion can only be identified in this model if $M_{\text{string}} \geq v_{\sigma_1}, v_{\tilde \sigma_1} \geq 10^9$ GeV, such that the corresponding decay constant meets the phenomenological constraints of the axion window. This consideration suggests a hierarchy between the {\it vev}s for the Standard Model singlets $\sigma_1$ and $\tilde \sigma_1$ on the one hand and the Higgses on the other hand:
\begin{equation} \label{Eq:HierarchyAxionHiggs}
  v_{\sigma_1}, v_{\tilde \sigma_1} \gg v_u, v_d \sim {\cal O}( 100 \text{ GeV}).
\end{equation}
As a consequence, the fractions of the CP-odd neutral Higgses eaten by the gauge boson $B_\mu$ are suppressed by at least a factor $10^{-7}$ with respect to the portions of the open string axions. It also implies that the CP-odd neutral Higgses can be safely disregarded in the identification process for the QCD axion. By focusing on the terms involving the open string axions $(\rho_1, \tilde \rho_1)$ residing inside the complex scalars $(\sigma_1, \tilde \sigma_1)$ and on the closed string axion $\xi$ in equation (\ref{Eq:KineticHiggs-axion}), 
\begin{equation}
\begin{array}{rcl}
{\cal L}_{\rm CP-odd} &=&  \frac{1}{2} (\partial_\mu \rho_1)^2 +   \frac{1}{2} (\partial_\mu \tilde \rho_1)^2 + B_\mu \left( v_{\sigma_1} q_{\sigma_1}   \partial^\mu \rho_1 + v_{\tilde \sigma_1} q_{\tilde \sigma_1}   \partial^\mu \tilde \rho_1\right) + \frac{q_{\sigma_1}^2 v_{\sigma_1}^2 + q_{\tilde\sigma_1}^2 v_{\tilde\sigma_1}^2 }{2} B_\mu B^\mu   \\
&& + \frac{1}{2} (\partial_\mu \xi  )^2 +  M_{\text{string}} B_\mu \partial^\mu  \xi + \frac{1}{2} M_{\text{string}}^2 B_\mu B^\mu,
 \end{array}
\end{equation}
one notices immediately that the axion eaten by the gauge field $B_\mu$ is a mixture of  $\xi$ with $\rho_1$ and $\tilde \rho_1$.  An $SO(3)$ rotation on the axions $(\zeta, \rho_1, \tilde \rho_1)$ allows to bring the action back to the standard St\"uckelberg form, similar to equation (\ref{Eq:StuckClosedOpenAxion}), and the axion $\zeta$ becoming the longitudinal mode of the $U(1)_b$ gauge boson is now identified as:
\begin{equation}\label{Eq:Def-zeta}
\zeta = \frac{M_{\text{string}}\,  \xi +  v_{\sigma_1} q_{\sigma_1} \, \rho_1 +  v_{\tilde \sigma_1} q_{\tilde \sigma_1} \, \tilde\rho_1}{\sqrt{M_{\text{string}}^2 +  q_{\sigma_1} ^2 v_{\sigma_1}^2  +  q_{\tilde \sigma_1}^2 v_{\tilde\sigma_1}^2}}.
\end{equation}
The two other linear combinations form orthogonal directions with respect to $\zeta$:
\begin{eqnarray}
\alpha_1 &=& \frac{  M_{\text{string}} ( q_{\sigma_1} v_{\sigma_1} \rho_1 + q_{\tilde \sigma_1} v_{\tilde \sigma_1} \tilde \rho_1) - (q_{\sigma_1}^2 v_{\sigma_1}^2 +q_{\tilde \sigma_1}^2 v_{\tilde \sigma_1}^2 )\, \xi}{\sqrt{(v_{\sigma_1}^2 q_{\sigma_1}^2 + v_{\tilde \sigma_1}^2 q_{\tilde \sigma_1}^2  ) [M_{\text{string}}^2 + (q_{\sigma_1}^2 v_{\sigma_1}^2 + q_{\tilde \sigma_1}^2 v_{\tilde \sigma_1}^2)] }}, \\
\alpha_2 &=&  \frac{-q_{\tilde \sigma_1} v_{\tilde \sigma_1} \, \rho_1 + q_{\sigma_1} v_{\sigma_1}\, \tilde \rho_1 }{\sqrt{v_{\sigma_1}^2 q_{\sigma_1}^2 + v_{\tilde \sigma_1}^2 q_{\tilde \sigma_1}^2}} ,
\end{eqnarray}
and remain massless CP-odd scalars as there are no couplings of the form $B_\mu\,  \partial^\mu \alpha_i$. As such the $U(1)_b$ will only manifest itself as a perturbative global symmetry once the gauge field acquired its mass via the St\"uckelberg mechanism. The field $\alpha_1$ describes a mixing between the closed and open string axions, whereas the field $\alpha_2$ only contains a mixing between the two open string axions. In models where $M_{\text{string}} \gg v_{\sigma_1}, v_{\tilde \sigma_1}$ the mixing between the closed string axion and the open string axions is very small, and the main constituent of the eaten axion $\zeta$ consists of the closed string axion $\xi$. 

The final step in identifying the QCD axion consists in evaluating the coupling to the topological QCD charge density and the values of the decay constants for the two CP-odd scalars $\alpha_i$. Using the elements discussed in appendix~\ref{A:ChiralRotation}, the coupling of the axions to the QCD anomaly term can be written as 
\begin{equation}
 {\cal L}_{\text{anom}}^{QCD} =   \frac{1}{32 \pi^2} \left[ \frac{\zeta}{f_\zeta}  +  \frac{\alpha_1}{f_{\alpha_1}}  +  \frac{\alpha_2}{f_{\alpha_2}}   \right] \; {\cal A}_{U(1)_b GG} \;  \Tr (G_{\mu \nu} \tilde G^{\mu \nu}),
\end{equation}
with ${\cal A}_{U(1)_b GG} $ the anomaly coefficient (${\cal A}_{U(1)_b GG} = 2N_{\text{generation}} = 2 \times 3$ for this model) and $(f_\zeta, f_{\alpha_1}, f_{\alpha_2})$ denoting the decay constants corresponding to $(\zeta, \alpha_1, \alpha_2)$:
\begin{equation} \label{Eq:AxionDecayConstantsZ6}
\begin{array}{lcl}
f_\zeta &=& \frac{\sqrt{M_{\text{string}}^2 +  \left( v^2_{\sigma_1} q_{\sigma_1}^2+ v_{\tilde\sigma_1}^2 q_{\tilde\sigma_1}^2 \right)}}{C_{\xi GG} + 1} ,\\
f_{\alpha_1} &=& \frac{ M_{\text{string}} \sqrt{q_{\sigma_1}^2 v_{\sigma_1}^2 + q_{\tilde\sigma_1}^2 v_{\tilde\sigma_1}^2} \sqrt{M_{\text{string}}^2 + \left( v^2_{\sigma_1} q_{\sigma_1}^2+ v_{\tilde\sigma_1}^2 q_{\tilde\sigma_1}^2 \right)} }{ \left(  M_{\text{string}}^2  - C_{\xi GG} (v_{\sigma_1}^2 q_{\sigma_1}^2 + v_{\tilde\sigma_1}^2 q_{\tilde\sigma_1}^2 )  \right)},\\
f_{\alpha_2} &=& \frac{f_{\rho_1} }{ f_{\tilde \rho_1}} \sqrt{f_{\rho_1}^2 + f_{\tilde \rho_1}^2} = \left| \frac{q_{\sigma_1} v_{\sigma_1} }{  q_{\tilde\sigma_1} v_{\tilde\sigma_1} } \right| \sqrt{q_{\sigma_1}^2 v_{\sigma_1}^2 + q_{\tilde\sigma_1}^2 v_{\tilde\sigma_1}^2} .
\end{array}
\end{equation}
The real dimensionless constant $C_{\xi GG}$ represents the coupling coefficient of the closed string axion $\xi$ to the QCD anomaly (see appendix~\ref{A:ChiralRotation}). The axion also couples to photons through the anomaly term of the electromagnetic field strength, as in equation (\ref{Eq:AnomalyAxionEM}). The model-dependent parameters $C_{\alpha_i \gamma \gamma}$ reads for the CP-odd scalars $\alpha_i$: 
\begin{equation} \label{Eq:AxionPhotonCouplingConstantZ6}
\begin{array}{ccl}
C_{\alpha_1 \gamma \gamma} &=& \frac{4 M_{\text{string}}^2 - C_{\xi \gamma \gamma} \left( v_{\sigma_1}^2 q_{ \sigma_1}^2  +v_{\tilde \sigma_1}^2 q_{\tilde \sigma_1}^2  \right)}{ M_{\text{string}}^2 - C_{\xi G G} \left( v_{\sigma_1}^2 q_{ \sigma_1}^2  +v_{\tilde \sigma_1}^2 q_{\tilde \sigma_1}^2  \right) }, \\
C_{\alpha_2 \gamma \gamma} &=& \frac{7 v_{\sigma_1}^2 q_{ \sigma_1}^2 + 11 v_{\tilde \sigma_1}^2 q_{\tilde \sigma_1}^2 }{v_{\tilde \sigma_1}^2 q_{\tilde \sigma_1}^2 },
\end{array}
\end{equation}
where $C_{\xi \gamma \gamma}$ represents the coupling coefficient of the closed string axion $\xi$ to the electromagnetic anomaly. Table~\ref{tab:ScalesDecayConstants} exhibits values of the decay constants $f_{\alpha_i}$ and coupling coefficients $C_{\alpha_i \gamma \gamma}$ for various values of the string scale and the saxion {\it vev}s $v_{\sigma_1}$ and $v_{\tilde \sigma_1}$. 
\mathtab{ {\small
\begin{array}{|c||c|c|c|}
\hline
\multicolumn{4}{|c|}{\text{ \bf Scales and axion decay constants } }\\
\hline \hline
 & \multicolumn{3}{|c|}{M_{\text{string}} \sim 10^{12} - 10^{16} \text{ GeV }}\\
\hline
v_{\sigma_1}& 10^{11} \text{ GeV }& 10^{10} \text{ GeV }&10^{9} \text{ GeV }    \\
v_{\tilde \sigma_1}& 10^{9} \text{ GeV }&10^{10} \text{ GeV } &10^{11} \text{ GeV }    \\
\hline
f_{\alpha_1}& 2 \cdot  10^{11}  \text{ GeV }&3 \cdot  10^{10}  \text{ GeV } & 2 \cdot  10^{11}  \text{ GeV } \\
f_{\alpha_2}& 2 \cdot  10^{13}  \text{ GeV } &3 \cdot  10^{10}  \text{ GeV } &2 \cdot  10^{9}  \text{ GeV }  \\
\hline
C_{\alpha_1\gamma \gamma}& 4 & 4 & 4  \\
C_{\alpha_2\gamma \gamma}& 70 \cdot 10^{3} & 18 & 11  \\
\hline
\end{array}}
}{ScalesDecayConstants}{Overview of the axion decay constants and the axion-photon coupling coefficients for the supersymmetric DFSZ model on $T^6/\Z_6$ depending on the scales $(M_{\text{string}}, v_{\sigma_1}, v_{\tilde \sigma_1})$ with closed string axion coefficients $C_{\xi GG} = 1$ and $C_{\xi \gamma \gamma} = 1$. For $M_{\text{string}} > v_{\sigma_1}, v_{\tilde \sigma_1}$, the saxions {\it vev}s set the scales for the decay constants, as can be inferred from equations (\ref{Eq:AxionDecayConstantsZ6}) and (\ref{Eq:AxionPhotonCouplingConstantZ6}).} 

From equation (\ref{Eq:AxionDecayConstantsZ6}) and table \ref{tab:ScalesDecayConstants}, one notices that the magnitude of the decay constant $f_{\alpha_1}$ is set by the largest of the two saxion {\it vev}s $v_{\sigma_1}$ and $v_{\tilde \sigma_1}$, and therefore falls into the axion window whenever the highest saxion {\it vev} lies in the axion window. As expected from the formula in (\ref{Eq:AxionPhotonCouplingConstantZ6}), the coupling coefficient $C_{\alpha_1 \gamma \gamma}$ for this CP-odd scalar remains a constant of order ${\cal O}(1)$, given that the string mass scale is assumed to be higher than the saxion {\it vev}s. Hence, the CP-odd scalar $\alpha_1$ can safely play the r\^ole of the QCD axion. The other CP-odd scalar $\alpha_2$ can be seen as an axion-like particle. That is to say, the decay constant $f_{\alpha_2}$ can be higher than the upper bound of the axion window, even though both saxion {\it vev}s lie in the axion window. Furthermore, the axion-photon coupling constant $C_{\alpha_2 \gamma \gamma}$ depends strongly on the largest saxion {\it vev}, such that the axion-photon interaction can be less suppressed than the axion-gluon interaction. Note that the ratio $f_{\alpha_2}/C_{\alpha_2 \gamma \gamma}$ is of the order $10^9$ GeV or smaller.

The QCD axion $\alpha_1$ plays the r\^ole of the pseudo-Goldstone boson arising from the spontaneous breaking of the global anomalous $U(1)_b$, left over after the St\"uckelberg mechanism. The spontaneous breaking of this symmetry is triggered by the non-vanishing {\it vev}s of the saxions $s_1$ and $\tilde s_1$ residing in the Standard Model singlet fields $\sigma_1$ and $\tilde \sigma_1$. Hence, an investigation of the vacuum structure for the saxions is in order as well. Under the hierarchy assumption of equation~(\ref{Eq:HierarchyAxionHiggs}), not all terms in the scalar potential of equation (\ref{Eq:FullScalarPotential}) contribute with the same order of magnitude. In this sense, the vacuum structure for the saxions is determined by those terms that only involve $\sigma_1$ and $\tilde \sigma_1$ and do not contain any of the Higgs fields. These terms originate partly from the F-term scalar potential, the $U(1)_b$ D-term potential and the soft supersymmetry breaking terms and can be combined into the following pure saxion potential:
\begin{eqnarray}
V_{\rm saxion} (\sigma_1, \tilde \sigma_1) =  \frac{g_2^2}{8} \left(2 |\sigma_1|^2 - 2 |\tilde \sigma_1|^2 \right)^2 + \hat m^2_{\sigma_1}  |\sigma_1 |^2  +   \hat  m^2_{\tilde \sigma_1}  |\tilde \sigma_1 |^2  - m_{12}^2\, \sigma_1 \tilde\sigma_1 - \ov m_{12}^{2}\, \sigma_1^{\dagger} \tilde \sigma_1^{\dagger},
\end{eqnarray}
where we introduced the following notation:
\begin{equation}
 \hat m^2_{\sigma_1}  \equiv m^2_{\sigma_1}  + |\tilde \kappa|^2, \qquad  \hat m^2_{\tilde \sigma_1}  \equiv m^2_{\tilde \sigma_1}  + |\tilde \kappa|^2.
\end{equation}
Next, we impose that the vacuum configuration for the singlets $\sigma_1$ and $\sigma_2$ as given in equation (\ref{Eq:MinimumPotential}) minimise the saxion potential and do not break supersymmetry spontaneously, i.e.~$V_{\rm saxion} (\langle \sigma_1 \rangle, \langle \tilde \sigma_1 \rangle) = 0$. These considerations can be recast into the following three constraint equations:
\begin{eqnarray}
2 g_b^2 (v_{\sigma_1}^2 - v_{\tilde \sigma_1}^2) + \hat m^2_{\sigma_1} -   \frac{v_{\tilde \sigma_1}}{v_{\sigma_1}} {\rm Re}\left(m_{12}^2 e^{i (\phi_1 + \tilde \phi_1)}\right)   = 0, \label{Eq:AxionVacuumCon1} \\
-2 g_b^2 (v_{\sigma_1}^2 - v_{\tilde\sigma_1}^2) + \hat m^2_{\tilde\sigma_1} -   \frac{v_{\sigma_1}}{v_{\tilde\sigma_1}} {\rm Re}\left(m_{12}^2 e^{i (\phi_1 +\tilde \phi_1)}\right)   = 0, \label{Eq:AxionVacuumCon2} \\
 g_b^2 (v_{\sigma_1}^2 - v_{\tilde \sigma_1}^2)^2 + \hat m^2_{\sigma_1} v_{\sigma_1}^2 + \hat m^2_{\sigma_2} v_{\sigma_2}^2 - 2  v_{\sigma_1} v_{\tilde\sigma_1} {\rm Re}\left(m_{12}^2 e^{i (\phi_1 + \phi_2)}\right)   = 0. \label{Eq:AxionVacuumCon3}
\end{eqnarray}
In order for the potential to be bounded from below, even along the D-flat direction $|\sigma_1| = |\tilde \sigma_1| $, the following inequality has to hold at all scales:
\begin{equation}
 \hat m^2_{\sigma_1} +  \hat m^2_{\tilde \sigma_1} = 2 |\kappa|^2 + m^2_{\sigma_1}  + m^2_{\tilde \sigma_1}   >  2\, {\rm Re}\left(m_{12}^2\right).
\end{equation}
However, we also need to make sure that the saxion vacuum configuration breaks spontaneously the Peccei-Quinn symmetry. To this end, we focus on the quadratic terms of the saxion potential along the D-flat directions 
\begin{eqnarray}
V_{\rm saxion}^{\rm quad.} = (\sigma_1^{\dagger}, \tilde\sigma_1) \left( \begin{array}{cc} \hat m_{\sigma_1}^2 & - \ov m_{12}^{2} \\ - m_{12}^2 & \hat m_{\tilde \sigma_1}^2 \end{array} \right) \left( \begin{array}{c} \sigma_1 \\ \tilde\sigma_1^{\dagger}  \end{array}\right), 
\end{eqnarray}
where the hermitian mass matrix can be diagonalised containing the following two eigenvalues
\begin{eqnarray}
\lambda_\pm = \frac{1}{2} \left[ \hat m_{\sigma_1}^2 + \hat m_{\tilde\sigma_1}^2 \pm \sqrt{4 |m_{12}|^4 + ( \hat m^2_{\sigma_1}- \hat m^2_{\tilde\sigma_1})^2}    \right].
\end{eqnarray}
In order for the singlets $\sigma_1$ and $\sigma_2$ to acquire a non-zero {\it vev}, one of the eigenvalues has to be negative, implying the following relation between the mass parameters:
\begin{eqnarray}
|m_{12}|^4 > m_{\sigma_1}^2 m_{\tilde\sigma_1}^2.
\end{eqnarray} 
This relation, which is equivalent to demanding a negative determinant for the mass matrix, should hold at and below the scale at which the global $U(1)_b \simeq U(1)_{PQ}$ symmetry is broken.

\subsubsection{The Higgs scalar potential}\label{Sss:HiggsSector}

The picture developed in the previous subsections distinguishes three symmetry breaking processes, occurring at different energy scales. At $M_{\text{string}}$, the local $U(1)_b$ symmetry acquires a mass by eating a stringy axion, such that the symmetry survives as a global perturbative Peccei-Quinn symmetry. This global symmetry is spontaneously broken when the saxions $s_1$ and $\tilde s_1$ develop non-vanishing {\it vev}s $v_{\sigma_1}$ and $v_{\tilde \sigma_1}$ at energy-scales between $10^9$ and \mbox{$10^{12}$ GeV}. At energy scales of the order of 100 GeV, the CP-even neutral components of the Higgses develop a non-zero {\it vev}, such that the electroweak symmetry is broken via the  Brout-Englert-Higgs mechanism.

Before determining the constraint equations, which express that the vacuum configuration in (\ref{Eq:MinimumPotential}) minimise the scalar potential for the Higgses, we integrate out the degrees of freedom of the massive $U(1)_b$ gauge boson and of the saxions, supported by the indicative hierarchy of energy scales in (\ref{Eq:HierarchyAxionHiggs}). The procedure of integrating out the heavy $U(1)_b$ is briefly outlined in appendix~\ref{A:Integration}, with the r\^ole of $U(1)_a$ there now played by the electroweak gauge symmetry $SU(2)\times U(1)_Y$. The effect of integrating out $U(1)_b$ can be captured by additional prefactors in the K\"ahler potential for the Higgses:
\begin{equation}  
\begin{array}{lcl}
K^{\text{SUSY}} &=& \omega_-\,  {H_u^{(1)}}^\dagger e^{2 g_2 V^{SU(2)}}  e^{2 g_Y  V^{Y} Y}H_u^{(1)} + \omega_+ \, {H_d^{(1)}}^\dagger  e^{2 g_2 V^{SU(2)}}  e^{2 g_Y  V^{Y} Y} H_d^{(1)} \\
&&+ \omega_+\, {H_u^{(2)}}^\dagger  e^{2 g_2 V^{SU(2)}}  e^{2 g_Y  V^{Y} Y} H_u^{(2)} + \omega_-\, {H_d^{(2)}}^\dagger  e^{2 g_2 V^{SU(2)}}  e^{2 g_Y  V^{Y} Y} H_d^{(2)},
\end{array}
\end{equation}
where we introduced the notation (remember  $q_{H_u^{(1)}} = q_{H_d^{(2)}} =  -1 = -q_{H_u^{(2)}} =- q_{H_d^{(1)}}$ under $U(1)_b$):
\begin{equation}
\omega_\pm \equiv 1 \pm 2 z + 2  z^2
\qquad
\text{with}
\quad 
z= \frac{g_2}{2 \, M_{\text{string}}} \langle U_{\zeta} + U_{\zeta}^{\dagger} \rangle
,
\end{equation}
with $U_{\zeta}$ the combination of open and closed string saxions which complexify the axion $\zeta$ defined in equation~(\ref{Eq:Def-zeta}). In the Type IIA string models at hand, the relevant closed string saxion is a complex structure modulus.

Upon integrating out the heavy vector multiplet, the $U(1)_b$ D-term contribution to the scalar potential will no longer be present, but the D-term potential associated to the  
 electroweak symmetry is slightly altered due to the prefactors in the K\"ahler potential:  
 \begin{equation}
\begin{array}{rcl}
V_D^{\text{eff}} &=& \frac{(g_Y^2 + g_2^2)}{8} \left( \omega_-  \left|H_u^{(1)}\right|^2 - \omega_- \left|H_d^{(2)}\right|^2  + \omega_+ \left|H_u^{(2)}\right|^2 - \omega_+  \left|H_d^{(1)}\right|^2 \right)^2 \\
&& + \frac{g_2^2}{2} \left( \omega_- \omega_+  \left|{H_u^{(1)}}^\dagger H_u^{(2)} \right|^2 + \omega_- \omega_+  \left|{H_d^{(1)}}^\dagger H_d^{(2)} \right|^2 + \omega_- \omega_+  \left|{H_u^{(1)}}^\dagger H_d^{(1)} \right|^2   \right. \\
&& \hspace{0.5in} + \omega_-^2  \left|{H_u^{(1)}}^\dagger H_d^{(2)} \right|^2 + \omega_+^2  \left|{H_u^{(2)}}^\dagger H_d^{(1)} \right|^2  + \omega_- \omega_+  \left|{H_u^{(2)}}^\dagger H_d^{(2)} \right|^2 \\
&& \hspace{0.5in} \left. - \omega_- \omega_+  \left|H_u^{(1)}\right|^2  \left|H_u^{(2)}\right|^2 - \omega_- \omega_+ \left|H_d^{(1)}\right|^2  \left|H_d^{(2)}\right|^2   \right). 
\end{array}
\end{equation}
It is noteworthy that the vacuum configuration in (\ref{Eq:MinimumPotential}) can still correspond to a minimum of the effective D-term potential.

The prefactors $\omega_\pm$ will also appear in the F-term contributions to the scalar potential, where we have to integrate out the degrees of freedom of the saxions. Replacing\footnote{Due to the coupling between the Higgses and the chiral superfields $\Sigma_1$ and $\tilde \Sigma_1$, the superpotential is not separable into light and heavy superfields. In this respect, one ought to integrate out the heavy chiral superfields by ensuring the stationarity of the superpotential, as argued in e.g.~\cite{Brizi:2009nn}. However, as the coupling (\ref{Eq:4ptCouplingDFSZ}) of the Higgses to the heavy superfields $\Sigma_1$ and $\tilde \Sigma_1$ is string-scale suppressed with respect to the supersymmetric mass term (\ref{Eq:3ptCouplingDFSZ}) for the heavy superfields, the replacement procedure is still assumed to be valid. } the superfields $\Sigma_1$ and $\tilde \Sigma_1$  by their {\it vev}s in the superpotential (\ref{Eq:FullSuperpotentialDFSZ}) yields the effective superpotential
\begin{equation}
{\cal W}_{\text{DFSZ}}^{\text{eff}} = \mu_1 \langle \Sigma_1\rangle H_d^{(1)} \cdot H_u^{(2)} + \mu_2 \langle \tilde \Sigma_1 \rangle H_d^{(2)} \cdot H_u^{(1)},
\end{equation}  
leading to the effective F-term potential
\begin{equation}
V_{F}^{\text{eff}} = \frac{|\mu_1|^2 v_{\sigma_1}^2}{ 2\omega_+} \left( \left|H_d^{(1)}\right|^2 + \left|H_u^{(2)}\right|^2   \right) +  \frac{|\mu_2|^2 v_{\tilde \sigma_1}^2}{ 2\omega_-} \left( \left|H_d^{(2)}\right|^2 + \left|H_u^{(1)}\right|^2   \right).
\end{equation}
Integrating out the heavy superfields $\Sigma_1$ and $\tilde \Sigma_1$ in the same way, one finds that the soft supersymmetry breaking potential terms  reduce to:
\begin{equation}
\begin{array}{ccl}
V_{\text{soft}}^{\text{eff}} &=& m^2_{H_u^{(1)}}  \left|H_u^{(1)}\right|^2 + m^2_{H_u^{(2)}}  \left|H_u^{(2)}\right|^2 + m^2_{H_d^{(1)}}  \left|H_d^{(1)}\right|^2 + m^2_{H_d^{(2)}}  \left|H_d^{(2)}\right|^2  \\
&&+ \left(c_1 \frac{v_{\sigma_1}}{\sqrt{2}} e^{i\, \phi_1} H_d^{(1)} \cdot H_u^{(2)} + h.c. \right) + \left(c_2 \frac{v_{\tilde \sigma_1}}{\sqrt{2}}  e^{i\, \tilde \phi_1}  H_d^{(2)} \cdot H_u^{(1)}  + h.c. \right)\\
&& + ( m_{11}^2 e^{i\, \frac{\xi_1}{f_{\xi_1}}}  e^{i\, \delta_v  \frac{\xi_2}{f_{\xi_2}}}  H_d^{(1)} \cdot  H_u^{(1)} + h.c. ) + (m_{22}^2 e^{i\, \frac{\xi_1}{f_{\xi_1}}}  e^{i\, \delta_v   \frac{\xi_2}{f_{\xi_2}}} H_d^{(2)} \cdot  H_u^{(2)} + h.c. ).
\end{array}
\end{equation} 
The additional axion terms in the last line might seem unexpected, but they are an unequivocal consequence of the fact that the axions $\xi_1$ and $\xi_2$ still correspond to massless excitations at the electroweak breaking scale. As such they cannot be integrated out and remain present in the scalar potential. Using the toolbox of appendix \ref{A:ChiralRotation} we have chosen a configuration where the axions appear in the third line instead of the second line, with the dimensionless constant $ \delta_v  = \frac{v_{ \sigma_1}^2 - v_{\tilde \sigma_1}^2}{ 2 v_{\tilde \sigma_1}^2}$ measuring the difference between the saxion {\it vev}s.    

In summary, integrating out the heavy degrees of freedom (massive $U(1)_b$ gauge boson and the saxions) in the scalar potential (\ref{Eq:FullScalarPotential}) leads to an effective potential that only depends on the degrees of freedom of the Higgs doublets and the massless axions:
\begin{equation}
V^{\text{eff}}_{\text{Higgs}} = V_{F}^{\text{eff}}   + V_D^{\text{eff}} + V_{\text{soft}}^{\text{eff}}.
\end{equation}
However, in the analysis of the vacuum configuration for this effective potential we neglect the interactions between the Higgses and the massless axions, as they are suppressed by the decay constants. In a more inclusive analysis, these interactions would lead to mixings between the axions and the CP-odd neutral components of the Higgses. 

In order to determine the conditions for which the Higgs fields develop a non-zero {\it vev}, as indicated in equation (\ref{Eq:MinimumPotential}), it suffices to investigate small deviations from the supposed vacuum configuration of the effective potential along the neutral Higgs directions. The quartic interactions of the D-term ensure that the Higgs potential is bounded from below, except along the directions where the D-terms vanish, i.e.~$|h_u^{0,(1)}| = |h_d^{0,(2)}| = |h_u^{0,(2)}| = |h_d^{0,(1)}|$. Requiring that the Higgs potential is also bounded from below along these D-flat directions, amounts to imposing the following relation among the parameters:
\begin{equation}\label{Eq:HiggsBounded}
\sum_{i=1}^2 m_{H_u^{(i)}}^2 + \sum_{i=1}^2 m_{H_d^{(i)}}^2  +  \frac{|\mu_1|^2}{ \omega_+}  v_{ \sigma_1}^2 +  \frac{|\mu_2|^2}{ \omega_-}  v_{\tilde \sigma_1}^2 > \sqrt{2} |c_1| v_{\sigma_1} + \sqrt{2} |c_2| v_{\tilde \sigma_1} + 2 |m_{11}|^2  + 2 |m_{22}|^2. 
\end{equation}
Note that we used the rescaling freedom of the arguments in the Higgs-doublets to eliminate the phases of the complex soft supersymmetry breaking parameters $c_1$, $c_2$, $m_{11}^2$ and $m_{22}^2$. Along the D-flat directions, the Higgs potential reduces to its quadratic part:
\begin{equation}\label{Eq:MassmatrixNeutralHiggs}
\begin{array}{lcl}
V_{\rm Higgs}^{0, \rm quad.} &=& \left( h_u^{0,(1)},  h_d^{0,(2)\dagger},  h_u^{0,(2)} ,  h_d^{0,(1)\dagger} \right) M^2_{\text{Higgs}} 
 \left( \begin{array}{c} h_u^{0,(1)\dagger} \\  h_d^{0,(2)} \\  h_u^{0,(2)\dagger}  \\ h_d^{0,(1)}    \end{array} \right), 
\end{array}
\end{equation}
with the Higgs mass matrix $M^2_{\text{Higgs}}$ given by
\begin{equation}
M^2_{\text{Higgs}} = \left( \begin{array}{cccc} m^2_{H_u^{(1)}}  + \frac{|\mu_2|^2}{2 \omega_-}  v_{\tilde \sigma_1}^2   & - |c_2| \frac{v_{\tilde \sigma_1} }{\sqrt{2}}  & 0 & -|m_{11}|^2 \\  - |c_2| \frac{v_{\tilde \sigma_1} }{\sqrt{2}}  &  m^2_{H_d^{(2)}}  + \frac{|\mu_2|^2}{2 \omega_-}  v_{\tilde \sigma_1}^2& - |m_{22}|^2& 0 \\ 0&-|m_{22}|^2&m^2_{H_u^{(2)}} + \frac{|\mu_1|^2}{2 \omega_+}  v_{ \sigma_1}^2 & - |c_1| \frac{v_{\sigma_1}}{\sqrt{2}}\\  - |m_{11}|^2& 0 & - |c_1| \frac{v_{\sigma_1}}{\sqrt{2}}  &  m^2_{H_d^{(1)}}  + \frac{|\mu_1|^2}{2 \omega_+}  v_{ \sigma_1}^2 \end{array} \right).
\end{equation}
In order for the neutral Higgses to acquire non-zero {\it vev}s, two of the eigenvalues of the Higgs mass matrix $M^2_{\text{Higgs}}$ have to be negative. For the sake of the argument, let us assume that the following relations among the soft supersymmetry breaking parameters are valid near the electroweak symmetry breaking scale:
\begin{equation}
m^2_{H_u^{(1)}} =  m^2_{H_d^{(2)}}, \qquad  m^2_{H_u^{(2)}} =  m^2_{H_d^{(1)}}, \qquad |m_{11}|^2= |m_{22}|^2.
\end{equation}
Under these assumptions, the condition that the Higgs mass matrix has two negative eigenvalues can be recast into one single inequality: 
\begin{equation}\label{Eq:HiggsNegativeEigenvalue}
|m_{11}|^4 > \left(m^2_{H_u^{(1)}}  + \frac{|\mu_2|^2}{2 \omega_-}  v_{\tilde \sigma_1}^2  + |c_2|  \frac{v_{\tilde \sigma_1} }{\sqrt{2}}  \right) \left( m^2_{H_u^{(2)}} + \frac{|\mu_1|^2}{2 \omega_+}  v_{ \sigma_1}^2 +  |c_1| \frac{v_{\sigma_1}}{\sqrt{2}}  \right).
\end{equation}
It is important to stress that this inequality only has to be satisfied at and below the electroweak symmetry breaking scale, while condition (\ref{Eq:HiggsBounded}) is valid at all energy scales.
In order for the {\it vev}s given in equation (\ref{Eq:MinimumPotential}) to represent a minimum for the effective Higgs potential, the following four constraint equations have to be imposed:
\begin{eqnarray}
\frac{|\mu_2|^2}{\omega_-}  v_{\tilde \sigma_1}^2  + \frac{g_Y^2 + g_2^2}{2} \left( \omega_- + \omega_+ \right) \omega_- (v_u^2 - v_d^2)  + 2 m^2_{H_u^{(1)}} -  \left( |c_2| \frac{ v_{\tilde \sigma_1}}{\sqrt{2}} +  |m_{11}|^2 \right) \frac{v_d}{v_u}  = 0, \label{Eq:HiggsConstraintwithM1}\\
\frac{|\mu_1|^2}{\omega_+}  v_{\sigma_1}^2  + \frac{g_Y^2 + g_2^2}{2}  \left( \omega_- + \omega_+ \right) \omega_+ (v_u^2 - v_d^2)  + 2 m^2_{H_u^{(2)}} -  \left( |c_1| \frac{v_{\sigma_1}}{\sqrt{2}} + |m_{22}|^2 \right) \frac{v_d}{v_u}    = 0,\label{Eq:HiggsConstraintwithM2}\\
\frac{|\mu_1|^2}{\omega_+}  v_{\sigma_1}^2  - \frac{g_Y^2 + g_2^2}{2}   \left( \omega_- + \omega_+ \right) \omega_+  (v_u^2 - v_d^2) + 2 m^2_{H_d^{(1)}} -  \left( |c_1| \frac{ v_{\sigma_1}}{\sqrt{2}} +  |m_{11}|^2 \right) \frac{v_u}{v_d}    = 0,\label{Eq:HiggsConstraintwithM3}\\
\frac{|\mu_2|^2}{\omega_-}  v_{\tilde \sigma_1}^2  - \frac{g_Y^2 + g_2^2}{2} \left( \omega_- + \omega_+ \right) \omega_- (v_u^2 - v_d^2)  + 2 m^2_{H_d^{(2)}} -  \left(|c_2| \frac{  v_{\tilde \sigma_1}}{\sqrt{2}} + |m_{22}|^2 \right) \frac{v_u}{v_d}    = 0.\label{Eq:HiggsConstraintwithM4}
\end{eqnarray} 
 
The final step in analysing the Higgs potential consists in investigating the mass spectrum for the various massive Higgs particles that are not eaten by the electroweak gauge bosons. Given the complexity of the Higgs potential, the analysis here is limited to a qualitative discussion, while a full quantitative discussion is postponed to future work~\cite{HoneckerStaessens:2014}. The model at hand contains  two up-type and two down-type Higgs doublets, which corresponds to four complex charged Higgses, four real CP-even Higgses and four real CP-odd Higgses. Upon electroweak symmetry breaking, a complex charged Higgs is eaten by the $W^\pm$ bosons, while the $Z^0$ absorbs a real neutral CP-odd Higgs as the longitudinal mode. This effectively leaves 13 massive Higgs particles: three complex charged, four real neutral CP-even and three real neutral CP-odd Higgses.

\subsubsection{A supersymmetric DFSZ Model on $T^6/\Z_6'$ with hidden $USp(6)$ }\label{Ss:Z6p}

A global five-stack D6-brane model with hidden $USp(6)_h$  gauge group was constructed in~\cite{Gmeiner:2008xq} on the $T^6/\Z_6'$ orbifold, with complex structure $U^{(3)} = \frac{1}{2}$ for the two-torus only experiencing a $\Z_2$ action. The non-chiral spectrum and beta-function coefficients can also be found in~\cite{Gmeiner:2009fb}, where the 1-loop gauge treshold corrections were explicitly computed. Tables~\ref{tab:5stack-Z6p-Chiral} and \ref{tab:5stack-Z6p-NonChiral} in appendix~\ref{A:Spectra} give an overview of the massless chiral and vector-like matter states after the right-symmetric group $USp(2)_c$ is spontaneously broken to $U(1)_c$, under a continuous displacement $\sigma_c^2$ of (or by turning on a Wilson line $\tau_c^2$ along) the $c$-brane on the two-torus without $\Z_2$ action.

Here, we are mostly interested in the Higgs-sector of the model and in the appearance of massless antisymmetric states of $U(2)_b$ that can serve as axion superfields. In the left-right symmetric phase, six Higgs doublet pairs $(H_u, H_d)$ arise as $(\ov \2, \2)$ chiral states of $U(2)_b \times USp(2)_c$ in the $bc=bc'$-sector, and three similar Higgs doublet pairs arise from the $b(\theta c)$-sector. The $b(\theta^2 c)$-sector gives rise to a non-chiral pair $[(\2,\2) +h.c.]$ of $U(2)_b \times USp(2)_c$. Under the spontaneous breaking of $USp(2)_c$ to a massless Abelian $U(1)_c$ gauge symmetry the chiral Higgs-doublet pairs decompose into nine chiral up-type Higgs-doublets $H_u^{(i)} = (\ov \2)^{(1)}$ 
and nine chiral down-type Higgs-doublets $H_d^{(i)} = (\ov \2)^{(-1)}$ 
of $U(2)_b^{(\times U(1)_c)}$ (with $i\in \{1, \ldots, 9 \}$). The non-chiral pair splits up into a non-chiral pair $[ \2^{(-1)} + h.c.]$ of $U(2)_b^{(\times U(1)_c)}$  and a second non-chiral pair $[ \2^{(1)} + h.c.]$, such that there exist two more up-type Higgs doublets $H_u^{(10,11)}$ and two more down-type Higgs doublets $H_d^{(10,11)}$, as listed in table~\ref{tab:5stack-Z6p-NonChiral}.

There is a crucial difference between the Higgses emerging as chiral states and those emerging as non-chiral states. Namely, the spontaneous breaking of $USp(2)_c$ induces a mass for the Higgs states realised in a non-chiral sector, analogously to the discussion for the Higgs sector of the $T^6/\Z_6$ model in section~\ref{Ss:Z6}. The Higgs states residing in the chiral sector on the other hand remain massless under a continuous displacement of (or a Wilson line along) the $c$-brane. By displacing also brane $d$ such that $\sigma^2_d \neq \sigma^2_c$ and $\sigma^2_{d}\neq 0$, various other vector-like matter states acquire a mass, as indicated in the right column of table~\ref{tab:5stack-Z6p-NonChiral} by the subscript $m$ of the multiplicity. In that right column we also find the candidates for the open string axion superfields $\Sigma_{0 \ldots 3}, \tilde \Sigma_{0 \ldots 3}$ arising as vector-like matter states in the $b (\theta^2 b)'$ sector, These states are left massless if the $b$-brane and it orientifold image $b'$ are not displaced with respect to each other ($\sigma^2_b = \sigma^2_{b'}=0$).

As a next step, we investigate the embedding of a supersymmetric DFSZ model with this global D6-brane model and focus on the perturbatively allowed couplings in the superpotential between the relevant massless states. A thorough investigating of the cubic Yukawa couplings has already been done in~\cite{Honecker:2012jd} (up to the subtlety due to D6-branes parallel along the two-torus without $\Z_2$ action mentioned in footnote~\ref{Fn:Existence-Yukawas}, which might also enforce higher order couplings with `adjoints' of the $U(1)$'s involved), so that we can mainly focus here on the couplings between the Higgses and the axion superfields. Disregarding the massive Higgs states from the non-chiral sector, all massless `chiral' Higgses can be used to form a three-point coupling with an antisymmetric superfield $\Sigma_i$ of $U(2)_b$, carrying a charge $+2$ under  $U(1)_b \simeq U(1)_{PQ}$. The resulting superpotential for this model is thus captured by:
\begin{equation}\label{Eq:DFSZSuperpotentialZ6p}
{\cal W}_{\text{DFSZ}} = \mu_{ijk} H_u^{(i)} \cdot H_d^{(j)} \Sigma_k +  {\cal W}_{\text{quarks}} + {\cal W}_{\text{leptons}},
\end{equation}
where the indices $(i,j,k)$ refer to the fixed points at which the various states are localized and the terms involving the quarks and leptons can be found in~\cite{Honecker:2012jd}.

The superpotential in (\ref{Eq:DFSZSuperpotentialZ6p}) fits into the class of models described in~\cite{Coriano:2008xa}, with a single axion superfield coupled to one up-type Higgs and one down-type Higgs. Hence, the Higgs-axion potential for this model can be further analysed in the same manner as discussed in~\cite{Coriano:2008xa}, from which we immediately infer that the $\rho$-parameter will no longer be 1 at tree-level, imposing a lower bound on the string mass scale. The identification of the St\"uckelberg particle eaten by the $U(1)_b$ gauge symmetry is completely analogous to the discussion in \ref{Sss:GSMechanismOpenStringAxions} and the model does not exhibit an axion-like particle. 
The most exciting aspect of this model is however not its ability to realise a supersymmetric DSFZ model, but rather its potential to realise a gaugino condensate due to the occurrence of a hidden gauge group $USp(6)_h$. For a $USp(2N)$ gauge factor with $N$ large, the beta-function coefficient becomes more negative, increasing the likelihood of having a strongly coupled gauge theory in the hidden sector. We will investigate the possibility for gaugino condensation and its implications in section~\ref{Sss:SoftSUSYbreakingModels}.   

\subsubsection{The value of the string scale $M_{\text{string}}$}\label{Sss:StringScale}

In previeous section, we explored the possible existence of lower bounds on the $M_{\text{string}}$ using the $\rho$-parameter. In this section, we explore bounds on the string mass scale by virtue of the relation between $M_{\text{string}}$ and the gauge coupling for the D-brane gauge theories. 
The weakness of four dimensional gravity relates the string coupling $g_{\text{string}} \equiv e^{\phi_{10}}$ and the compact volume $\text{Vol}_6$ to the Planck and string scale, see e.g.~\cite{Blumenhagen:2006ci},
\begin{equation}\label{Eq:mass-ratios}
\frac{M^{2}_{\text{Planck}}}{M_{\text{string}}^2} = \frac{4\pi}{g^2_{\text{string}}} \, \frac{\text{Vol}_6}{\alpha^{\prime 3}}
\stackrel{\text{examples}}{=}
\frac{4 \pi \; v_1 v_2 v_3 }{g^2_{\text{string}}}  
,
\end{equation}
with $v_1, v_2, v_3$ the volume per two-torus in units of $\alpha'$.
 
At tree level, the gauge couplings are determined by the string coupling and the volume of the 3-cycle on which the fractional D6-brane $a$ is wrapped,
\begin{equation}
\frac{4 \pi}{g_{SU(N_a),{\rm tree}}^2} =  2\pi \Re \left({\rm f}_{SU(N_a)}^{\text{tree}} \right)
= \frac{1}{2} \, \frac{1}{g_{\text{string}}} \frac{\text{Vol}_{3,a}}{\ell_s^3}
\qquad
\text{with}
\qquad
\ell_s \equiv 2\pi \sqrt{\alpha'}
,
\end{equation}
and similar expressions for $USp(2N)$ and $U(1)$ gauge symmetries, see e.g.~\cite{Honecker:2011sm} for details. 
For the two examples of sections~\ref{Ss:Z6} and~\ref{Ss:Z6p}, the tree-level gauge couplings are given by 
\begin{equation}
2\pi \Re ({\rm f}^{\text{tree}}_G) = \frac{4 \pi}{g_{G,{\rm tree}}^2} =  \frac{\sqrt{v_1 v_2 v_3}}{8 \pi^3 3^{1/4} \; g_{\text{string}}} \times 
\left\{
\begin{array}{cc}
\sqrt{2}  \times 
\left\{\begin{array}{cr}  1    & G=SU(3)_a, SU(2)_b \\  \frac{7}{6} & U(1)_Y \\  \frac{8}{3} & U(1)_{B-L}  \\   \frac{1}{2} & USp(2)_e \end{array} \right. & \Z_6
\\
\frac{1}{\sqrt{3}}  \times 
\left\{\begin{array}{cr} 1 & G=SU(3)_a\\ 6 & SU(2)_b \\ \frac{19}{18} & U(1)_Y \\ \frac{11}{9} & U(1)_{B-L} \\ \frac{1}{2} & USp(6)_h \end{array} \right. & \Z_6'
\end{array}
\right.
\propto 
\frac{M_{\text{Planck}}}{M_{\text{string}}}
,
\end{equation}
which naively implies that only a high string scale is phenomenologically acceptable.
However, in~\cite{Honecker:2012qr} it was noted that if the one-loop corrections become negative and sizable, large cancellations among tree-level and one-loop contributions to the gauge couplings 
can occur and consequently $M_{\text{string}}$ can be decoupled from $M_{\text{Planck}}$. Using the asymptotic approximations in the geometric regime $v > 1$,
\begin{equation}
\begin{aligned}
-\frac{1}{2\pi} \ln(\eta(iv)) \stackrel{v\to \infty}{\longrightarrow} & \; \frac{v}{24}
,
\\
-\frac{1}{4\pi} \,  \ln \Bigl(e^{-\frac{\pi \sigma^2 v}{4}}\frac{ |\vartheta_1 (\frac{\tau - i \sigma v}{2},i v)|}{\eta (i v)} \Bigr) \stackrel{v\to \infty}{\longrightarrow}& \; \frac{\big[3 (1-\sigma)^2 -1 \bigr] \, v}{48}
 - \delta_{\sigma,0} \frac{\ln [2 \sin (\frac{\pi \tau}{2})]}{4\pi} 
 ,
\end{aligned}
\end{equation}
with errors highly suppressed as  $\ln(1 \pm e^{-\pi v})$,  the one-loop corrections to the gauge kinetic functions take the following form for the $T^6/\Z_6$ model:
\begin{equation}
\begin{aligned}
2\pi \Re (\delta{\rm f}^{\text{loop}}_{SU(3)_a}) \stackrel{v_i\to \infty}{\longrightarrow} & 
 \frac{3 \ln(2)}{2   \pi} + \frac{5\, v_1-\tilde{v}_1}{8}  + \frac{2 \, v_2 - \tilde{v}_2}{8}  + \frac{v_3}{6} \\
& + \frac{ 3 (\sigma_{aa'}^3)^2 - 2 (\sigma_{ab'}^3)^2 + (\sigma_{ac}^3)^2 + (\sigma_{ac'}^3)^2+2 (\sigma_{ad}^3)^2 + (\sigma_{ae}^3)^2 + (\sigma_{ae'}^3)^2}{16} v_3
,
\\
2\pi \Re (\delta{\rm f}^{\text{loop}}_{SU(2)_b})  \stackrel{v_i\to \infty}{\longrightarrow}& 
\frac{3 \, \ln(2)}{4 \pi} +  \frac{5\, v_1 - \tilde{v}_1}{8}  + \frac{2 \, v_2 - \tilde{v}_2}{8}    + \frac{v_3}{6} \\
 & + \frac{3 (\sigma_{ab'}^3 )^2 + 2 (\sigma_{bb'}^3)^2 + (\sigma_{bc}^3)^2 + (\sigma_{bc'}^3)^2 + (\sigma_{be}^3)^2  + (\sigma_{be'}^3)^2   }{16} v_3
,
\\
2\pi \Re (\delta{\rm f}^{\text{loop}}_{USp(2)_e})  \stackrel{v_i\to \infty}{\longrightarrow}& 
- \frac{\ln 2}{ \pi} +  \frac{2 \, v_1 - 3\, v_2}{16 }  +  \frac{2 + 3  (\sigma^3_{ae})^2 +2  (\sigma^3_{be})^2}{16 } v_3 - \frac{ \tilde{v}_3  }{12 }
,
\end{aligned}
\end{equation}
where $\tilde{v}_i \equiv 2 \, v_i$ per tilted two-torus stems from M\"obius strip contributions, and at least for $\sigma^i \tau^i \neq 0$ has to be treated with caution due to the caveat that  compared to the previous results~\cite{Gmeiner:2009fb,Honecker:2011sm} used in the gauge threshold computation, a conjectured sign factor $(-1)^{\sigma^i\tau^i}$ in the formula for the beta function coefficient was found in~\cite{Forste:2010gw,Honecker:2012qr} and therefore the corresponding one-loop gauge correction might have to be modified as well. Such a configuration with $\sigma^i\tau^i=1$ occurs for the $USp(2)_e$ gauge  factor along $T^2_{(2)}$.
The one-loop corrections to the $SU(3)_a$ and $SU(2)_b$ gauge couplings are not affected by this caveat, and they are
 always positive. Phenomenologically acceptable values of the gauge couplings $g_3$ and $g_2$ can thus only be achieved for $M_{\text{string}}$ high.
  Interestingly, the one-loop correction to the `hidden' group $USp(2)_e$ contains a negative off-set and a negative $v_2$ dependence, both of which are favourable for engineering strong coupling in this `hidden' sector.

For the $T^6/\Z_6'$ model, the asymptotics of the one-loop corrections are given by
\begin{equation}
\begin{aligned}
2\pi \Re (\delta{\rm f}^{\text{loop}}_{SU(3)_a})  \stackrel{v_{i}\to \infty}{\longrightarrow}& 
\frac{v_1 + 2 \, \tilde{v}_1}{12}    + \frac{v_3}{8} +  \frac{1}{4 \, \pi} \ln \left(\frac{v_1v_3}{v_2^2} \right)+ \frac{4\, \ln(2)}{3 \pi}  
,
\\
2\pi \Re (\delta{\rm f}^{\text{loop}}_{SU(2)_b})  \stackrel{v_i\to \infty}{\longrightarrow}& 
 \frac{7 \, v_1 + 4 \tilde{v}_1}{16}   + \frac{3 \, v_3}{4} + \frac{6 + 8 (\sigma_{bb'}^2)^2 + (\sigma_{bc}^2)^2 + (\sigma_{bc'}^2)^2 + 2 (\sigma_{bd}^2)^2 + (\sigma_{bd'}^2)^2}{16} v_2 +  \frac{35 \, \ln(2)}{8 \pi} 
,
 \\
 2\pi \Re (\delta{\rm f}^{\text{loop}}_{USp(6)_h})  \stackrel{v_i\to \infty}{\longrightarrow}& 
  - \frac{v_1 + 4 \tilde{v}_1}{48}   - \frac{v_2}{4}  - \frac{v_3}{8} +  \frac{1}{8 \, \pi} \ln \left(\frac{v_1v_3}{v_2^2} \right)
   + \frac{5 \, \ln(2)}{24 \, \pi} 
,
\end{aligned}
\end{equation}
with again $\tilde{v}_1 \equiv 2 v_1$ but the caveat that stacks $a$ and $b$ have $\sigma^1\tau^1=1$.
For $v_2^2 \gg v_1v_3$, the one-loop correction to $SU(3)_a$ can become negative, and  the one-loop correction to the `hidden' $USp(6)_h$ factor contains negative contributions from all three two-torus 
volumes. This D6-brane configuration is thus naively consistent with lowering $M_{\text{string}}$ to some intermediary scale around $10^{12}$ GeV or even as low as $10^{6}$ GeV, while achieving gaugino condensation in a strongly coupled sector as displayed in table~\ref{tab:StringMassScalevsViGs}. 

\mathtabfix{
\begin{array}{|c|c|c||c|c|c||c|c|c|}
\hline \multicolumn{9}{|c|}{\text{\bf $M_{\text{string}}$ as a function of $v_i$ and $g_{\text{string}}$ }}\\
\hline \hline
\multicolumn{3}{|c||}{g_{\text{string}}=0.1}&\multicolumn{3}{|c||}{g_{\text{string}}=0.01} & \multicolumn{3}{|c|}{g_{\text{string}}=0.001}\\
\hline \hline
v_1 v_3 & v_{2, \rm max}^2 & M_{\text{string}}&v_1 v_3 &  v_{2, \rm max}^2 & M_{\text{string}}& v_1 v_3 &  v_{2, \rm max}^2 & M_{\text{string}}\\
\hline 10^8 & 9.7 \times 10^9 & 1.6 \times 10^{10} \text{ GeV }& 10^6 & 1.5 \times 10^{10} & 1.6 \times 10^{10}  \text{ GeV }& 10^2 & 1.5 \times 10^6 & 1.6 \times 10^{12} \text{ GeV } \\
10^{10} & 1.5 \times 10^{14} & 2.8 \times 10^{9}  \text{ GeV }& 10^8 &  1.6 \times10^{14}& 1.5 \times 10^{8}  \text{ GeV }& 10^4 & 1.6 \times 10^{10} & 1.5 \times 10^{10} \text{ GeV } \\
10^{12} & 1.5 \times 10^{18} & 2.8 \times 10^{8}   \text{ GeV }& 10^{10} & 1.6 \times10^{18} & 1.5 \times 10^{6}  \text{ GeV }& 10^6 & 1.6  \times 10^{14} & 1.5 \times 10^{8}  \text{ GeV }\\
\hline
\end{array}
}{StringMassScalevsViGs}{Estimate for the lower bound on the string mass scale $M_{\text{string}}$ in the $T^6/\Z_6'$ model as a function of the string coupling constant $g_{\text{string}}$ and of the bulk K\"ahler moduli $(v_1, v_2, v_3)$ measured in units of $\alpha'$, based on the gauge couplings for gauge groups $SU(3)_a$ and $USp(6)_h$. The maximal value $v_{2, \rm max}^2$ in dependence of $v_1v_3$ originates from the constraint $2\pi \Re({\rm f}^{\text{tree}}_G +\delta{\rm f}^{\text{loop}}_G)>0$, i.e. that the gauge couplings are real valued.}

\subsubsection{Soft supersymmetry breaking}\label{Sss:SoftSUSYbreakingModels}

As a last topic, we investigate the hidden gauge sectors of the supersymmetric DFSZ models discussed in sections~\ref{Ss:Z6} and \ref{Ss:Z6p} and verify whether the gauge groups can become strongly coupled and form a gaugino condensate. The model on $T^6/\Z_6$ is characterised by a hidden $USp(2)_e$ gauge group and an explicit computation using the expressions from~\cite{Honecker:2011sm} yields the corresponding beta-function coefficient:
\begin{equation}
b_{USp(2)_e} =  -1 +  3_m^{(ae)} + 2_m^{(be)},
\end{equation}
where $3_m^{(ae)}$ and $2_m^{(be)}$ denote the contributions from the vector-like matter states charged under the hidden gauge group. Under a displacement of the D-brane stacks $a$ and $b$, these states become massive as discussed in section~\ref{Ss:Z6}, and $USp(2)_e$ remains unbroken provided that brane $e$ is left untouched. Displacing brane $e$ would spontaneously break the $USp(2)_e$ gauge group down to an Abelian massless gauge symmetry, which clearly eliminates all chances to have a strongly coupled gauge group. Hence, in the D-brane set-up described in section~\ref{Ss:Z6}, the beta function coefficient of the enhanced  gauge group  $USp(2)_e$ is negative, as the massive vector-like matter states do not contribute to the beta-function. As a result, the hidden gauge group can be strongly coupled at energy scales smaller than $M_{\text{string}}$ and induce a gaugino condensate.   

The $\Z_6$ orbifold group acting on the factorisable six-torus $T^6$ freezes the bulk complex structure moduli on all three two-tori $T^2_{(i)}$ ($U^{(i)} = e^{i\, \pi/3}$ with $i = 1, 2, 3$). The gaugino condensate  will therefore have to couple to (one of) the $h^{21 (\text{twisted})} = 5$ complex structure moduli associated to the $\Z_2$-twisted sector in a non-perturbative superpotential of the form:
\begin{equation}\label{Eq:SuperpotentialGauginoCondensation}
{\cal W}_0 \supset \Lambda^3_c \, A_{USp(2N)_x}  \,   e^{\frac{8 \pi^2}{b_{USp(2N)_x}} f_{USp(2N)_x} (\tilde U^{(i)})},
\end{equation}
with the gaugino condensation scale $\Lambda_c$ defined as in section \ref{Ss:Soft-SUSY}, $A_{USp(2N)_x} $ a dimensionless constant, and $f_{USp(2N)_x}(\tilde U^{(i)})$ the gauge kinetic function of the $USp(2N)_x$ gauge group depending on the twisted complex structure moduli $\tilde U^{(i)}$. 

For the global model on the $T^6/\Z_6'$ orbifold the hidden gauge group is also an enhanced symplectic gauge group, namely $USp(6)_h$. Due to the negative value of the beta-function coefficient:
\begin{equation}
b_{USp(6)_{h}} = -4,
\end{equation}
the hidden gauge factor $USp(6)_h$ will become  strongly coupled below the string scale. In this case, the $USp(2)_h$ group cannot be broken by any continuous Wilson line or displacement, and gaugino condensation can be used to generate a non-perturbative superpotential for the complex structure moduli. 
Only two two-tori have a frozen complex structure modulus due to the $\Z_6'$ action of the orbifold group on the factorisable six-torus $T^6$. The complex structure modulus for the third two-torus is, however, dynamically stabilised by virtue of supersymmetry of each D6-brane ($U^{(3)} =~\frac{1}{2}$). Also for this orbifold the non-perturbative superpotential arising from the gaugino condensation will involve at least one of the $h^{21 (\Z_2-\text{twisted})} = 4$ complex structure moduli associated to the $\Z_2$-twisted sector, as in equation (\ref{Eq:SuperpotentialGauginoCondensation}).

Despite the fact that both global models allow for a strongly coupled hidden gauge group, a couple of comments are in order here. First of all, one has to make sure that the non-perturbative superpotential from the gaugino condensation does not involve the complex structure modulus whose CP-odd scalar partner is eaten by the $U(1)_b$ gauge boson or by any other anomalous $U(1)$ direction. As discussed in appendix \ref{A:Integration} those complex structure moduli are already stabilised at $M_{\text{string}}$.  

A second comment involves the geometric meaning of the twisted complex structure moduli,  which describe possible deformations of the $\Z_2$-fixed lines of the orbifold action. At the orbifold point,  a $\Z_2$-fixed line is geometrically represented as a co-dimension two singularity on $T^4/\Z_2$ combined with a one-dimensional hypersurface along the $\Z_2$-invariant two-torus. From a physics viewpoint, the singularities indicate vanishing {\it vev}s for the twisted complex structure moduli. Resolving the $\Z_2$-fixed lines allows for non-zero {\it vev}s~\cite{BlaszczykHoneckerKoltermann:2014}, implying that a full discussion of gaugino condensation in these models requires to go over to the resolved Calabi-Yau phase of  the toroidal orbifold.

Last but not least, in this section we have only briefly investigated the necessary conditions for gaugino condensation, but a conclusive analysis about the stabilisation of the complex structure modulus requires an in-depth discussion of the scalar potential for all moduli fields derived from the full superpotential ${\cal W}_0$. In this respect, it is important to be aware that the gauge treshold corrections, as discussed in the previous section, introduce a dependence on the K\"ahler moduli in the non-perturbative superpotential~(\ref{Eq:SuperpotentialGauginoCondensation}). Furthermore, it is known~\cite{Ibanez:2012zz} that a single non-perturbative correction of the form (\ref{Eq:SuperpotentialGauginoCondensation}) is not capable of stabilising the modulus $\tilde U^{(i)}$ at a finite value. Instead, one might need to consider various strongly coupled hidden gauge groups, all with a gaugino condensate, resulting in a racetrack-type superpotential.

\section{Discussion and Conclusions}\label{S:Conclusions}

This article presents a methodic approach to a supersymmetrised version of the DFSZ axion model which can be explicitly realised in {\it global} D-brane model building scenarios. The starting point of the approach is the Higgs-axion potential of the original DFSZ model, whose particle content (two Higgs doublets and one Standard Model singlet scalar) and shape suggest a straightforward generalisation to supersymmetric field theory. Through a nifty combination of F-terms, D-terms and soft supersymmetry breaking terms, a full supersymmetric version of the DFSZ model is put forward, where the supersymmetric interaction between the axion superfield and the Higgs superfields is realised as a cubic renormalisable coupling. 

The embedding of this supersymmetric DFSZ model into Type II string theory follows through the identification of the $U(1)_{PQ}$ symmetry as one of the anomalous massive $U(1)$ gauge symmetries inherent to D-brane model building. The models considered in this article all use the $U(1)_b$ symmetry of the left-symmetric D-brane stack $U(2)_b$ as the Peccei-Quinn symmetry, implying that the axion superfield is realised by an antisymmetric representation under $U(2)_b$. In order to test the full proposal for this new supersymmetric version of the DFSZ model, we turn to the framework of intersecting D6-brane models on toroidal orbifolds. As it turns out, two globally consistent (i.e.~vanishing RR tadpoles and satisfied K-theory constraints) models that exhibit the necessary features to realise the proposed ideas, have already been constructed in earlier works by one of the authors of this paper.

The Higgs sector of the intersecting D6-brane model on the $T^6/\Z_6$ orbifold arises from the non-chiral sector and contains two up-type Higgses and two down-type Higgses, due to {\it ab initio} original left-right symmetry of the model. An investigation of the Yukawa couplings for the three generations of quarks and leptons shows that all four Higgses have to be involved. The supersymmetric DFSZ model then has to be constructed with the four Higgs-~and two axion-superfields. A detailed analysis of the intricate Higgs-axion potential and the symmetry breaking mechanisms in this model yields explicit mass-relations for the electroweak gauge bosons similar to the ones in the Minimal Supersymmetric Standard Model, see table~\ref{tab:OverviewTableEWGaugeBosonMasses}. The presence of two axion-superfields does not only provide a suitable candidate for the QCD axion, but also implies the existence of an axion-like particle, whose characteristics fit nicely to explain the anomalous transparency of the universe with respect to TeV photons.       

\mathtab{ {\small
\begin{array}{|c|c|c|}
\hline \multicolumn{3}{|c|}{ \text{ \bf Electroweak Gauge Boson Masses}}\\
\hline
\hline \text{ SM } & \text{ MSSM } & \text{ DFSZ model on $T^6/\Z_6$ }\\
\hline \hline
m_{W}^2 =  \frac{g_2^2 }{4} v^2 & m_{W}^2 = \frac{g_2^2 }{4} (v_u^2 + v_d^2)  & m_{W}^2 = \frac{g_2^2 }{4} ( 2 v_u^2 + 2 v_d^2) \\
m_{Z}^2 = \frac{(g_Y^2 + g_2^2) }{4}  v^2&m_{Z}^2 = \frac{(g_Y^2 + g_2^2) }{4} (v_u^2 + v_d^2)  & m_{Z}^2 = \frac{(g_Y^2 + g_2^2)}{4} ( 2 v_u^2 + 2 v_d^2) \\
v \simeq 246 \text{ GeV }&\sqrt{v_u^2 + v_d^2} \simeq 246 \text{ GeV } & \sqrt{2(v_u^2 + v_d^2)} \simeq 246 \text{ GeV } \\
\rho = 1 & \rho = 1 & \rho = 1\\
\hline
\end{array}}
}{OverviewTableEWGaugeBosonMasses}{Comparison of the electroweak gauge boson masses for the Standard Model (SM), the Minimal Supersymmetric Standard Model (MSSM) and the supersymmetric DFSZ model on $T^6/\Z_6$. The factor 2 in the latter model is an immediate consequence of the presence of four Higgses, as compared to two Higgses in the MSSM and just one in the SM.}

We propose a method to handle the complicated Higgs-axion potential, based on the separation of scales between the Peccei-Quinn symmetry breaking scale and the electroweak breaking scale. By integrating out the heavy degrees of freedom (massive $U(1)_b$ gauge boson and saxions), the resulting Higgs-axion potential is much simpler to handle and the constraint equations for the Higgs-vacuum configuration are derived. A full computation of the Higgs mass spectrum, which goes beyond the scope of this article, is definitely worthwhile to obtain a better understand of the Higgs parameter space. Also other phenomenological properties of this supersymmetric DFSZ model deserve further attention, such as the interactions between the axion and its superpartner (axino) on the one hand and the particles charged under the Standard Model gauge group on the other hand, and the implications for cosmology with an axion and an axino dark matter candidate, amongst others.  Discussions of this nature will be postponed to future work~\cite{HoneckerStaessens:2014}.   

A different situation occurs for the intersecting D6-brane model on the $T^6/\Z_6'$ orbifold, where the Higgses can be realised in the chiral part of the massless spectrum. The supersymmetric version of the DFSZ model in this configuration only requires one axion superfield, such that the axion-Higgs potential falls into a class of models that was already studied before in the literature in the context of supersymmetric field theory. The interest in this model stems from its large hidden gauge group $USp(6)$ with negative beta-function coefficient. If the hidden gauge group has a strongly coupled regime, soft supersymmetry breaking finds a natural explanation through gravity mediated gaugino condensation. In this way, this article offers a concrete relation between the hidden sector in globally consistent D-brane models and the soft supersymmetry breaking terms introduced by hand in the supersymmetric DFSZ model.     

From the string theory side, the two models discussed here form the perfect Guinea Pigs to explore the limits of our understanding of intersecting D6-brane model building on toroidal orbifolds with fractional and rigid three-cycles. In particular, perturbative $n$-point couplings resulting from pointlike worldsheet instantons with parallel D-brane boundaries deserve full attention, as they are ubiquitous for toroidal orbifolds with an orbifold group containing a $\Z_2$ subsymmetry. Another so far poorly studied aspect involves the physical implications from deforming the $\Z_2$-fixed lines on which the fractional D-branes wrap~\cite{BlaszczykHoneckerKoltermann:2014}. Nonetheless, transitioning from the orbifold point to the resolved Calabi-Yau manifold is necessary for the moduli in the twisted sector to be non-vanishing. Such a transition would also enlarge the number of plausible backgrounds on which one can study D-brane model building using intersecting D6-branes.

This article also touches upon the connections between soft supersymmetry breaking, moduli stabilisation and the presence of a non-trivial hidden sector. In the context of Type IIA superstring 
theory it is extremely difficult to find backgrounds where all these aspects can be considered simultaneously. Other D-brane model building scenarios in Type IIB superstring compactifications have shown to be more suitable to address questions related to moduli stabilisation and soft supersymmetry breaking by fluxes. It might thus be interesting to see if the supersymmetric DFSZ model can be realised in these frameworks and whether they can offer new perspectives in bridging the gap between particle physics phenomenology on the one hand and moduli stabilisation mechanisms on the other hand.

\noindent
{\bf Acknowledgements:} 
The authors would like to thank Alain Dirkes for collaboration at initial stages of this project.\\
This work is partially supported by the {\it Cluster of Excellence `Precision Physics, Fundamental Interactions and Structure of Matter' (PRISMA)} DGF no. EXC 1098,
the DFG research grant HO 4166/2-1 and the Research Center {\it `Elementary Forces and Mathematical Foundations' (EMG)} at JGU Mainz.

\appendix
\section{The effective action for axions}\label{A:ChiralRotation}

One of the defining properties of an axion is its linear coupling to the anomalous topological charge density of a gauge theory suppressed by the axion decay constant, as in equation (\ref{Eq:StandardActionAxion}) for the QCD axion. Yet an axion can also interact with fermions and bosons. In order to understand~\cite{Kim:1986ax} these interactions one has to take into account how the generating functional for a (Dirac) fermion $\psi$ charged under a local gauge symmetry with field strength $F_{\mu \nu}$ transforms under a chiral rotation:
\begin{equation}
\psi \longrightarrow e^{i\, \alpha \gamma^5} \psi \qquad \text{ or } \qquad \left\{\begin{array}{rcl} \psi_L& \longrightarrow & e^{-i\, \alpha} \psi_L \\ \psi_R& \longrightarrow & e^{i\, \alpha} \psi_R \end{array} \right..
\end{equation}
Starting with the gauge-invariant kinetic term for the Dirac fermion, one notices that the term is shifted by an axial-vector coupling under a chiral rotation:
\begin{equation}
i\, \ov \psi \slashed{D} \psi \longrightarrow i\, \ov \psi \slashed{D} \psi - \ov \psi \gamma^\mu \gamma^5 \psi \partial_\mu \alpha ,
\end{equation} 
while the fermionic path integral measure transforms as follows:\footnote{The trace is considered for the fundamental representation,
$\Tr_F (T_a T_b) = \frac{1}{2} \delta_{ab}$, 
and in case of an Abelian symmetry, the normalisation enforces an additional factor 2 in the denominator on the r.h.s..}
\begin{equation}
{\cal D} \psi {\cal D} \ov \psi \longrightarrow {\cal D} \psi {\cal D} \ov \psi e^{i\, \int d^4 x\, \alpha (x) \frac{1}{16 \pi^2}\Tr(F_{\mu \nu} \tilde F^{\mu \nu})} .
\end{equation}
The total effect of the chiral rotation thus corresponds to a shift of the lagrangian ${\cal L} \rightarrow {\cal L} + \Delta {\cal L}$ with 
\begin{equation}\label{Eq:ShiftedLagrangianAxion}
\Delta {\cal L} =  - \ov \psi \, \gamma^\mu \gamma^5 \partial_\mu \alpha \, \psi +  i \frac{1}{16 \pi^2}  \alpha (x) \,\varepsilon^{\mu \nu \rho \sigma} \Tr (G_{\mu \nu} G_{\rho \sigma}).
\end{equation}

Chiral rotations come in handy when the fermion acquires its mass through a Yukawa coupling to a Higgs field $H$: 
\begin{equation}\label{Eq:YukawaEffectiveAxion}
{\cal L}_{\psi \psi H} = f_\psi \, \ov \psi_L H \psi_R + f_\psi^*\, \ov \psi_R H^{\dagger} \psi_L,
\end{equation}
that in turn interacts with an axion $\alpha$ through terms in the potential of the model, for instance of the form
\begin{equation}
V \supset g (C_i) H e^{i \alpha/ f_\alpha}  +  g^* (C^{\dagger}_i) H^{\dagger} e^{-i \alpha/ f_\alpha},
\end{equation}
where $g(C_i)$ represents a polynomial expression of  spectator scalar fields $C_i$ required by gauge invariance. By a local rescaling of the Higgs field, the interaction between the axion and the Higgs field can be eliminated from the potential. However, in order to avoid that the axion pops up in the Yukawa coupling (\ref{Eq:YukawaEffectiveAxion}) one can perform a chiral rotation of the fermions, by which the Lagrangian is shifted as discussed above in equation (\ref{Eq:ShiftedLagrangianAxion}). Furthermore, the local rescaling of the Higgs fields will also introduce a derivative coupling of the Higgs to the axion by virtue of the kinetic terms for the Higgs field.  

In conclusion, the most generic action for an axion $\alpha$ coupled to a charged fermion $\psi$ and a Higgs field $H$ takes the following form (see also~\cite{Kim:2008hd}):
\begin{equation} \label{Eq:EffactionAxion}
 \begin{array}{rcl}
 {\cal L}_{\text{axion}} &=& \frac{1}{2} (\partial_\mu \alpha)  (\partial^\mu \alpha) + \frac{c_h}{f_{\alpha}} (\partial^\mu \alpha) \left( H^\dagger i D_\mu H - i (D_\mu H)^\dagger  H \right) + \frac{c_\psi}{f_{\alpha}}    \ov \psi \, \gamma^\mu \gamma^5 \, \psi (\partial_\mu \alpha)   \\
&& +  C_{\alpha FF} \frac{\alpha}{f_\alpha} \frac{1}{16 \pi^2} \Tr F_{\mu \nu} \tilde F^{\mu \nu}+ f_\psi H \ov \psi_L \psi_R e^{i c_m \alpha/ f_{\alpha}} + f_\psi^* H^{\dagger} \ov \psi_R \psi_L e^{-i c_m \alpha / f_{\alpha}}.
\end{array}
\end{equation}
This action can be generalised to accommodate the particle content and gauge structure of the Standard Model, in which case the fermionic part has to be extended to all quarks and left-handed leptons, and the anomalous term with field strength $F_{\mu \nu}$ has to be taken into consideration for the QCD field strength $G_{\mu \nu}$, the weak field strength $W_{\mu \nu}$ and for the hypercharge field strength $Y_{\mu \nu}$. The parameters  $c_{h}$, $c_\psi$, $c_m$, $ C_{\alpha GG}$,  $C_{\alpha WW}$ and $ C_{\alpha YY}$ are model-dependent constants characterizing the type of axion model:
 \begin{equation*}
\hspace{-0.2in} \begin{array}{ll}
 \text{ PQWW: } & c_h = 0, c_\psi = 0, c_m\neq 0, \\
 \text{ KSVZ: } & c_h = 0, c_\psi = 0, c_m= 0  \leadsto \text{ pair of fourth generation quarks induces anomaly } (C_{\alpha GG}  \neq 0),\\
 \text{ DFSZ: } & c_h \neq 0, c_\psi = 0, c_m\neq 0. \\ 
 \end{array}
 \end{equation*}
The parameters $C_{\alpha GG}$,  $C_{\alpha WW}$ and $C_{\alpha YY}$ are determined by the anomaly coefficients, which clearly depend on the particle content and the gauge structure of the model at hand. The parameter $c_m$ is correlated with a complex mass matrix for the quarks, which forms an additional source for CP violation, on top of the intricate QCD vacuum structure. As already suggested in the paragraph above equation (\ref{Eq:EffactionAxion}), one can eliminate the axion in the Yukawa couplings by virtue of a chiral rotation. As a result, the coefficient in front of the axion coupling to the QCD anomaly will be shifted, i.e.~$C_{\alpha GG} \rightarrow C_{\alpha GG} +c_m$.
 
An important observation to be drawn from this is that the form of the action depends on the energy scale at which the model is considered, as advocated  e.g. in~\cite{Georgi:1986df}.
At energies below the electroweak symmetry breaking scale, the weak gauge bosons have to be integrated out, and the anomalous coupling of the axion to the weak gauge bosons reduces to an anomalous coupling for the photon:
\begin{equation}\label{Eq:AnomalyAxionEM}
{\cal L}_{\alpha \gamma \gamma} = -\frac{g_{\alpha \gamma \gamma}}{4} \, \alpha \, F_{\mu \nu} \tilde F^{\mu \nu}
,
\end{equation}
where the coupling constant $g_{\alpha \gamma \gamma}$ consists of a model-dependent constant $C_{\alpha \gamma \gamma} = C_{\alpha WW} + C_{\alpha YY}$ and an additional term~\cite{Srednicki:1985xd} related to the mixing of the axion with the $\pi^0$- and $\eta$-mesons:
\begin{equation}
g_{\alpha \gamma \gamma} =\frac{e^2}{8 \pi^2 f_{\alpha}} \left( C_{\alpha \gamma \gamma} - \frac{2}{3} \frac{4 m_d m_s +m_u m_s  +m_u m_d}{m_d m_s +m_u m_s + m_u m_d}   \right) .
\end{equation}

The generic action (\ref{Eq:EffactionAxion}) is also valid for closed string axions $\xi$, in which case the only non-vanishing parameter is the coupling $C_{\xi FF}$ to the gauge anomaly term. The values for the parameters $C_{\xi FF}$ follow then from the dimensional reduction of the Chern-Simons action for the D-branes. In the models presented in section~\ref{Ss:Z2N} also axions originating from the open string sector were considered. As open string axions behave entirely as field theory axions, the whole discussion from above is equally valid.

\section{Integrating out a heavy vector multipet}\label{A:Integration}

In this appendix, we use the supersymmetric formulation~\cite{Kuzmin:2002nt,Kors:2004ri} of the St\"uckelberg mechanism of equation~(\ref{Eq:GS-coupling})
and integrate out a heavy vector multiplet $V_b$ in the presence of a light vector multiplet $V_a$
following the prescriptions of~\cite{Brizi:2009nn}. 
The model we consider consists of a St\"uckelberg multiplet $U$ and two chiral superfields $\Phi_1$ and $\Phi_2$ both charged under $U(1)_a \times U(1)_b$ but with opposite charges:
\begin{center}
\begin{tabular}{c|cc}
& $U(1)_a$ & $U(1)_b$\\
\hline $\Phi_1$ & $q_a$ & $q_b$\\
$\Phi_2$ & $-q_a$ & $-q_b$\\ 
\end{tabular}
\end{center}
The most general K\"ahler potential we can write down for this model (with global supersymmetry) with some chiral multiplet $U$ (containing a complex structure modulus in Type IIA string theory with D6-branes) reads:
\begin{equation}\label{Eq:KaehlerStuckelbergSit3}
K^{\text{SUSY}} = (M_{\text{string}} \, V_b + U + U^\dagger)^2 + \Phi_1^\dagger e^{2 g_a q_a V_a}  e^{2 g_b q_b V_b} \Phi_1 +  \Phi_2^\dagger e^{- 2 g_a q_a V_a}  e^{- 2 g_b q_b V_b} \Phi_2,
\end{equation}
with the most generic supergauge-invariant superpotential given by:
\begin{equation}\label{Eq:SuperpotentialBeforeIntegration}
{\cal W}  = m \Phi_1 \Phi_2.
\end{equation}
By integrating out the vector multiplet (up to order $M_{\text{string}}^{-1}$) through its equation of motion,
\begin{equation}\label{Eq:EOMVectormultipletIntegration}
V_b \simeq  - \frac{ (U + U^\dagger)}{M_{\text{string}}} + {\cal O}(M_{\text{string}}^{-2}),
\end{equation}
we obtain the following effective K\"ahler potential:
\begin{equation}
\begin{array}{rcl}
K^{\text{SUSY}}_{\text{eff}} &=& \Phi_1^\dagger  e^{2 g_a q_a V_a} \Phi_1 +  \Phi_2^\dagger  e^{-2 g_a q_a V_a} \Phi_2 \\
 &&  - \frac{2 g_b q_b}{M_{\text{string}}} (U+U^\dagger) \Phi_1^\dagger  e^{2 g_a q_a V_a} \Phi_1  + \frac{2 g_b^2 q_b^2}{M_{\text{string}}^2} (U+U^\dagger)^2 \Phi_1^\dagger  e^{2 g_a q_a V_a} \Phi_1   \\
&&  + \frac{2 g_b q_b}{M_{\text{string}}} (U+U^\dagger) \Phi_2^\dagger  e^{-2 g_a q_a V_a} \Phi_2  + \frac{2 g_b^2 q_b^2}{M_{\text{string}}^2} (U+U^\dagger)^2 \Phi_2^\dagger  e^{-2 g_a q_a V_a} \Phi_2    + {\cal O}(M_{\text{string}}^{-2} \Phi^4).
\end{array}
\end{equation}
In the next step, we determine the effective scalar potential for this theory by considering those terms in the Lagrangian containing the auxiliary fields and the scalar fields:
\begin{equation}
\begin{array}{rcl}
{\cal L}^{\text{eff}}_{F+D} &=& (1- 2 z +2 z^2)  |F_1|^2 + (1+ 2 z +2 z^2) |F_2|^2 + m F_1 \phi_2 + m F_2 \phi_1 + \ov m F_1^{\dagger} \phi_2^{\dagger} + \ov m F_2^{\dagger} \phi_1^{\dagger}\\
 &&+ \phi_1^{\dagger} \phi_1  \frac{D_a}{2} 2 g_a q_a (1 - 2 z + 2 z^2) - \phi_2^{\dagger} \phi_2  \frac{D_a}{2} 2 g_a q_a (1 + 2 z + 2 z^2) + \frac{D_a^2}{2},
 \end{array}
\end{equation}
where we introduced the following constant containing the {\it vev} of the (in Type IIA complex structure) modulus,
\begin{equation}
z= \frac{g_b q_b}{M_{\text{string}}} \langle U+U^{\dagger} \rangle.
\end{equation}
Due to supersymmetry, the (complex structure) modulus is stabilised with a mass $2 M_{\text{string}}$ when its corresponding axion is eaten by a gauge boson through the St\"uckelberg mechanism, cf.
equation~(\ref{Eq:GS-coupling}). Extracting the equations of motion for the auxiliary fields: 
\begin{equation}
\begin{array}{l}
F_1^{\dagger} = - \frac{m}{1-2 z + z^2} \phi_2 \qquad F_2^{\dagger} = - \frac{m}{1+2 z + z^2} \phi_1,\\
D_a = -g_a q_a \left( (1-2 z + 2 z^2) \phi_1^{\dagger} \phi_1 - (1+2 z + 2 z^2) \phi_2^{\dagger} \phi_2    \right),
\end{array}
\end{equation}
and inserting them back into ${\cal L}^{\text{eff}}_{F+D}$ yields the effective scalar potential.
\begin{equation}
V_{\text{eff}} = \frac{|m|^2}{1+2 z + 2 z^2} |\phi_1|^2 + \frac{|m|^2}{1-2z +2 z^2} |\phi_2|^2 + \frac{1}{2} g_a^2 q_a^2 \left[ (1-2 z + 2 z^2) |\phi_1|^2 - (1+2 z + 2 z^2) |\phi_2|^2 \right]^2
.
\end{equation}
The most important observation is that the parameters in the effective potential are different with respect to a model with only gauge symmetry $U(1)_a$:
\begin{itemize}
\item Although  both superfields $\Phi_1$ and $\Phi_2$ receive a supersymmetric mass by virtue of the superpotential (\ref{Eq:SuperpotentialBeforeIntegration}), the masses for the scalar fields $\phi_1$ and $\phi_2$ clearly differ upon integrating out a heavy vectormultiplet.
\item Also the quartic terms for $\phi_1$ and $\phi_2$ are no longer universal upon integrating out a heavy vectormultiplet.
\end{itemize}
A final observation concerns the contributions coming from the supercovariant field strength $W_\alpha = -\frac{1}{4} \ov D^2 D_\alpha V_b$, associated with the heavy vectormultiplet $V_b$. As the multiplet $U$ satisfies the conditions for a chiral superfield, the field strength $W_\alpha$ will vanish upon integrating out the vector multiplet. The superspace action involving the field strength only contributes non-trivially, if a term of the order $M_{\text{string}}^{-2}$ is taken into account upon integrating out the vector multiplet by its equation of motion (\ref{Eq:EOMVectormultipletIntegration}).

\section{Spectra of the $T^6/\Z_6$ and $T^6/\Z_6'$ Standard Models}\label{A:Spectra}

In this appendix, the chiral and non-chiral spectra for the supersymmetric Standard Models constructed on $T^6/\Z_6$~\cite{Honecker:2004kb,Gmeiner:2009fb} and $T^6/\Z_6'$~\cite{Gmeiner:2008xq,Gmeiner:2009fb} are briefly summarized.
The lower index $m$ of the multiplicity of vector-like matter indicates that these states acquire a mass if some relative continuous displacement or Wilson line along the $T^2$ without $\Z_2$ action is switched on.
The non-chiral spectrum in table~\ref{tab:5stack-Z6p-NonChiral} contains the corrections of multiplicities in the $bb'$ and $ch$  sector introduced in~\cite{Honecker:2012jd} as well as the change $3 \times (\ov{\3}_{\Anti}) \to 2 \times (\ov{\3}_{\Anti}) + ({\bf 6}_{\Sym})$
in the $aa'$ sector due to the conjectured subtlety~\cite{Forste:2010gw,Honecker:2012qr} in the sign factor for D6-brane configurations with non-vanishing displacement and Wilson line parallel to some O6-plane.

\mathtabfix{
\begin{array}{|c||c|c|||c||c|c|}
\hline \multicolumn{6}{|c|}{\text{\bf Chiral SM spectrum  on the AAB lattice of } T^6/\Z_6} \\
\hline \hline
\text{\bf Matter} & \text{\bf Sector} & U(3)_a\times U(2)_b \times USp(2)_e{}^{\bigl( \times U(1)_c \times  U(1)_d\bigr)}_{\times U(1)_Y \times U(1)_{B-L}}
& \text{\bf Matter} & \text{\bf Sector} & U(3)_a\times U(2)_b \times USp(2)_e{}^{\bigl( \times U(1)_c \times  U(1)_d\bigr)}_{\times U(1)_Y \times U(1)_{B-L}}
 \\ 
\hline 
Q_L & ab' & 3 (\3,\2;\1)^{(0,0)}_{\frac{1}{6},\frac{1}{3}} 
&  L & bd' & 3 (\1, \ov{\2};\1)^{(0,-1)}_{-\frac{1}{2},-1} \\
d_R & ac & 3 (\ov{\3}, \1;\1)^{(1,0)}_{\frac{1}{3},-\frac{1}{3}} 
&  \nu_R & cd & 3 (\1, \1;\1)^{(-1,1)}_{0,1} \\ 
u_R & ac' & 3 (\ov{\3}, \1;\1)^{(-1,0)}_{-\frac{2}{3},-\frac{1}{3}}  
& e_R & cd' & 3 (\1, \1;\1)^{(1,1)}_{1,1}\\
& be = be' & 3 (\1, \ov{\2};{\bf 2})^{(0,0)}_{0,0}
& & &  
 \\\hline
\end{array}
}{5stackLRSChiral}{Chiral spectrum of a global D6-brane model on $T^6/\Z_6$.}

\mathtabfix{
\begin{array}{|c||c|c|||c||c|c|}
\hline \multicolumn{6}{|c|}{\text{\bf Non-chiral SM spectrum  on the AAB lattice of } T^6/\Z_6} \\
\hline \hline
\text{\bf Matter} & \text{\bf Sector} & U(3)_a\times U(2)_b  \times USp(2)_e{}^{\bigl(\times U(1)_c \times U(1)_d\bigr)}_{\times U(1)_Y \times  U(1)_{B-L}} 
& \text{\bf Matter} & \text{\bf Sector} & U(3)_a\times U(2)_b  \times USp(2)_e{}^{\bigl(\times U(1)_c \times U(1)_d\bigr)}_{\times U(1)_Y \times  U(1)_{B-L}} 
\\\hline
A_{0 \ldots 3}& aa & 4 ({\bf 9_\Adj},\1;\1)^{(0,0)}_{0,0} 
& & aa' &  (1_m+3) [ (\ov{\3}_{\Anti},\1;\1)^{(0,0)}_{\frac{1}{3},\frac{2}{3}} + h.c.]  \\
B_{0 \ldots 3} & bb  & 4  (\1,{\bf 4_\Adj};\1)^{(0,0)}_{0,0} 
& \Sigma_{0 \ldots 3}, \tilde{\Sigma}_{0 \ldots 3} & bb' & (1_m+3) [(\1,{\bf 1_\Anti};\1)^{(0,0)}_{0,0} + h.c.] \\
& cc  &  (\1,\1;\1)^{(0_{\Adj},0)}_{0,0} 
& & ac & 1_m [(\ov{\3},\1;\1)^{(1,0)}_{\frac{1}{3},-\frac{1}{3}} + h.c.] \\ 
D_{0 \ldots 3} & dd  & 4 (\1,\1,;\1)^{(0,0_\Adj)}_{0,0} 
& & ac' & 1_m [(\ov{\3},\1;\1)^{(-1,0)}_{-\frac{2}{3},-\frac{1}{3}} + h.c.]\\
& ee  &  (\1,\1;{\bf 3_\Adj})^{(0,0)}_{0,0} 
& & ae=ae' & 1_m  [({\bf 3},\1;{\bf 2})^{(0,0)}_{\frac{1}{6},\frac{1}{3}} +h.c.]\\
& ab & 3 [({\bf 3},{\bf \ov 2};\1)^{(0,0)}_{\frac{1}{6},\frac{1}{3}} +h.c.]
& & ab' & 1_m [({\bf 3}, {\bf 2};\1)^{(0,0)}_{\frac{1}{6},\frac{1}{3}} + h.c.]\\ 
& ad' & 3 [(\3,\1;\1)^{(0,1)}_{\frac{2}{3},\frac{4}{3}}+h.c.] 
& & ad & 2_m [(\ov{\3},\1;\1)^{(0,1)}_{\frac{1}{3},\frac{2}{3}}+h.c.]\\
H^{(1)}_d + H^{(1)}_u &bc & 1_m [ (\1,\2;\1)^{(-1,0)}_{-\frac{1}{2},0}  + h.c.]  
& & be=be' & 1_m  [(\1,{\bf 2};{\bf  2})^{(0,0)}_{0,0}  + h.c. ]\\
H^{(2}_u + H^{(2)}_d  & bc' &  1_m [ (\1,\2;\1)^{(1,0)}_{\frac{1}{2},0}  + h.c.]   
& & cc'&  1_m[ (\1,\1;\1)^{(2,0)}_{1,0} + c.c.] 
\\\hline
\end{array}
}{5stackLRSNonChiral}{Non-chiral spectrum of a global D6-brane model on $T^6/\Z_6$. }

\mathtabfix{
\begin{array}{|c||c|c|||c||c|c|}
\hline \multicolumn{6}{|c|}{\text{\bf Chiral SM spectrum  on the ABa lattice of } T^6/\Z_6'} \\
\hline \hline
\text{\bf Matter} & \text{\bf Sector} &  U(3)_a\times U(2)_b \times USp(6)_h{}^{\bigl( \times U(1)_c \times  U(1)_d\bigr)}_{\times U(1)_Y \times U(1)_{B-L}}
& \text{\bf Matter} & \text{\bf Sector} & U(3)_a\times U(2)_b \times USp(6)_h{}^{\bigl( \times U(1)_c \times  U(1)_d\bigr)}_{\times U(1)_Y \times U(1)_{B-L}} \\ 
\hline 
Q_L & ab' & 3 (\3,\2;\1)_{\frac{1}{6},\frac{1}{3}}^{(0,0)}
&  L_{1 \ldots 6} & bd & 6 (\1,\2;\1)_{-\frac{1}{2},-1}^{(0,-1)}\\
d_R & ac & 3 (\ov{\3},\1;\1)_{\frac{1}{3},-\frac{1}{3}}^{(1,0)}
& \ov{L}_{1 \ldots 3} & bd' & 3 (\1,\2;\1)_{\frac{1}{2},1}^{(0,1)}\\
u_R & ac' & 3 (\ov{\3},\1;\1)_{-\frac{2}{3},-\frac{1}{3}}^{(-1,0)}
& H^{(1 \ldots 9)}_d & bc'  & 9 (\1,\ov{\2};\1)_{-\frac{1}{2},0}^{(-1,0)}\\
e_R & cd ' & 3 (\1,\1;\1)_{1,1}^{(1,1)}
& H^{(1\ldots 9)}_u & bc & 9 (\1,\ov{\2};\1)_{\frac{1}{2},0}^{(1,0)}\\
\nu_R & cd & 3 (\1,\1;\1)_{0,1}^{(-1,1)}& & & 
\\\hline
\end{array}
}{5stack-Z6p-Chiral}{Chiral spectrum of a global D6-brane model with hidden $USp(6)_h$ on $T^6/\Z_6'$.}

\mathtabfix{
\begin{array}{|c||c|c|||c||c|c|}
\hline \multicolumn{6}{|c|}{\text{\bf Non-chiral SM spectrum  on the ABa lattice of } T^6/\Z_6'} \\
\hline \hline
\text{\bf Matter} & \text{\bf Sector} &  U(3)_a\times U(2)_b \times USp(6)_h{}^{\bigl( \times U(1)_c \times  U(1)_d\bigr)}_{\times U(1)_Y \times U(1)_{B-L}}
& \text{\bf Matter} & \text{\bf Sector} & U(3)_a\times U(2)_b \times USp(6)_h{}^{\bigl( \times U(1)_c \times  U(1)_d\bigr)}_{\times U(1)_Y \times U(1)_{B-L}}
 \\ 
\hline 
& aa &2 ({\bf 9}_{\Adj},\1;\1)_{0,0}^{(0,0)}
& & bh=bh' & [(\1,\2;{\bf 6})_{0,0}^{(0,0)} + h.c.]
\\
B_{0 \ldots 9} & bb & 10 (\1,\4_{\Adj};\1)_{0,0}^{(0,0)}
& & ch=ch' & 2[(\1,\1;{\bf 6})^{(1,0)}_{\frac{1}{2},0} + h.c. \Bigr]
\\
& cc & 4 (\1,\1;\1)_{0,0}^{(0_{\Adj},0)}  
& H_u^{(10)} + H_d^{(10)} &  bc & 1_{m} [(\1,\ov{\2};\1)_{\frac{1}{2},0}^{(1,0)}+ h.c.]  
\\
& dd & 10 (\1,\1;\1)_{0,0}^{(0,0_{\Adj})}  
& H_d^{(11)} + H_u^{(11)} & bc' & 1_{m}[ (\1,\ov{\2};\1)_{-\frac{1}{2},0}^{(-1,0)}+ h.c.]
\\
& hh & 2 (\1,\1;{\bf 15}_{\Anti})_{0,0}^{(0,0)}
& & bd  &  2_{m} [(\1,\2;\1)_{-\frac{1}{2},-1}^{(0,-1)}+ h.c.]
\\
& ab' & [(\3,\2;\1)_{\frac{1}{6},\frac{1}{3}}^{(0,0)}+h.c.]
&  & bd' & 1_{m} [(\1,\2;\1)_{\frac{1}{2},1}^{(0,1)} + h.c.]
\\
& ad & 3 [(\ov{\3},\1;\1)_{\frac{1}{3},\frac{2}{3}}^{(0,1)} + h.c.] 
 & & cc' & 1_{m} (\1,\1;\1)_{1,0}^{(2_{\Sym},0)}+ h.c.]
 \\
& ad' &  3 [(\ov{\3},\1;\1)_{-\frac{2}{3},-\frac{4}{3}}^{(0,-1)} + h.c.]
 & & cd & 1_{m} [ (\1,\1;\1)_{0,-1}^{(1,-1)}+ h.c.]
\\
&aa' &   [(2 \times \ov{\3}_{\Anti} + {\bf 6}_{\Sym} ,\1;\1)_{\frac{1}{3},\frac{2}{3}}^{(0,0)} + h.c.]
 & & cd' & 1_{m}[ (\1,\1;\1)_{1,1}^{(1,1)}+ h.c.]
 \\
 & bb' & 6 [(\1,\3_{\Sym};\1)_{0,0}^{(0,0)} + h.c.]
 &\Sigma_{0 \ldots 3}, \tilde{\Sigma}_{0 \ldots 3} &bb' & 4_{m} [(\1,\1_{\Anti};\1)_{0,0}^{(0,0)}+ h.c.]
\\\hline
\end{array}
}{5stack-Z6p-NonChiral}{Non-chiral spectrum of a global D6-brane model with hidden $USp(6)_h$ on $T^6/\Z_6'$.}


\addcontentsline{toc}{section}{References}
\bibliographystyle{ieeetr}
\bibliography{refs_Axions}

\begin{thebibliography}{10}

\bibitem{Peccei:1977hh}
R.~Peccei and H.~R. Quinn, ``{CP Conservation in the Presence of Instantons},''
  {\em Phys.Rev.Lett.}, vol.~38, pp.~1440--1443, 1977.

\bibitem{Peccei:1977ur}
R.~Peccei and H.~R. Quinn, ``{Constraints Imposed by CP Conservation in the
  Presence of Instantons},'' {\em Phys.Rev.}, vol.~D16, pp.~1791--1797, 1977.

\bibitem{Weinberg:1977ma}
S.~Weinberg, ``{A New Light Boson?},'' {\em Phys.Rev.Lett.}, vol.~40,
  pp.~223--226, 1978.

\bibitem{Wilczek:1977pj}
F.~Wilczek, ``{Problem of Strong p and t Invariance in the Presence of
  Instantons},'' {\em Phys.Rev.Lett.}, vol.~40, pp.~279--282, 1978.

\bibitem{Vogel:2013bta}
J.~Vogel, F.~Avignone, G.~Cantatore, J.~Carmona, S.~Caspi, {\em et~al.},
  ``{IAXO - The International Axion Observatory},'' arXiv:1302.3273
  [physics.ins-det].

\bibitem{Bahre:2013ywa}
R.~B{\"a}hre, B.~D{\"o}brich, J.~Dreyling-Eschweiler, S.~Ghazaryan,
  R.~Hodajerdi, {\em et~al.}, ``{Any light particle search II ÑTechnical Design
  Report},'' {\em JINST}, vol.~8, p.~T09001, 2013.

\bibitem{Baker:2013zta}
K.~Baker, G.~Cantatore, S.~Cetin, M.~Davenport, K.~Desch, {\em et~al.}, ``{The
  quest for axions and other new light particles},'' {\em Annalen Phys.},
  vol.~525, pp.~A93--A99, 2013.

\bibitem{Graham:2013gfa}
P.~W. Graham and S.~Rajendran, ``{New Observables for Direct Detection of Axion
  Dark Matter},'' {\em Phys.Rev.}, vol.~D88, p.~035023, 2013.

\bibitem{Irastorza:2013kda}
I.~Irastorza, F.~Avignone, G.~Cantatore, J.~Carmona, S.~Caspi, {\em et~al.},
  ``{Future axion searches with the International Axion Observatory (IAXO)},''
  {\em J.Phys.Conf.Ser.}, vol.~460, p.~012002, 2013.

\bibitem{Ringwald:2013via}
A.~Ringwald, ``{Ultralight Particle Dark Matter},'' arXiv:1310.1256 [hep-ph].

\bibitem{Abbott:1989jw}
L.~Abbott and M.~B. Wise, ``{Wormholes and global symmetries},'' {\em
  Nucl.Phys.}, vol.~B325, p.~687, 1989.

\bibitem{Coleman:1989zu}
S.~R. Coleman and K.-M. Lee, ``{Wormholes made without massless matter
  fields},'' {\em Nucl.Phys.}, vol.~B329, p.~387, 1990.

\bibitem{Kallosh:1995hi}
R.~Kallosh, A.~D. Linde, D.~A. Linde, and L.~Susskind, ``{Gravity and global
  symmetries},'' {\em Phys.Rev.}, vol.~D52, pp.~912--935, 1995.

\bibitem{Banks:2010zn}
T.~Banks and N.~Seiberg, ``{Symmetries and Strings in Field Theory and
  Gravity},'' {\em Phys.Rev.}, vol.~D83, p.~084019, 2011.

\bibitem{Banks:1988yz}
T.~Banks and L.~J. Dixon, ``{Constraints on String Vacua with Space-Time
  Supersymmetry},'' {\em Nucl.Phys.}, vol.~B307, pp.~93--108, 1988.

\bibitem{Hellerman:2010fv}
S.~Hellerman and E.~Sharpe, ``{Sums over topological sectors and quantization
  of Fayet-Iliopoulos parameters},'' {\em Adv.Theor.Math.Phys.}, vol.~15,
  pp.~1141--1199, 2011.

\bibitem{Blumenhagen:2006ci}
R.~Blumenhagen, B.~K{\"o}rs, D.~L{\"u}st, and S.~Stieberger,
  ``{Four-dimensional String Compactifications with D-Branes, Orientifolds and
  Fluxes},'' {\em Phys.Rept.}, vol.~445, pp.~1--193, 2007.

\bibitem{Ibanez:2012zz}
L.~E. Ib\'{a}\~{n}ez and A.~M. Uranga, ``{String theory and particle physics:
  An introduction to string phenomenology},'' 2012.

\bibitem{Blumenhagen:2009qh}
R.~Blumenhagen, M.~Cveti\v{c}, S.~Kachru, and T.~Weigand, ``{D-Brane Instantons
  in Type II Orientifolds},'' {\em Ann.Rev.Nucl.Part.Sci.}, vol.~59,
  pp.~269--296, 2009.

\bibitem{Kim:2009cp}
J.~E. Kim and H.~P. Nilles, ``{Axionic dark energy and a composite QCD
  axion},'' {\em JCAP}, vol.~0905, p.~010, 2009.

\bibitem{Kim:2013pja}
J.~E. Kim, ``{Modeling small dark energy scale with quintessential pseudoscalar
  boson},'' arXiv:1311.4545 [hep-ph].

\bibitem{Coriano:2008xa}
C.~Coriano, M.~Guzzi, A.~Mariano, and S.~Morelli, ``{A Light Supersymmetric
  Axion in an Anomalous Abelian Extension of the Standard Model},'' {\em
  Phys.Rev.}, vol.~D80, p.~035006, 2009.

\bibitem{Higaki:2011bz}
T.~Higaki and R.~Kitano, ``{On Supersymmetric Effective Theories of Axion},''
  {\em Phys.Rev.}, vol.~D86, p.~075027, 2012.

\bibitem{Arias:2012az}
P.~Arias, D.~Cadamuro, M.~Goodsell, J.~Jaeckel, J.~Redondo, {\em et~al.},
  ``{WISPy Cold Dark Matter},'' {\em JCAP}, vol.~1206, p.~013, 2012.

\bibitem{Chatzistavrakidis:2012bb}
A.~Chatzistavrakidis, E.~Erfani, H.~P. Nilles, and I.~Zavala, ``{Axiology},''
  {\em JCAP}, vol.~1209, p.~006, 2012.

\bibitem{Bae:2013hma}
K.~J. Bae, H.~Baer, and E.~J. Chun, ``{Mixed axion/neutralino dark matter in
  the SUSY DFSZ axion model},'' {\em JCAP}, vol.~1312, p.~028, 2013.

\bibitem{Choi:2009jt}
K.-S. Choi, H.~P. Nilles, S.~Ramos-Sanchez, and P.~K. Vaudrevange,
  ``{Accions},'' {\em Phys.Lett.}, vol.~B675, pp.~381--386, 2009.

\bibitem{Cicoli:2012aq}
M.~Cicoli, J.~P. Conlon, and F.~Quevedo, ``{Dark Radiation in LARGE Volume
  Models},'' {\em Phys.Rev.}, vol.~D87, no.~4, p.~043520, 2013.

\bibitem{Higaki:2012ar}
T.~Higaki and F.~Takahashi, ``{Dark Radiation and Dark Matter in Large Volume
  Compactifications},'' {\em JHEP}, vol.~1211, p.~125, 2012.

\bibitem{Conlon:2013txa}
J.~P. Conlon and M.~C.~D. Marsh, ``{Searching for a 0.1-1 keV Cosmic Axion
  Background},'' {\em Phys.Rev.Lett.}, vol.~111, p.~151301, 2013.

\bibitem{Higaki:2013lra}
T.~Higaki, K.~Nakayama, and F.~Takahashi, ``{Moduli-Induced Axion Problem},''
  {\em JHEP}, vol.~1307, p.~005, 2013.

\bibitem{Angus:2013zfa}
S.~Angus, J.~P. Conlon, U.~Haisch, and A.~J. Powell, ``{Loop corrections to
  Delta $N_{eff}$ in large volume models},'' {\em JHEP}, vol.~1312, p.~061,
  2013.

\bibitem{Higaki:2013qka}
T.~Higaki, K.~Nakayama, and F.~Takahashi, ``{Cosmological constraints on
  axionic dark radiation from axion-photon conversion in the early Universe},''
  {\em JCAP}, vol.~1309, p.~030, 2013.

\bibitem{Higaki:2011me}
T.~Higaki and T.~Kobayashi, ``{Note on moduli stabilization, supersymmetry
  breaking and axiverse},'' {\em Phys.Rev.}, vol.~D84, p.~045021, 2011.

\bibitem{Gao:2013rra}
X.~Gao and P.~Shukla, ``{F-term Stabilization of Odd Axions in LARGE Volume
  Scenario},'' arXiv:1307.1141 [hep-th].

\bibitem{Berenstein:2012eg}
D.~Berenstein and E.~Perkins, ``{Open string axions and the flavor problem},''
  {\em Phys.Rev.}, vol.~D86, p.~026005, 2012.

\bibitem{Honecker:2004kb}
G.~Honecker and T.~Ott, ``{Getting just the supersymmetric standard model at
  intersecting branes on the Z(6) orientifold},'' {\em Phys.Rev.}, vol.~D70,
  p.~126010, 2004.

\bibitem{Honecker:2004np}
G.~Honecker, ``{Chiral N=1 4-D orientifolds with D-branes at angles},'' {\em
  Mod.Phys.Lett.}, vol.~A19, pp.~1863--1879, 2004.

\bibitem{Gmeiner:2009fb}
F.~Gmeiner and G.~Honecker, ``{Complete Gauge Threshold Corrections for
  Intersecting Fractional D6-Branes: The Z6 and Z6' Standard Models},'' {\em
  Nucl.Phys.}, vol.~B829, pp.~225--297, 2010.

\bibitem{Gmeiner:2008xq}
F.~Gmeiner and G.~Honecker, ``{Millions of Standard Models on Z-prime(6)?},''
  {\em JHEP}, vol.~0807, p.~052, 2008.

\bibitem{Honecker:2012jd}
G.~Honecker and J.~Vanhoof, ``{Yukawa couplings and masses of non-chiral states
  for the Standard Model on D6-branes on T6/Z6'},'' {\em JHEP}, vol.~1204,
  p.~085, 2012.

\bibitem{Honecker:2012fn}
G.~Honecker and J.~Vanhoof, ``{Towards the field theory of the Standard Model
  on fractional D6-branes on T6/Z6': Yukawa couplings and masses},'' {\em
  Fortsch.Phys.}, vol.~60, pp.~1050--1056, 2012.

\bibitem{Zavattini:2005tm}
E.~Zavattini {\em et~al.}, ``{Experimental observation of optical rotation
  generated in vacuum by a magnetic field},'' {\em Phys.Rev.Lett.}, vol.~96,
  p.~110406, 2006.

\bibitem{Ehret:2010mh}
K.~Ehret, M.~Frede, S.~Ghazaryan, M.~Hildebrandt, E.-A. Knabbe, {\em et~al.},
  ``{New ALPS Results on Hidden-Sector Lightweights},'' {\em Phys.Lett.},
  vol.~B689, pp.~149--155, 2010.

\bibitem{Barth:2013sma}
K.~Barth, A.~Belov, B.~Beltran, H.~BrŠuninger, J.~Carmona, {\em et~al.},
  ``{CAST constraints on the axion-electron coupling},'' {\em JCAP}, vol.~1305,
  p.~010, 2013.

\bibitem{Kim:1979if}
J.~E. Kim, ``{Weak Interaction Singlet and Strong CP Invariance},'' {\em
  Phys.Rev.Lett.}, vol.~43, p.~103, 1979.

\bibitem{Shifman:1979if}
M.~A. Shifman, A.~Vainshtein, and V.~I. Zakharov, ``{Can Confinement Ensure
  Natural CP Invariance of Strong Interactions?},'' {\em Nucl.Phys.},
  vol.~B166, p.~493, 1980.

\bibitem{Dine:1981rt}
M.~Dine, W.~Fischler, and M.~Srednicki, ``{A Simple Solution to the Strong CP
  Problem with a Harmless Axion},'' {\em Phys.Lett.}, vol.~B104, p.~199, 1981.

\bibitem{Zhitnitsky:1980tq}
A.~Zhitnitsky, ``{On Possible Suppression of the Axion Hadron Interactions. (In
  Russian)},'' {\em Sov.J.Nucl.Phys.}, vol.~31, p.~260, 1980.

\bibitem{Coriano':2005js}
C.~Coriano, N.~Irges, and E.~Kiritsis, ``{On the effective theory of low scale
  orientifold string vacua},'' {\em Nucl.Phys.}, vol.~B746, pp.~77--135, 2006.

\bibitem{Bardeen:1977bd}
W.~A. Bardeen and S.-H. Tye, ``{Current Algebra Applied to Properties of the
  Light Higgs Boson},'' {\em Phys.Lett.}, vol.~B74, p.~229, 1978.

\bibitem{Kim:1983dt}
J.~E. Kim and H.~P. Nilles, ``{The mu Problem and the Strong CP Problem},''
  {\em Phys.Lett.}, vol.~B138, p.~150, 1984.

\bibitem{Cerdeno:1998hs}
D.~Cerdeno and C.~Munoz, ``{An introduction to supergravity},'' {\em PoS},
  vol.~CORFU98, p.~011, 1998.

\bibitem{Kaplunovsky:1993rd}
V.~S. Kaplunovsky and J.~Louis, ``{Model independent analysis of soft terms in
  effective supergravity and in string theory},'' {\em Phys.Lett.}, vol.~B306,
  pp.~269--275, 1993.

\bibitem{Brignole:1997dp}
A.~Brignole, L.~E. Ibanez, and C.~Munoz, ``{Soft supersymmetry breaking terms
  from supergravity and superstring models},'' [hep-ph/9707209].

\bibitem{Ferrara:1982qs}
S.~Ferrara, L.~Girardello, and H.~P. Nilles, ``{Breakdown of Local
  Supersymmetry Through Gauge Fermion Condensates},'' {\em Phys.Lett.},
  vol.~B125, p.~457, 1983.

\bibitem{Nilles:1983ge}
H.~P. Nilles, ``{Supersymmetry, Supergravity and Particle Physics},'' {\em
  Phys.Rept.}, vol.~110, pp.~1--162, 1984.

\bibitem{Svrcek:2006yi}
P.~Svrcek and E.~Witten, ``{Axions In String Theory},'' {\em JHEP}, vol.~0606,
  p.~051, 2006.

\bibitem{Grimm:2004ua}
T.~W. Grimm and J.~Louis, ``{The Effective action of type IIA Calabi-Yau
  orientifolds},'' {\em Nucl.Phys.}, vol.~B718, pp.~153--202, 2005.

\bibitem{Conlon:2006tq}
J.~P. Conlon, ``{The QCD axion and moduli stabilisation},'' {\em JHEP},
  vol.~0605, p.~078, 2006.

\bibitem{Cicoli:2012sz}
M.~Cicoli, M.~Goodsell, and A.~Ringwald, ``{The type IIB string axiverse and
  its low-energy phenomenology},'' {\em JHEP}, vol.~1210, p.~146, 2012.

\bibitem{BerasaluceGonzalez:2011wy}
M.~Berasaluce-Gonzalez, L.~E. Ib\'{a}\~{n}ez, P.~Soler, and A.~M. Uranga,
  ``{Discrete gauge symmetries in D-brane models},'' {\em JHEP}, vol.~1112,
  p.~113, 2011.

\bibitem{Anastasopoulos:2012zu}
P.~Anastasopoulos, M.~Cveti\v{c}, R.~Richter, and P.~K. Vaudrevange, ``{String
  Constraints on Discrete Symmetries in MSSM Type II Quivers},'' {\em JHEP},
  vol.~1303, p.~011, 2013.

\bibitem{Ibanez:2012wg}
L.~Ib\'{a}\~{n}ez, A.~Schellekens, and A.~Uranga, ``{Discrete Gauge Symmetries
  in Discrete MSSM-like Orientifolds},'' {\em Nucl.Phys.}, vol.~B865,
  pp.~509--540, 2012.

\bibitem{Honecker:2013hda}
G.~Honecker and W.~Staessens, ``{To Tilt or Not To Tilt: Discrete Gauge
  Symmetries in Global Intersecting D-Brane Models},'' {\em JHEP}, vol.~1310,
  p.~146, 2013.

\bibitem{Honecker:2013kda}
G.~Honecker and W.~Staessens, ``{D6-Brane Model Building and Discrete
  Symmetries on $T^ 6/(Z_2 \times Z_6 \times \Omega{\cal R})$ with Discrete
  Torsion},'' {\em PoS}, vol.~Corfu2012, p.~107, 2013.

\bibitem{Camara:2011jg}
P.~G. Camara, L.~E. Ib\'{a}\~{n}ez, and F.~Marchesano, ``{RR photons},'' {\em
  JHEP}, vol.~1109, p.~110, 2011.

\bibitem{HoneckerStaessens:2014}
G.~Honecker and W.~Staessens, ``{work in progress},''

\bibitem{Berenstein:2010ta}
D.~Berenstein and E.~Perkins, ``{A viable axion from gauged flavor
  symmetries},'' {\em Phys.Rev.}, vol.~D82, p.~107701, 2010.

\bibitem{Brizi:2009nn}
L.~Brizi, M.~Gomez-Reino, and C.~A. Scrucca, ``{Globally and locally
  supersymmetric effective theories for light fields},'' {\em Nucl.Phys.},
  vol.~B820, pp.~193--212, 2009.

\bibitem{Honecker:2011sm}
G.~Honecker, ``{Kaehler metrics and gauge kinetic functions for intersecting
  D6-branes on toroidal orbifolds - The complete perturbative story},'' {\em
  Fortsch.Phys.}, vol.~60, pp.~243--326, 2012.

\bibitem{Honecker:2012qr}
G.~Honecker, M.~Ripka, and W.~Staessens, ``{The Importance of Being Rigid:
  D6-Brane Model Building on $T^6/Z_2 x Z_6'$ with Discrete Torsion},'' {\em
  Nucl.Phys.}, vol.~B868, pp.~156--222, 2013.

\bibitem{Forste:2010gw}
S.~F{\"o}rste and G.~Honecker, ``{Rigid D6-branes on $T^6/(Z_2 x Z_{2M} x
  \Omega R)$ with discrete torsion},'' {\em JHEP}, vol.~1101, p.~091, 2011.

\bibitem{BlaszczykHoneckerKoltermann:2014}
M.~Blasczcyk, G.~Honecker, and I.~Koltermann, ``{in preparation},''

\bibitem{Kim:1986ax}
J.~E. Kim, ``{Light Pseudoscalars, Particle Physics and Cosmology},'' {\em
  Phys.Rept.}, vol.~150, pp.~1--177, 1987.

\bibitem{Kim:2008hd}
J.~E. Kim and G.~Carosi, ``{Axions and the Strong CP Problem},'' {\em
  Rev.Mod.Phys.}, vol.~82, pp.~557--602, 2010.

\bibitem{Georgi:1986df}
H.~Georgi, D.~B. Kaplan, and L.~Randall, ``{Manifesting the Invisible Axion at
  Low-energies},'' {\em Phys.Lett.}, vol.~B169, p.~73, 1986.

\bibitem{Srednicki:1985xd}
M.~Srednicki, ``{Axion Couplings to Matter. 1. CP Conserving Parts},'' {\em
  Nucl.Phys.}, vol.~B260, p.~689, 1985.

\bibitem{Kuzmin:2002nt}
S.~Kuzmin and D.~McKeon, ``{The Supersymmetric Stuckelberg mass and overcoming
  the Fayet-Iliopoulos mechanism for breaking symmetry},'' {\em
  Mod.Phys.Lett.}, vol.~A17, pp.~2605--2610, 2002.

\bibitem{Kors:2004ri}
B.~K{\"o}rs and P.~Nath, ``{A Supersymmetric Stueckelberg U(1) extension of the
  MSSM},'' {\em JHEP}, vol.~0412, p.~005, 2004.

\end{thebibliography}

\end{document}